\documentclass[a4paper,11pt]{article}
\usepackage{graphicx}
\pdfoutput=1 % if your are submitting a pdflatex (i.e. if you have
\usepackage{mathrsfs}
\usepackage[T1]{fontenc} % if needed
\usepackage{bbold}
\usepackage{url}
\usepackage[english]{babel}
\usepackage[utf8]{inputenc}
\usepackage{amsmath}

\usepackage{amsfonts}
\usepackage{amssymb}
\usepackage{mathtools}
\usepackage{placeins}
\usepackage{float}
\usepackage{tabularx}
\usepackage{adjustbox}
\usepackage{caption}
\usepackage{subcaption}
\DeclareUnicodeCharacter{2212}{-}
\usepackage{jcappub} % for details on the use of the package, please
\title{\boldmath Constraining the quintessential \texorpdfstring{$\alpha$}{a}-attractor inflation through dynamical horizon exit method}
\author{Arunoday Sarkar,}%\note{Also at Some University.}}
\author{Buddhadeb Ghosh}
\affiliation{Centre of Advanced Studies, Department of Physics, The University of Burdwan,\\Burdwan 713 104, India}
\emailAdd{adsarkar@scholar.buruniv.ac.in}
\emailAdd{bghosh@phys.buruniv.ac.in}
\abstract{In the present paper, we perform a sub-Planckian quantum mode analysis of linear cosmological perturbation in the inflaton field over a classical quasi de-Siter metric background by dynamical horizon exit (DHE) method. In this way, we probe the inflationary regime of a quintessential $\alpha$-attractor model by analysing the COBE/Planck normalized power spectra, spectral indices, tensor to scalar ratio, number of e-folds, running of the spectral index and inflationary Hubble parameter in $k$-space. We compare our results with ordinary $\alpha$-attractor $E$ and $T$ models and with that of Planck-2018 results. Our estimated values of $n_s$ and $r$ lie within $68\%$ CL with respect to Planck data for $k=0.001 - 0.009$ Mpc$^{-1}$ for all values of $\alpha$. The $\alpha$ values, obtained in our calculations satisfy various post inflationary constraints regarding preheating and reheating, reported in current literature. We observe that quintessence sets an upper bound of $\alpha=4.3$ and thereby restricts the model from becoming of the power law type, making it more efficacious than ordinary $\alpha$-attractors in explaining both inflation and dark energy. A striking observation in our analyses is that, unlike in our previous study, we find a continuous values of $\alpha$ within $\frac{1}{10}\leq \alpha\leq 4.3$ for the specified $k$ range. At the end, we have shown that the model parameters constrained in this work give a very small vacuum density $\sim 10^{-117}-10^{-115} M_P^4$ which is an essential criterion for current and future dark energy observations of the universe.}

\begin{document} 
\maketitle
\flushbottom

%\tableofcontents

\section{Introduction}
\label{sec:intro}
Cosmological observations \cite{SupernovaSearchTeam:1998fmf,SupernovaCosmologyProject:1998vns} from the redshifts of distant type Ia supernovae and the analysis of cosmic microwave background (CMB) anisotropies and polarizations \cite{WMAP:2012nax,Planck:2015fie,Planck:2018vyg,BICEP2:2015xme} reveal that our universe is currently in a state of late time accelerating expansion, in contrast to the early inflationary one. It is widely believed that an unknown form of energy, called the `\textit{dark energy}' (DE) \cite{Huterer:1998qv} occupying \cite{Planck:2018vyg} approximately $69\%$ of the density budget is responsible for that. A usual way of understanding this exotic energy is to introduce a non-zero positive cosmological constant ($\Lambda$) in $\Lambda$-cold dark matter ($\Lambda$CDM) model\footnote{also called `concordance model'.}. But regarding its extreme fine tuning \textit{viz.,} $\Lambda\sim 10^{-120}$ (in the unit of Planck mass) there are few radical inconsistencies with general relativity (GR) and quantum gravity (QG)\footnote{In fact a positive $\Lambda$ gives rise to a disagreement with the swampland conjecture \cite{Vafa:2005ui,Obied:2018sgi,Agrawal:2018own,Roupec:2018mbn,Akrami:2018ylq,Kehagias:2018uem} in string theory program so far as obtaining a de-Sitter vacuum is concerned.}\cite{Dimopoulos:2022wzo}. Towards realizing the solution of this `\textit{cosmological constant problem}', some early attempts find that at very large scale of our universe, either GR should incorporate the DE \cite{Copeland:2006wr} or it must be modified in order to be compatible with the observed expansion \cite{Clifton:2011jh}.\par 
Now, the first alternative replaces $\Lambda$ with a dynamical scalar field - the `\textit{quintessence}'\cite{Peebles:2002gy,Ratra:1987rm,Caldwell:1997ii}. Just like the slow-roll dynamics of the inflaton field, it can drive the accelerating expansion if its potential energy dominates over the kinetic energy significantly. But fundamentally this idea of quintessence\footnote{Specifically this type of quintessence is called `thawing quintessence' \cite{Dimopoulos:2022wzo}.} is plagued by the `\textit{coincidence problem}' in the context of its initial conditions \cite{Dimopoulos:2021xld,Dimopoulos:2022wzo}. In a nutshell, this quintessence field should have well defined initial conditions quite similar to the inflationary attractors or it should have a coherent connection with the inflaton field. The later possibility appears to be more convincing in various ways \cite{Peebles:1998qn,Peloso:1999dm,Sen:2000ym,Kaganovich:2000fc,Yahiro:2001uh,Martin:2004ba,Barenboim:2005np,Rosenfeld:2005mt,Cardenas:2006py,BuenoSanchez:2006fhh,Membiela:2006rj,Rosenfeld:2006hs,Neupane:2007mu,Bastero-Gil:2009wdy,Piedipalumbo:2011bj,Wetterich:2014gaa,Hossain:2014xha,Hossain:2014coa,Hossain:2014ova,Geng:2015fla,WaliHossain:2014usl,Haro:2015ljc,deHaro:2016ftq,deHaro:2016cdm,Guendelman:2016kwj,Rubio:2017gty,Ahmad:2017itq,Haro:2018zdb,Bettoni:2018pbl,Selvaganapathy:2019bpm,Lima:2019yyv,Kleidis:2019ywv,Haro:2019peq,Benisty:2020xqm,Benisty:2020vvm,deHaro:2021swo,Dimopoulos:2021xld,Tian:2021cqq,AresteSalo:2021wgb,Akrami:2020zxw,Garcia-Garcia:2019ees,Garcia-Garcia:2018hlc,Akrami:2017cir,Kepuladze:2021tsb,Dimopoulos:2017zvq,Geng:2017mic,Agarwal:2017wxo,AresteSalo:2017lkv,DeHaro:2017abf,Haro:2019gsv,deHaro:2019oki,Benisty:2020qta,Shokri:2021zqw,Bettoni:2021qfs,Jaman:2022bho,Jesus:2021bxq,Fujikura:2022udt,Karciauskas:2021fdu,Basak:2021cgk,AresteSalo:2020yxl} which combine inflation and quintessence within a single unified framework - called the `\textit{quintessential inflation}'. In this model, the inflaton field plays a dual role corresponding to early (inflationary) and late time (quintessence/DE) expansions of the universe. The only difference is, unlike the inflation, the quintessential inflaton field does not oscillate at the bottom of the potential, rather it undergoes through a flat tail like part of the potential followed by the kination period \cite{Spokoiny:1993kt,Pallis:2005hm,Pallis:2005bb,Gomez:2008js} surviving until today to become DE. The post inflationary particle production takes place via different mechanisms \textit{viz.,} instant preheating \cite{Campos:2002yk,Dimopoulos:2017tud}, curvaton reheating \cite{Feng:2002nb,BuenoSanchez:2007jxm,Matsuda:2007ax}, gravitational reheating \cite{Chun:2009yu}, Ricci reheating \cite{Dimopoulos:2018wfg,Opferkuch:2019zbd} and warm quintessential inflation \cite{Dimopoulos:2019gpz,Rosa:2019jci,Gangopadhyay:2020bxn}.\par Recently, quintessential forms of $\alpha$-attractors have been found to be quite suitable for DE observations as well as for inflation. The relevant cosmological parameters, kination period and gravitational waves in these models have been discussed in \cite{Dimopoulos:2017zvq}. Various interacting and non-interacting single and multi-field models are studied in \cite{Akrami:2017cir,Garcia-Garcia:2018hlc,Garcia-Garcia:2019ees,Akrami:2020zxw} for CMB and stage-IV large scale structure (LSS) surveys. The associated preheating \cite{Dimopoulos:2017tud} and reheating \cite{AresteSalo:2021lmp,Salo:2021vdv}, and their subsequent modifications have been analysed in current literature \cite{Salo:2021vdv,Salo:2021piz,deHaro:2022vxc,deHaro:2022ukj}.
\par In an earlier work \cite{Sarkar:2021ird} we performed a microscopic mode analysis of linear cosmological perturbations by the novel dynamical horizon exit (DHE) method of $\alpha$-attractor $E$ and $T$ models in the classical quasi de-Sitter background to constrain the exponent $n$ and the multi-functional parameter $\alpha$. The self-consistent calculations reveal that the calculated cosmological parameters agree with Planck-2018 data very well in the momentum range $0.001$ Mpc$^{-1}$ - $0.009$ Mpc$^{-1}$. Our results in the $k\rightarrow 0$ limit match with the Planck data in the $n_s$ - $r$ parametric plane within $68\%$ CL for $\alpha = 1,5,10$ of $E$ model and $\alpha= 1,6,10$ of $T$ model and that for $\alpha=15$ of both $E$ and $T$ models within $95\%$ CL for $n=2$. Similar results are obtained for $n=4$ of $E$ model for $\alpha=1,6,11$ and that of $T$ model for $\alpha=1,4,9$ within $68\%$ CL. The observations of discrete transitions in $\alpha$-attractors from low to high values of $\alpha$ are important results in our analysis\footnote{In fact we first observed the double-pole behaviour of $\alpha$-attractors in $k$-space through a sub-Planckian quantum mode analysis.}. In this respect, our proposed formalism provide a new way of looking microscopically the efficacy of the $\alpha$-attractors in explaining the cosmological predictions regarding inflation, CMB $B$ mode and LSS data.\par Now, so far as the current and future DE observations \cite{Garcia-Garcia:2019ees,Akrami:2017cir,Garcia-Garcia:2018hlc,Akrami:2020zxw} are concerned, it will be worthwhile to look again into these aspects in the quintessential versions of $\alpha$-attractors in conjunction with their parameter estimations. In these models the inflaton field plays the roles for inflation and quintessence in a unified way so that it can fix the appropriate initial conditions for the late time expansion of the universe. Therefore it is quite understandable that in these models the momentum-dependent cosmological parameters should carry the imprints of the dynamical quintessence in contrast to the $\alpha$-attractors without quintessence. Hence, a systematic analysis of such modifications can be helpful in understanding the large scale dynamics of our universe. This picture has recently been rekindled in the context of Hubble tension (see \cite{Brissenden:2023yko} as for example). After all, a suitable inflaton potential should accommodate all possible microscopic elements to satisfy today's observational bounds including dark energy.\par Motivated by these ideas, in the present paper we continue our investigations towards perturbative mode analysis in $k$-space involving a specific model of quintessential $\alpha$-attractors. Our examining process mainly center around obtaining the solutions of three coupled non-linear differential equations developed in \cite{Sarkar:2021ird} for the inflaton field, its first order perturbation and the Bardeen potential as metric perturbation in sub-Planckian $k$ limit within the inflationary regime of the concerned model. These solutions are utilised to figure out the mode-dependent scalar and tensor power spectra, spectral indices, running of the spectral index, tensor to scalar ratio, number of e-folds and inflationary Hubble parameter. \par The present paper is organised as follows. In section \ref{sec:our model} we describe our model in details with its different regions of interests and their variations with the model parameters. Section \ref{sec:formalism} briefly describes the DHE method and the setting up of the $k$ space evolution equations. The numerical and graphical analyses of our calculations are discussed in section \ref{sec: result}. In this section we also compare our results with that of $\alpha$-attractor $E$ and $T$ models, highlight the new outcomes and show a calculation of amplitude of the small DE-potential part (which is the vaccum density today) of our model with the help of parameters estimated in section \ref{sec: result}. Finally we conclude our study in section \ref{sec:conclusion} by elucidating the main results and some future prospects in the light of ongoing and forthcoming cosmological surveys. 
\section{Description of the model}
\label{sec:our model}
We consider a simple Lagrangian\footnote{Some authors like in \cite{Dimopoulos:2017zvq,Dimopoulos:2017tud} prefer to keep a `$-1$' in the potential part of the Lagrangian in order to make the vacuum density zero due to some unknown symmetry. On the other hand, some literature \cite{AresteSalo:2021wgb,Guendelman:2002js,Guendelman:2014bva} show that the omission of this `$-1$' facilitates to highlight a non-zero vacuum density in the asymptotically flat region of the effective potential dynamically generated by the quintessence and can lead to a more successful model of the universe. We agree to this latter viewpoint which says that a non-vanishing vacuum density is required to explain the current accelerating expansion of the universe. That is why, we do not consider this numerical factor in our Lagrangian.},
\begin{equation}
    \mathcal{L}_{\mathrm{noncan}}(\chi, \partial\chi)=\left[1-\frac{\chi^2}{(\sqrt{6\alpha})^2}\right]^{-2}\frac{\partial_\mu\chi\partial^\mu\chi}{2}M^2_P-V_0 e^{-n}\left[\exp\left[n\left(1-\frac{\chi}{\sqrt{6\alpha}}\right)\right]\right]
    \label{eq:Eq1}
\end{equation}
of the non-canonical inflaton field $\chi$ measured in the unit of reduced Planck mass $M_P = \frac{1}{\sqrt{8\pi G}} = 2.43\times 10^{18}$ GeV. $\alpha$, $n$ are dimensionless positive constants and $V_0$ is a fixed scale of the inflationary energy density. The celebrated $\alpha$ parameter originating from the $\alpha$-attractors \cite{Kallosh:2013yoa,Ferrara:2013rsa,Kallosh:2013pby,Kallosh:2013maa,Kallosh:2015lwa,Carrasco:2015pla,Kallosh:2013hoa,Maeda:2018sje,Kallosh:2014rga,Galante:2014ifa,Kallosh:2013tua,Ferrara:2013kca} is related to the inverse curvature of the $SU(1,1)/U(1)$ K\"{a}hler manifold
\begin{equation}
    \mathcal{R}_K = -\frac{2}{3\alpha}
    \label{eq:Eq2}
\end{equation}
in $\mathcal{N}=1$ minimal supergravity involving chiral inflaton multiplet. Determining the precise values of $\alpha$ has profound implications in understanding various theoretical phenomena like hyperbolic half-plane disk geometry \cite{Carrasco:2015rva,Kallosh:2015zsa,Carrasco:2015uma}, maximal supersymmetry \cite{Kallosh:2017ced}, modified gravity \cite{Odintsov:2016vzz}, string theory \cite{Scalisi:2018eaz,Kallosh:2017wku,Kallosh:2021vcf} \textit{etc.} and also several experimental aspects like tensor to scalar ratio ($r$) \cite{Kallosh:2019hzo}, $B$ mode targets \cite{Kallosh:2019eeu,Ferrara:2016fwe}, potential reconstruction \cite{Planck:2018jri} and many other CMB constraints \cite{Planck:2018vyg}. (See \cite{Sarkar:2021ird} for more details.) In this work we explore the regimes of operation of the exponent $n$ and the parameter $\alpha$ while the $\alpha$-attractor is equipped with the element of quintessence.\par Parameter $\alpha$, present in the Lagrangian of Eq. (\ref{eq:Eq1}) has another valuable role to play. The quadratic pole structure in the kinetic part restricts $\chi$ within $\chi = \pm \sqrt{6\alpha}$ making the field excursion sub-Planckian for which no issues regarding the super-Planckian field effects (like the `\textit{fifth force problem}' \cite{Wetterich:2004ff,Dimopoulos:2017zvq,Dimopoulos:2022wzo}) and radiative corrections \cite{Linde:2017pwt,Kallosh:2016gqp} arise. Such type of advantage is in fact a generic feature of $\alpha$-attractors \cite{Chojnacki:2021fag,Kallosh:2016gqp}. A striking property is that almost all the models derived from minimal supergravity theory consist of kinetic poles, for which they are often referred as `pole inflation' \cite{Pallis:2022cnm,Karamitsos:2019vor,Dias:2018pgj,Saikawa:2017wkg}. These poles make the theory safe from the issues arising from infinite field excursion. Interestingly such poles are not generally found in the potentials motivated by string theory\footnote{The precise structure of $\alpha$-attractors arises from the poles in the K\"{a}hler potential of $\mathcal{N}=1$ supergravity theory. But no such poles are found in the K\"{a}hler potential in type IIB string theory. In \cite{Let:2022fmu} we have encountered this aspect in deriving a slow-roll inflaton potential from string compactification scenario. In such models singularities of orbifold types are smoothed by the implication of large volume scenario (LVS) to incorporate required quantum corrections on an orientifold. This may be a possible reason of absence of poles in string-motivated potentials.}, for which trans-Plackian problems (such as Transplanckian Censorship Conjecture (TCC)) \cite{Bedroya:2019tba,Brandenberger:2021pzy,Cai:2019dzj} arise. On the other hand, the TCC is a byproduct of swampland distance conjecture \cite{Brahma:2019vpl,Bedroya:2019snp,Andriot:2020lea,Saito:2019tkc,Draper:2019utz} which states that de-Sitter inflation ($\Lambda>0$) is inconsistent with string theory \cite{Obied:2018sgi,Dasgupta:2019gcd,Garg:2018reu}. So these issues of obtaining a successful model for inflation from stringy framework is not yet settled. Interestingly, various quantum corrections of perturbative and non-perturbative origins have recently paved the way \cite{Let:2022fmu} to a possible realization  of stringy  inflation (see also \cite{Dasgupta:2018rtp} for related discussions).\par  
Now we choose a canonical field 
\begin{equation}
    \psi = -\sqrt{6\alpha} M_P \tanh^{-1}{\left(\frac{\chi}{\sqrt{6\alpha}}\right)}
    \label{eq:Eq3}
\end{equation}
to get the following canonical form of the Lagrangian
\begin{equation}
    \mathcal{L}_{\mathrm{can}}(\psi,\partial\psi)=\frac{1}{2}\partial_\mu\psi\partial^\mu\psi - V(\psi)
    \label{eq:Eq4}
\end{equation}
where
\begin{equation}
    V(\psi)= V_0 e^{-n}\left[\exp\left[n\left(1+\tanh{\frac{\psi}{\sqrt{6\alpha}M_P}}\right)\right]\right]
\end{equation} is the required quintessential inflaton potential. In Eq. (\ref{eq:Eq3}), we have taken unconventionally, a negative sign to choose the inflationary regime on the positive side and quintessential part on the negative side of the field $\psi$. Such consideration will be helpful for our later calculations. We can make the potential more convenient in form by choosing a new scale $M=(e^n V_0)^{1/4}$ as
\begin{equation}
    V(\psi)=e^{-2n}M^4\left[\exp\left[n\left(1+\tanh{\frac{\psi}{\sqrt{6\alpha}M_P}}\right)\right]\right].
    \label{eq:Eq6}
\end{equation}
\begin{figure}[H]
	\centering
	\includegraphics[width=0.8\linewidth]{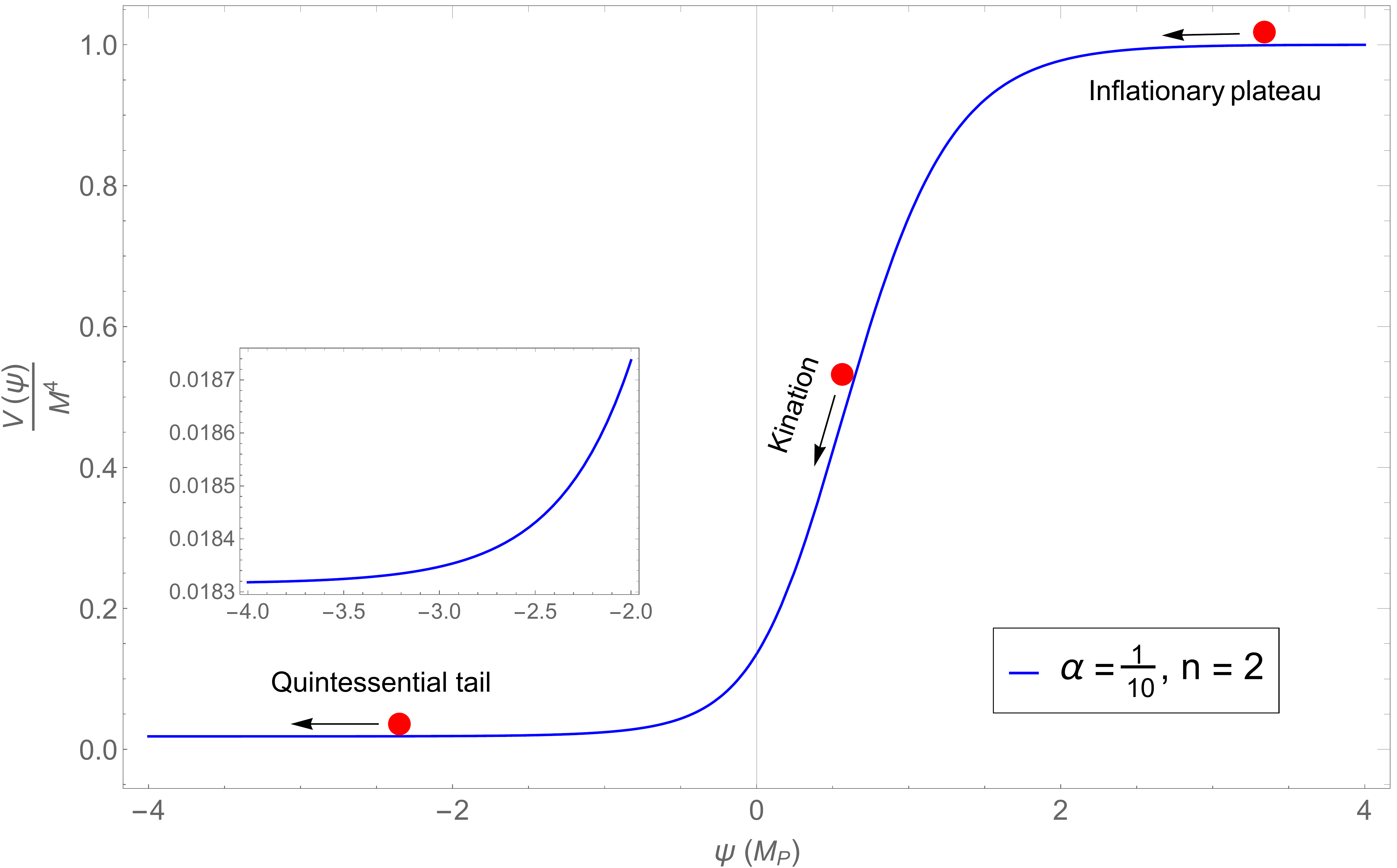}
	\caption{Schematic diagram of the quintessential inflaton potential for $\alpha=\frac{1}{10}$ and $n=2$ (as for example; exact values will be illustrated in section \ref{sec: result}.) normalised by $M^4$ plotted against $\psi$ in the unit of $M_P$. $\psi>0$ and $\psi<0$ regions correspond to the inflationary slow-roll plateau and quintessential flat tail respectively. The inflaton field (the red ball) rolls over the potential from the regimes of inflation to quintessence smoothly through the period of kination, where the field is allowed to free-fall. At the bottom of the potential the inflaton field is non-oscillatory and survives until today to become dark energy. Standard model particles are produced at the potential minima in the thermal bath of the hot big bang by non-standard ways.}
	\label{fig:Fig1}
\end{figure}
The potential in Eq. (\ref{eq:Eq6}), shown in Figure \ref{fig:Fig1}, has two plateaus corresponding to positive and negative values of $\psi$. $\psi>0$ region signifies the inflationary regime where the inflaton field condensate slowly rolls over the potential with minimal expense of kinetic energy (due to very large Hubble friction). The potential energy dominates over the kinetic energy until it reaches a cliff (shown in the middle of the Figure \ref{fig:Fig1}) where the inflation ends. Then it enters into a new phase, called \textit{kination} which is absent in the usual theories of inflation. Actually it acts like a linking bridge between inflation and quintessence. In this zone the dynamics of the field $\psi$ is governed by the equation of motion of a particle in free-fall. Because here the kinetic energy becomes surplus over the potential one, such that the Hubble friction can not resist. The total energy density $\rho_\psi$ falls rapidly with the scale factor $a$ ($\rho_\psi\propto a^{-6}$) \cite{Dimopoulos:2022wzo} such that the field remains frozen \cite{Dimopoulos:2021xld} until it reappears as quintessence in $\psi<0$ region. In this tail-like part, the potential energy again becomes dominant but then it is much much smaller ($\sim 10^{-12}$ GeV) than that of inflationary one ($\sim 10^{16}$ GeV). Such a huge drop of energy is an essential requirement for the model building of quintessential inflation. And today we observe this faint energy density as dark energy, the driving force of the current accelerating expansion of the universe.\par Now the required initial conditions for quintessential field dynamics are already embedded during inflation via quantum fluctuations inside the Hubble horizon and at the same time the associated modes exit which take part in the late time expansion of the universe as classical density perturbations after horizon re-entry. Consequently the scalar field can smoothly traverse from inflation to quintessence through kination, overcoming the barrier of coincidence problem. These initial conditions are actually fixed by the shape factors of the inflaton potential \textit{viz.,} the exponent $n$ and the parameter $\alpha$ and are quantified by the estimations of the cosmological parameters. To explore these facets is the main theme of the present paper.
\begin{figure}[H]
    \begin{subfigure}{0.5\linewidth}
  \centering
   \includegraphics[width=70mm,height=55mm]{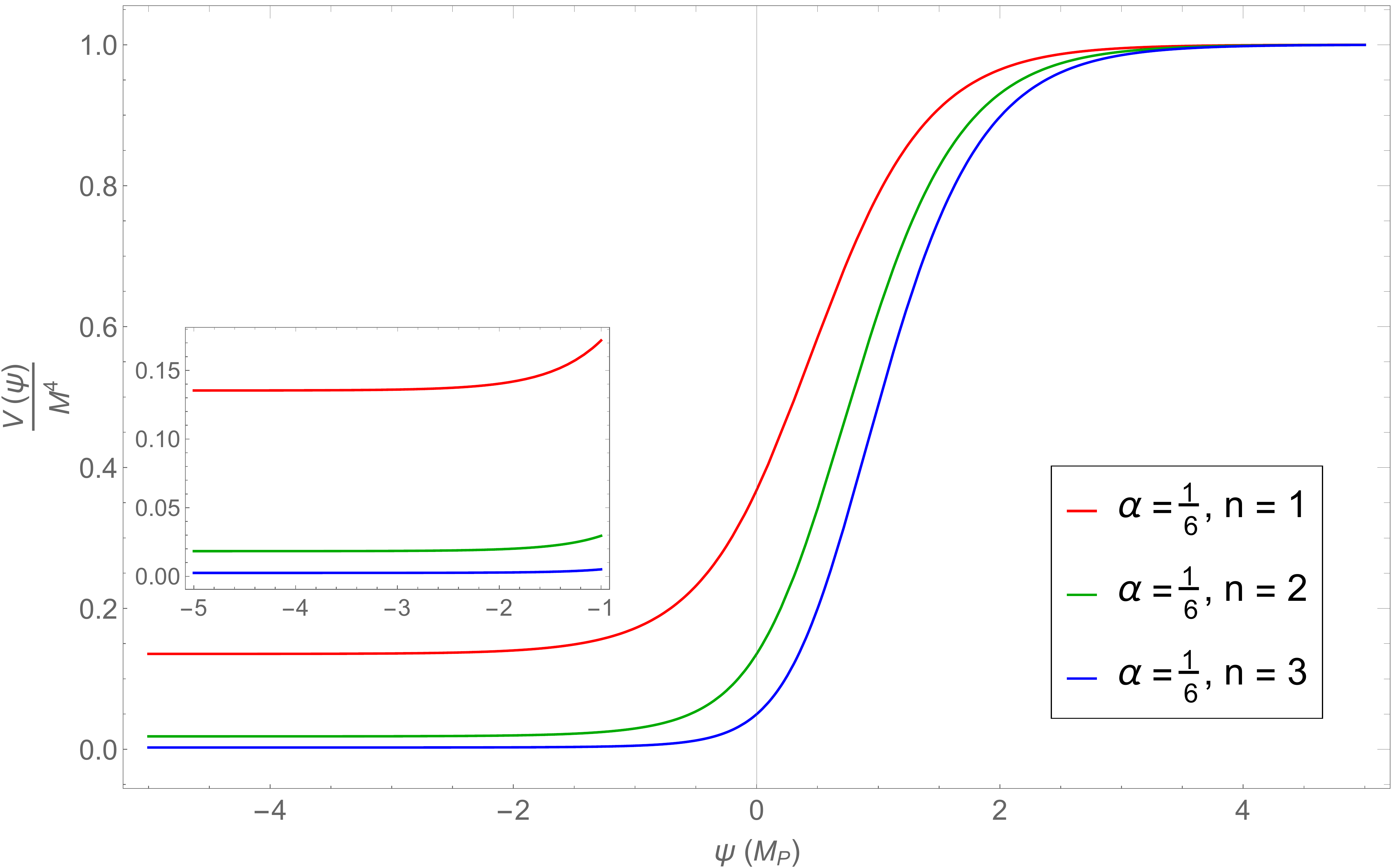}
   \subcaption{}
   \label{fig:Fig2a}
\end{subfigure}
\begin{subfigure}{0.5\linewidth}
  \centering
   \includegraphics[width=70mm,height=55mm]{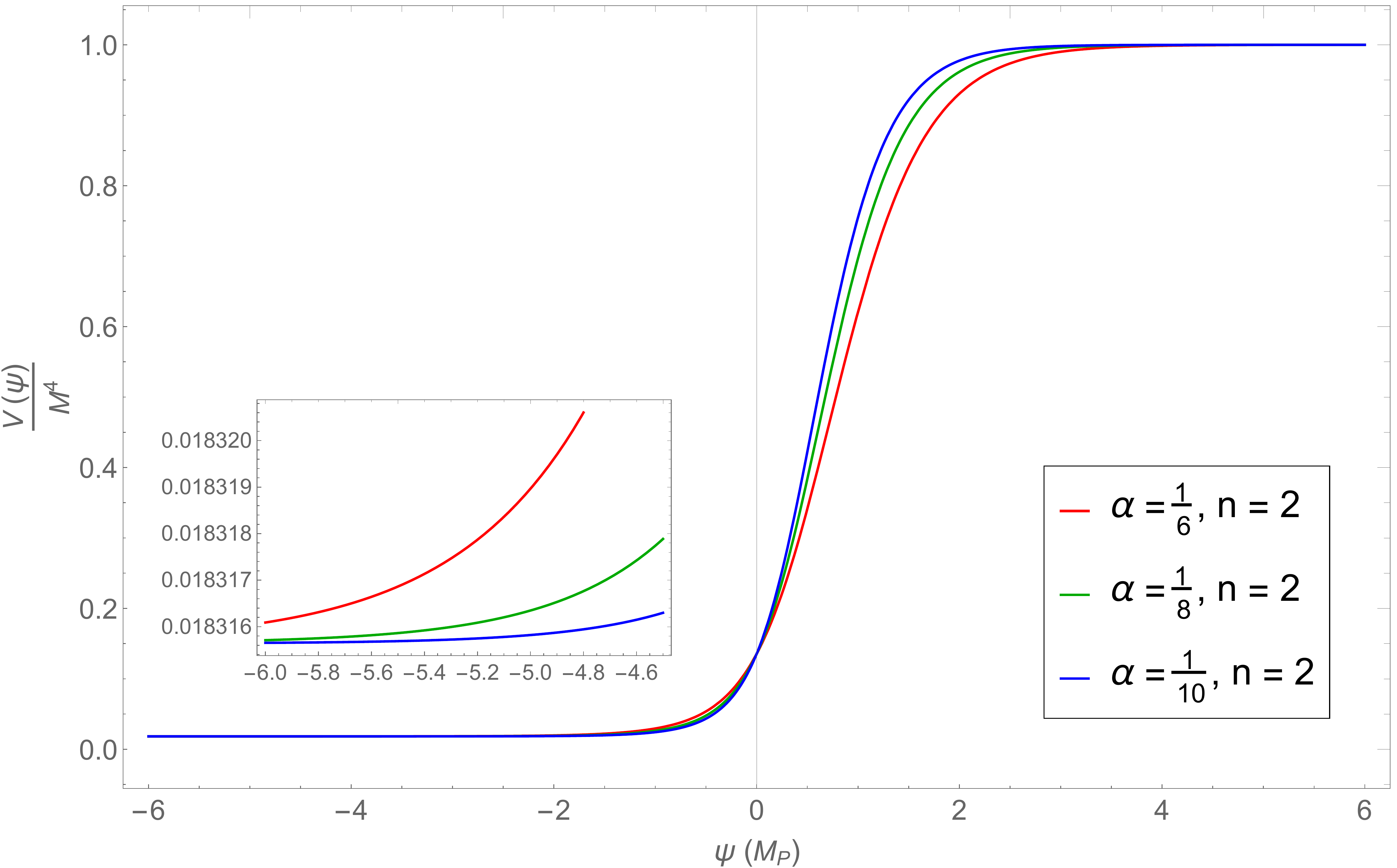}
   \subcaption{}
    \label{fig:Fig2b}
\end{subfigure}
     \caption{Changes in shape and energy scale of the potential with the variations of $n$ and $\alpha$. Figure (\ref{fig:Fig2a}) shows that the amplitude of the potential towards inflation remains almost fixed whereas that in the quintessential tail part varies with the increase in $n$ for a fixed value of $\alpha$ (here $\alpha=\frac{1}{6}$). Figure (\ref{fig:Fig2b}) shows that, as $\alpha$ increases for a fixed value of $n$ (here $n=2$) the potential loses its plateau form and starts to become polynomial type. This aspect will be more clear in the next figure.}
    \label{fig:Fig2}
\end{figure}
The potential in Eq. (\ref{eq:Eq6}) has two asymptotic limits indicating two plateaus (see Figure \ref{fig:Fig1}) corresponding to $\psi\rightarrow +\infty$ and $\psi\rightarrow -\infty$ as
\begin{enumerate}
    \item \begin{equation}
    \lim_{\psi\rightarrow +\infty}\tanh{\frac{\psi}{\sqrt{6\alpha}M_P}}=+1 \implies \lim_{\psi\rightarrow +\infty}V(\psi)=e^{-2n}M^4 e^{2n}=M^4
    \label{eq:limit1}
\end{equation}
\item \begin{equation}
    \lim_{\psi\rightarrow -\infty}\tanh{\frac{\psi}{\sqrt{6\alpha}M_P}}=-1 \implies \lim_{\psi\rightarrow -\infty}V(\psi)=e^{-2n}M^4=e^{-n}V_0 = V_{\Lambda}.
    \label{eq:limit2}
    \end{equation}
\end{enumerate}
The $M^4$ factor signifies the inflationary energy scale ($M\sim 10^{16}$ GeV) determined by the COBE/Planck normalization of the amplitude of scalar perturbation. And the $V_{\Lambda}$ is the vacuum density of the present universe which is $\sim 10^{-120}$ $M_P^4$ or $ {V_{\Lambda}}^{1/4}\sim 10^{-12}$ GeV according to Planck-2018 \cite{Planck:2018vyg,Planck:2018jri}. This vacuum energy density acts as the dark energy in the present universe.\par
Figures (\ref{fig:Fig2a}) and (\ref{fig:Fig2b}) depict the dependencies of the normalized potential with the parameters $n$ and $\alpha$ respectively. When $n$ increases from $n=1$ to $n=3$ the inflationary plateau described by the limit in Eq. (\ref{eq:limit1}) being $n$-independent remains same, while the quintessential tail part dictated by the limit in Eq. (\ref{eq:limit2}) proportional to $e^{-2n}$ varies with $n$.  Now when $n$ is fixed in figure (\ref{fig:Fig2b}), the plateaus remain almost constant in their respective values (here $e^{-4}\approx0.018$ and $1$), the only change appears with the increase in $\alpha$. The potential slowly becomes distorted towards a simple chaotic one which we shall demonstrate more clearly in the next figure.
\begin{figure}[H]
	\centering
	\includegraphics[width=0.8\linewidth]{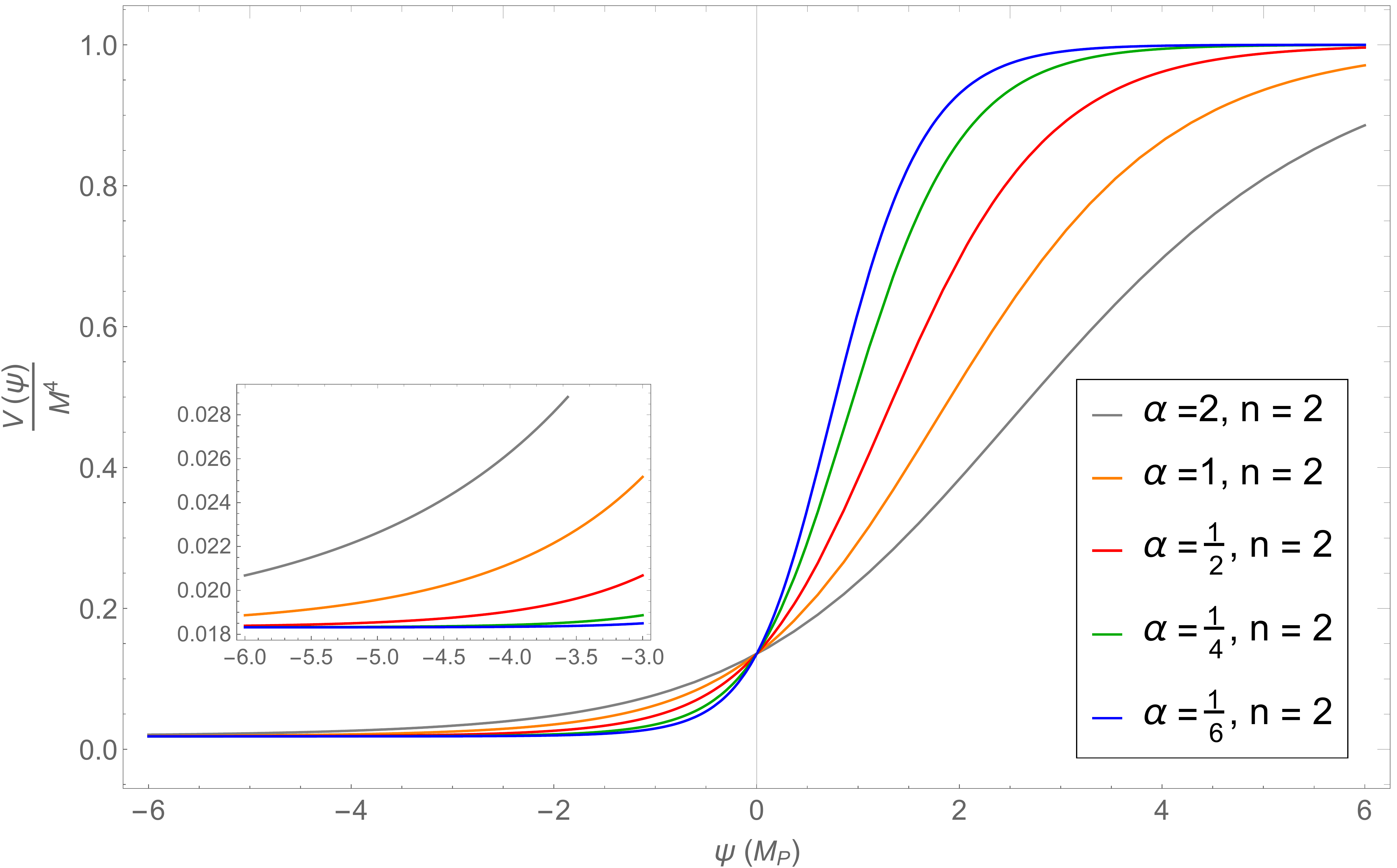}
	\caption{Variation of the potential with $\alpha$ for a fixed value of $n$.}
	\label{fig:Fig3}
\end{figure}
Figure \ref{fig:Fig3} shows how the potential becomes polynomial type as we increase the values of $\alpha$ from $\frac{1}{6}$ to $2$ keeping $n$ fixed at $2$. From figures (\ref{fig:Fig2b}) and \ref{fig:Fig3} it is understandable that the potential is almost identical as long as $\alpha$ remains within $\frac{1}{6}$ and beyond that it is like a simple power law potential and after $\alpha =1$ it drastically changes its nature. This behaviour is quite similar to that of the original $\alpha$-attractors, where, $\alpha\rightarrow 0$ and $\alpha\rightarrow \infty$ are two attractor points. The only difference is that, in the case of $\alpha$-attractors the transition takes place after $\alpha =1$ and, here, in its quintessential version, it starts after $\alpha =\frac{1}{6}$. That shows, that the fractional values of $\alpha$ are more favourable than the integer ones in case of quintessential $\alpha$-attractors.\footnote{Such fractional $\alpha$ values are also supported by $\mathcal{N}=8$ supergravity \cite{Kallosh:2017ced}.} We shall verify these aspects in Section \ref{sec: result} through $k$-space analyses.\par 
Now, for the sake of simplicity we would like to extract the inflationary regime by approximating the potential in large $\psi$ limit ($\psi\gg 0$). Then from Eq. (\ref{eq:Eq6}) we get,
\begin{equation}
    \tanh{\left(\frac{\psi}{\sqrt{6\alpha}M_P}\right)}\approx 1-2e^{-\frac{2\psi}{\sqrt{6\alpha}M_P}},
\end{equation}
\begin{equation}
    \exp\left[n\left(1+\tanh{\frac{\psi}{\sqrt{6\alpha}M_P}}\right)\right]\approx \exp\left[2n\left(1-e^{-\frac{2\psi}{\sqrt{6\alpha}M_P}}\right)\right]\approx e^{2n}\left(1-2ne^{-\frac{2\psi}{\sqrt{6\alpha}M_P}}\right).
\end{equation} and 
\begin{equation}
    V(\psi)\approx M^4\left(1-2ne^{-\frac{2\psi}{\sqrt{6\alpha}M_P}}\right)=M^4\left(1-2ne^{-\sqrt{\frac{2}{3\alpha}}\frac{\psi}{M_P}}\right)=V_{\mathrm{inf}}(\psi).
    \label{eq:Eq7}
\end{equation}
Similarly in $\psi\ll 0$ limit we can obtain the quintessential portion by 
\begin{equation}
    \tanh{\left(\frac{\psi}{\sqrt{6\alpha}M_P}\right)}\approx -1+2e^{\frac{2\psi}{\sqrt{6\alpha}M_P}},
\end{equation}
\begin{equation}
    \exp\left[n\left(1+\tanh{\frac{\psi}{\sqrt{6\alpha}M_P}}\right)\right]\approx \exp\left[2n\left(e^{\frac{2\psi}{\sqrt{6\alpha}M_P}}\right)\right]\approx \left(1+2ne^{\frac{2\psi}{\sqrt{6\alpha}M_P}}\right)=\left(1+2n e^{\sqrt{\frac{2}{3\alpha}}\frac{\psi}{M_P}}\right)
\end{equation}
as
\begin{equation}
    V(\psi)\approx e^{-2n}M^4\left(1+2n e^{\sqrt{\frac{2}{3\alpha}}\frac{\psi}{M_P}}\right)=V_{\Lambda}\left(1+2n e^{\sqrt{\frac{2}{3\alpha}}\frac{\psi}{M_P}}\right) =V_{\mathrm{Quintessence}}(\psi).
    \label{eq:Eq8}
\end{equation}
Eqs. (\ref{eq:Eq7}) and (\ref{eq:Eq8}) describe that on two extreme limits of $\psi$ ($\psi\rightarrow +\infty$ and $\psi\rightarrow -\infty$), the inflationary regime saturates at $M^4$ while the quintessential tail dies down to $V_{\Lambda}$ asymptotically, respectively.
 The extrapolation of $\psi$ towards $\psi=\pm\infty$ is not threatening due the non-canonical poles (at $\pm\sqrt{6\alpha}$) in the Lagrangian of Eq. (\ref{eq:Eq1}) described above. On switching from non-canonical to canonical field the poles are stretched to infinity giving rise to the inflationary plateau and the quintessential tail. The inflaton field $\psi$ therefore remains sub-Planckian, which is an indispensable criterion for our analysis described in the next section. 
\section{Formalism}
\label{sec:formalism}
In this section we briefly recapitulate the calculational framework developed in \cite{Sarkar:2021ird} and explain the required cosmological parameters within the model described earlier. Our conventions are of standard practice: $c=\hbar=1$, $M_P=1$, mode momenta are measured in the unit of Mpc$^{-1}$. Over dot and over dash represent derivatives \textit{w.r.t.} time and momentum respectively.
\subsection{\boldmath DHE and mode equations in \texorpdfstring{$k$}{k}-space}
\label{sec:k space}
Cosmological inflation is the birthplace of all the large scale structures we observe today. These are the gravitationally amplified versions of the quantum fluctuations of the inflaton field, produced inside the co-moving Hubble horizon. Along the superluminal expansion of space-time \textit{vis-\`{a}-vis} shrinking of the Hubble sphere, these fluctuating modes of momenta $k$'s of the perturbations exit the horizon dynamically satisfying a relation $k=aH$ over the slowly evolving quasi de-Sitter metric background \cite{Baumann:2009ds} of scale factor $a$ and Hubble parameter $H$, and re-enter as classical density perturbations after inflation ends. As a result, all modes do not exit the horizon at once, rather a finite and non-zero random time gap always exists among their instants of leaving the horizon. This idea we term as dynamical horizon exit (DHE). For details, see \cite{Sarkar:2021ird}.\par Now, no such explicit functional dependency exists between $t$ and $k$. However, as described above, over a quasi de-Sitter fluctuating background, we can treat the DHE condition as a one-to-one mapping between $t$ and $k$ as $f:k\rightarrow t$ such that $k=f(t)=a(t)H(t)$. Here the sets $t$ and $k$ are collections of points indicating the finite number of modes exiting the horizon. But mathematically the differences of $t$'s or $k$'s of the modes are infinitesimally small, for which we can consider the $t$ and $k$ spaces as continuous. Using this idea, we can transform any $t$-derivative into $k$-derivative by the equation\footnote{By carefully looking into the Eq. (\ref{eq:DHE}) we can realize that such expression can only be feasible in expanding universe where $\Ddot{a}\neq 0$.}
\begin{equation}
    \Dot{k}=\lim_{\delta t\rightarrow 0}\frac{f(t+\delta t)-f(t)}{\delta t}=\frac{d(aH)}{dt}=\Ddot{a}
\end{equation}
as
\begin{equation}
    \frac{1}{\Ddot{a}(t)}\frac{\partial}{\partial t}\equiv \frac{\partial}{\partial k}.
    \label{eq:DHE}
\end{equation}
\par Now we expand the inflaton field $\psi(t,\Vec{r})$ by Fourier transform as
\begin{equation}
    \psi(t,\Vec{r})=\int\frac{d^3 k}{(2\pi)^3}\left[\psi(k,t)b(\Vec{k})e^{i\Vec{k}\cdot\Vec{r}}+\psi^{*}(k,t)b^{*}(\Vec{k})e^{-i\Vec{k}\cdot\Vec{r}}\right]
\end{equation}
where $\psi(k,t)$'s are the linearly independent classical mode functions, also called the Fourier modes. In quantum domain the inflaton scalar field behaves as an operator (indicated by the hat symbol) acting on the Bunch-Davies vacuum (BD) \cite{Bunch:1978yq,Birrell:1982ix,Kundu:2011sg} and thus we obtain
\begin{equation}
 \hat{\psi}(t,\Vec{r})=\int\frac{d^3 k}{(2\pi)^3}\left[\psi(k,t)\hat{b}(\Vec{k})e^{i\Vec{k}\cdot\Vec{r}}+\psi^{*}(k,t)\hat{b}^{\dagger}(\Vec{k})e^{-i\Vec{k}\cdot\Vec{r}}\right]
\end{equation}
subjected to the equal time commutation
\begin{equation}
    \left[\hat{b}(\Vec{k}),\hat{b}^{\dagger}(\Vec{k'})\right]_{t=t'}=(2\pi)^3\delta (\Vec{k}-\Vec{k'}),
\end{equation}
$\psi (k,t)$ being the quantum mode function of momentum $k=|\Vec{k}|$.\par In the framework of linear cosmological perturbations, we split the inflaton field into its zeroth order ($\psi^{(0)} (t)$) and first order ($\delta\psi(t,k)$ parts as
\begin{equation}
    \psi(k,t)=\psi^{(0)} (t)+\delta\psi(k,t).
\end{equation}
The temporal part of $\psi(k,t)$ is governed by the following evolution equations \cite{Sarkar:2021ird}
\begin{equation}
    \Ddot{\psi}^{(0)}(t)+3H\Dot{\psi}^{(0)}(t)+\frac{\partial V(\psi^{(0)}(t))}{\partial\psi^{(0)}(t)}=0,
    \label{eq:modeEQ1}
\end{equation}
\begin{equation}
    \begin{split}
        &\delta\Ddot{\psi}(k,t)+3H\delta\Dot{\psi}(k,t)+\frac{\partial^2V(\psi^{(0)}(t))}{\partial\psi^{(0)}(t)^2}\delta\psi(k,t)+\frac{k^2}{a^2}\delta\psi(k,t)\\
        &=\Dot{\psi}^{(0)}(t)\left[\Dot{\Phi}_B(k,t)+\frac{\partial^2}{\partial t^2}\left(\frac{\Phi_B(k,t)}{H}\right)+\frac{k^2}{a^2}\left(\frac{\Phi_B(k,t)}{H}\right)\right]\\
        &-2\left[\Phi_B(k,t)+\frac{\partial}{\partial t}\left(\frac{\Phi_B}{H}\right)\right]\frac{\partial V(\psi^{(0)})(t)}{\partial\psi^{(0)}(t)},
    \end{split}
    \label{eq:modeEQ2}
\end{equation}
$\Phi_B$ being the gauge invariant Bardeen potential \cite{Baumann:2009ds} which acts like a metric perturbation in the background. Eqs. (\ref{eq:modeEQ1}) and (\ref{eq:modeEQ2}) are the most general equations of mode functions which are derived from the solutions of unperturbed and perturbed Einstein's field equations \cite{Baumann:2009ds} of gravity under spatially flat gauge. But when we go to apply these equations in co-moving DHE during inflation, we have to follow the derivative identity (\ref{eq:DHE}). In this respect, we can define a $k$ scale just like $t$ scale and transit between them using $k=aH$. Thus, the $t$-dependency of the mode function will be mapped to $k$-dependency and only the modes crossing the horizon will be taken into account in the mode equations. In this way the quantum mode function becomes a function of mode momentum $k$ only whose evolution should be governed by a set of ordinary differential equations (ODE) in $k$-space.\par That's exactly what we derived in \cite{Sarkar:2021ird} using (\ref{eq:DHE}) and the conditions of inflationary slow-roll. Here, we write these $k$-space evolution equations for $\psi^{(0)}(k)$, $\delta\psi(k)$ and $\Phi_B (k)$ as
\begin{equation}
    k^2\psi''^{(0)}+k^2F_1{\psi'^{(0)}}^2+4k\psi'^{(0)}+k^2\left(F_1\delta\psi'+F_2\psi'^{(0)}\delta\psi\right)\psi'^{(0)}+6F_1\left(1-2F_1\delta\psi\right)=0,
    \label{eq:modeKeqfinal1}
\end{equation}
\begin{equation}
    \begin{split}
        &k^2\delta\psi''+\left(k^2F_1\psi'^{(0)}+4k\right)\delta\psi'+\left(1+6\left(F_2+2F_1^2\right)\right)\delta\psi\\
        &=\Phi_B''k^3\psi'^{(0)}+\Phi_B'\left(2k^2\psi'^{(0)}-k^3F_1{\psi'^{(0)}}^2-12kF_1\right)\\
        &+\Phi_B\left(k\psi'^{(0)}-k^2{\psi'^{(0)}}^2F_1-k^3\psi'^{(0)}\left(F_2{\psi'^{(0)}}^2+F_1\psi''^{(0)}\right)-12F_1+12kF_1^2\psi'^{(0)}\right)
    \end{split}
     \label{eq:modeKeqfinal2}
\end{equation}
and
\begin{equation}
    \begin{split}
        &k^4{\psi'^{(0)}}^2\Phi_B''+\left(2k^4\psi'^{(0)}\psi''^{(0)}-2k^3{\psi'^{(0)}}^2-k^4F_1{\psi'^{(0)}}^3-2k\right)\Phi_B'\\
        &+\Bigl(6+k^3\left(2-3kF_1\psi'^{(0)}\right)\psi''^{(0)}\psi'^{(0)}
        -k^2\left(k^2F_2{\psi'^{(0)}}^2-3kF_1\psi'^{(0)}+4\right){\psi'^{(0)}}^2\Bigr)\Phi_B\\
        &=6k\left(k\psi''^{(0)}+kF_1{\psi'^{(0)}}^2+\psi'^{(0)}+F_2\psi'^{(0)}\right)\delta\psi+k^3\psi'^{(0)}\delta\psi''\\
        &+k\left(k^2\psi''^{(0)}+2k\psi'^{(0)}+6F_1\right)\delta\psi'
    \end{split}
      \label{eq:modeKeqfinal3}
\end{equation}
where 
\begin{equation}
    F_j = \frac{\partial^j}{\partial{\psi^{(0)}}^j}\ln{\sqrt{V\left(\psi^{(0)}\right)}},\quad j=1,2.
\end{equation}
The $F$-functions bear the signatures of compatibility of quantum fluctuations with the geometric perturbation $\Phi_B$, which control sensitively the self-consistent solutions of the non-linearly coupled ODE's (\ref{eq:modeKeqfinal1})-(\ref{eq:modeKeqfinal3}) for a particular initial condition, stated in section \ref{sec: result}. These solutions will be transcended to the mode-dependent cosmological parameters described below, which will ultimately be utilised to constrain the potential in the light of current experimental bounds.
\subsection{Cosmological parameters}
\label{sec:cosmoparam}
Now we shift our focus to the central theme of our discussion, the cosmological parameters coming out of the potential $V(\psi)$, which is the quintessential $\alpha$-attractor (see Eq. (\ref{eq:Eq7})).
\subsubsection{Number of e-folds}
The number of e-folds is defined as
\begin{equation}
    N(\psi)=\int_{\psi_{\mathrm{end}}}^{\psi}\frac{V(\Tilde{\psi})}{\frac{dV(\Tilde{\psi})}{d\Tilde{\psi}}}d\Tilde{\psi}
    \label{eq:efoldsDEF}
\end{equation}
where the $\psi_{\mathrm{end}}$ corresponds to the value of $\psi$ for which the first slow-roll parameter
\begin{equation}
    \epsilon_V(\psi_{\mathrm{end}})=\frac{1}{2}\left(\frac{\frac{dV(\psi)}{d\psi}}{V(\psi)}\right)_{\psi=\psi_{\mathrm{end}}}^2=1,
\end{equation}
which yields from (\ref{eq:Eq7})
\begin{equation}
    \psi_{\mathrm{end}}=\sqrt{\frac{3\alpha}{2}}\ln{\left(\frac{2n}{\sqrt{3\alpha}}+2n\right)}.
    \label{eq:psiEND}
\end{equation}
In general the upper limit of Eq. (\ref{eq:efoldsDEF}) is larger than the lower limit \textit{i.e.} $\psi\gg\psi_{\mathrm{end}}$ which can be verified by taking a simple example where we put $n=3$ and $\alpha=\frac{1}{6}$ in (\ref{eq:Eq7}). In this case $\psi_{\mathrm{end}}=1.3$ which is significantly small from the $\psi$ values of inflation (it is due to the infinitely long plateau. See figure \ref{fig:Fig2}).\par Now we calculate $N$ under the condition (\ref{eq:psiEND}) as
\begin{equation}
\begin{split}
    N(\psi)&=\frac{3\alpha}{4n}\left[\exp\left(\sqrt{\frac{2}{3\alpha}}\psi\right)-\exp\left(\sqrt{\frac{2}{3\alpha}}\psi_{\mathrm{end}}\right)\right]-\sqrt{\frac{3\alpha}{2}}\left(\psi-\psi_{\mathrm{end}}\right)\\
    &\approx \frac{3\alpha}{4n}\left[\exp\left(\sqrt{\frac{2}{3\alpha}}\psi\right)-\left(\frac{2n}{\sqrt{3\alpha}}+2n\right)\right]
\end{split}
\label{eq:finalEFOLDS}
\end{equation}
from which we can also write its mode-dependent form
\begin{equation}
    N(k)=\frac{3\alpha}{4n}\left[\exp\left(\sqrt{\frac{2}{3\alpha}}\psi(k)\right)-\left(\frac{2n}{\sqrt{3\alpha}}+2n\right)\right].
    \label{eq:modeEFOLDS}
\end{equation}
\subsubsection{Power spectra}
\label{subsec:POWERspecSUBSEC}
\begin{figure}[H]
	\centering
	\includegraphics[width=0.8\linewidth]{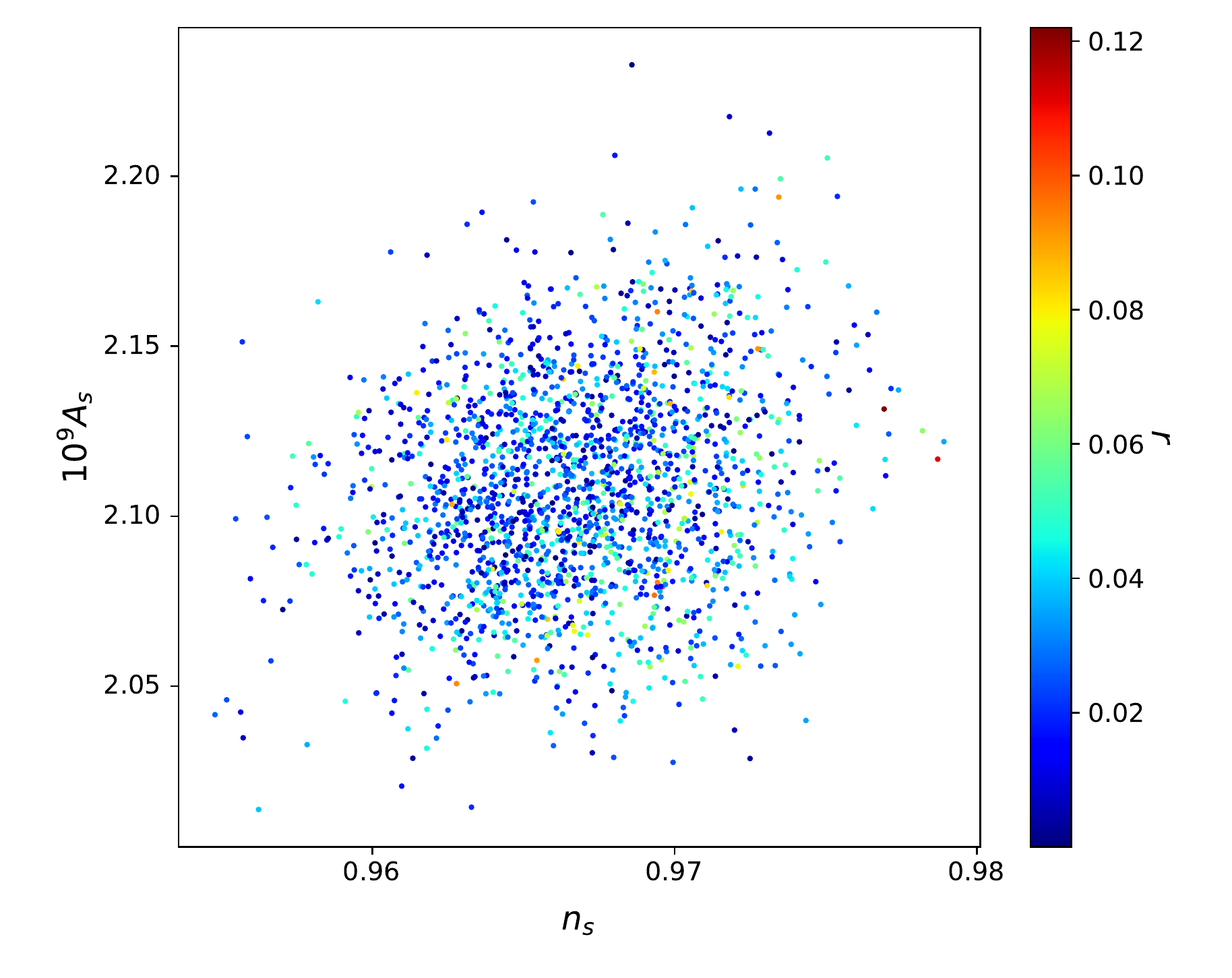}
	\caption{Planck-2018 results for $A_s$, $n_s$ and $r$ are regenerated using the experimental data obtained as the public chains from the Planck Legacy Archive (\url{https://pla.esac.esa.int/}) by running the GetDist (\url{https://getdist.readthedocs.io/en/latest/}) plotting utility. The best-fit results for $n_s=0.9649\pm 0.0042$ at $68\%$ CL and $r_{0.002}<0.064$ at $95\%$ CL are found around $A_s\approx 2.1\times 10^{-9}$.}
	\label{fig:AmpScalarPerturbation}
\end{figure}

The statistical properties of the quantum nuggets $\delta\psi_k$'s in quasi de-Sitter first order perturbation theory is determined by a two-point correlation\footnote{we are not considering any higher order correlations and non-Gaussian properties.} over a BD vacuum $|0\rangle$ (see \cite{Kundu:2011sg,Jiang:2016nok} for details)
\begin{equation}
    \langle\delta\psi_k\delta\psi_{k'}\rangle =\langle 0|\delta\psi_k\delta\psi_{k'}|0\rangle= \frac{4\pi^3}{k^3}\delta(\Vec{k}+\Vec{k'})H^2
\end{equation}
which is parameterized by the dimensionless inflationary scalar power spectrum defined as \cite{Baumann:2009ds}
\begin{equation}
    \Delta_s (k) = \left(\frac{H^2}{8\pi^2\epsilon}\right)_{k=aH}.
\end{equation}
Under slow-roll condition, this equation takes the form
\begin{equation}
    \Delta_s(k) = \frac{1}{12\pi^2}\frac{V(\psi(k))^3}{\left(\frac{dV(\psi(k))}{d\psi(k)}\right)^2}.
    \label{eq:ScalarPower1}
\end{equation}
Similarly for tensor perturbations having two orthogonal polarizations, the dimensionless tensor power spectrum is given by \cite{Baumann:2009ds}
\begin{equation}
    \Delta_t(k) = \left(\frac{2}{\pi^2}H^2\right)_{k=aH}
\end{equation} which under slow-roll, becomes
\begin{equation}
    \Delta_t(k) = \left(\frac{2}{3\pi^2}V(\psi(k))\right)_{k=aH}.
    \label{eq:TensorPower1}
\end{equation}
Now, we specialize the power spectra in our quintessential model. Using Eqs. (\ref{eq:Eq7}) and (\ref{eq:modeEFOLDS}) we can write
\begin{equation}
    \Delta_s(k) = \frac{M^4}{18\pi^2\alpha}\frac{\left(N(k)+\frac{\sqrt{3\alpha}}{2}\right)^3}{\left(N(k)+\frac{\sqrt{3\alpha}}{2}+\frac{3\alpha}{2}\right)}.
    \label{eq:modeScalarPower}
\end{equation}
Empirically, the scalar power spectrum is approximated by \cite{Baumann:2009ds}
\begin{equation}
    \Delta_s(k)=A_s(k_{*})\left(\frac{k}{k_{*}}\right)^{n_s(k_{*})-1+\frac{\alpha_s(k_{*})\ln{\frac{k}{k_{*}}}}{2}}
\end{equation} \textit{w.r.t.} a reference scale $k_{*}$ such that
\begin{equation}
    \Delta_s(k_{*})=A_s(k_{*}),
    \label{eq:EqualAMPscalar}
\end{equation}
$A_s(k_{*})$ being the amplitude of scalar perturbation, $n_s$ is scalar spectral index and $\alpha_s$ is called running of the spectral index. However, the exact mode behaviour of $\Delta_s(k)$ depends upon the specific model and the order of perturbation concerned. In our model we shall show its mode dependence through the solutions of $k$-space evolution equations. Planck-2018 results \cite{Planck:2018jri,Planck:2018vyg} have constrained $A_s(k_{*})=2.105\pm 0.030\times 10^{-9}$ for $k_{*}=0.002$ Mpc$^{-1}$ within $68\%$ CL with $TT+TE+EE+\mathrm{low}E$ polarizations and $\mathrm{lensing}+\mathrm{BAO}$ effects (see figure \ref{fig:AmpScalarPerturbation}). 
 The number of e-folds corresponding to $k_{*}$ in quintessential $\alpha$-attractor is considered to be $N_{*}\approx 63$ in \cite{Dimopoulos:2017zvq} which incorporates the reheating effects also. Now following Eq. (\ref{eq:EqualAMPscalar}) the scalar perturbation amplitude $A_s(k_{*})$ can also be related to $N_{*}$ as
 \begin{equation}
     A_s(k_{*}) = \frac{M^4}{18\pi^2\alpha}\frac{\left(N_{*}+\frac{\sqrt{3\alpha}}{2}\right)^3}{\left(N_{*}+\frac{\sqrt{3\alpha}}{2}+\frac{3\alpha}{2}\right)}.
 \end{equation}
 After some simplification and applying the approximation 
 \begin{equation}
     \exp\left[\frac{3\alpha}{2}\left(N_{*}+\frac{\sqrt{3\alpha}}{2}\right)^{-1}\right]\approx \left[1+\frac{3\alpha}{2}\left(N_{*}+\frac{\sqrt{3\alpha}}{2}\right)^{-1}\right]
 \end{equation}
 we get
 \begin{equation}
     M^4 = 9\pi^2 (2\alpha A_s(k_{*}))\left(N_{*}+\frac{\sqrt{3\alpha}}{2}\right)^{-2}\exp\left[\frac{3\alpha}{2}\left(N_{*}+\frac{\sqrt{3\alpha}}{2}\right)^{-1}\right],
     \label{eq:COBEnormalization}
 \end{equation}
whose right hand side is completely fixed for a given value of $\alpha$. A precise determination of this normalizing factor $M$ in Eqs. (\ref{eq:Eq6}) and (\ref{eq:Eq7}) determines the proper energy scale of the inflation which is according to Planck data ${V_{*}}^{1/4} = M<1.7\times 10^{16}$ GeV or ${V_{*}}^{1/4} = M<6.99\times 10^{-3}$ in reduced Planck unit signifying GUT inflation. This is actually the required COBE/Planck normalization condition for our quintessential $\alpha$-attractor potential, which we shall calculate from the above equation to analyse the cosmological parameters in the next section. Also, according to the Eq. (\ref{eq:modeScalarPower}) this condition limits the range of values of the scalar power spectrum which is an important criterion for the inflationary scenario.\par Similarly from Eqs. (\ref{eq:Eq7}) and (\ref{eq:modeEFOLDS}) we can express the tensor power spectrum in the present model as
\begin{equation}
    \Delta_t(k) = \frac{2}{3\pi^2}M^4\left[1-\frac{2n}{\frac{4nN(k)}{3\alpha}+\frac{2n}{\sqrt{3\alpha}}+2n}\right]=\frac{2}{3\pi^2}M^4\left[1-\frac{1}{\frac{2N(k)}{3\alpha}+\frac{1}{\sqrt{3\alpha}}+1}\right]
    \label{eq:modeTensorPOWER}
\end{equation} whose scale is again set by the factor $M$ according to the normalization condition (\ref{eq:COBEnormalization}).
\subsubsection{Spectral indices}
The $k$-dependent scalar and tensor spectral indices are defined by
\begin{equation}
    n_s(k) -1 =\frac{d\ln{\Delta_s(k)}}{d\ln{k}},
    \label{eq:LOGspectralIndex}
\end{equation}
\begin{equation}
    n_t(k) = \frac{d\ln{\Delta_t(k)}}{d\ln{k}}
    \label{eq:LogNT}
\end{equation} and the tensor-to-scalar ratio is given by
\begin{equation}
    r(t)=\frac{\Delta_t(k)}{\Delta_s(k)}.
    \label{eq:LogR}
\end{equation}
$n_{s,t}$ indicate the logarithmic scale dependencies of $\Delta_{s,t}$ respectively during inflation. Under slow-roll condition the spectral indices are expressed in terms of first and second slow-roll parameters $\epsilon_V$ and $\eta_V$ respectively as
\begin{equation}
    n_s(k) -1 = 2\eta_V(\psi(k))-6\epsilon_V(\psi(k)),
\end{equation}
\begin{equation}
    n_t(k) = -2\epsilon_V(\psi(k))
\end{equation} and 
\begin{equation}
    r(k)=16\epsilon_V(\psi(k)),
\end{equation}
where $\epsilon_V(\psi(k)) = \frac{1}{2}\left(\frac{\frac{dV(\psi(k))}{d\psi(k)}}{V(\psi(k))}\right)^2$ and $\eta_V(\psi(k))=\frac{\frac{d^2V(\psi(k))}{d\psi(k)^2}}{V(\psi(k))}$. We calculate these slow-roll parameters using Eqs. (\ref{eq:Eq7}) and (\ref{eq:modeEFOLDS}) in terms of $N(k)$ in the limit $\mathcal{O}(1/(N^2))\ll1$ as
\begin{equation}
    \epsilon_V(k)\approx \frac{3\alpha}{4N(k)^2}
\end{equation} and 
\begin{equation}
    \eta_V(k)\approx -\frac{1}{N(k)}+\frac{\sqrt{3\alpha}}{2N(k)^2}
\end{equation}
which result
\begin{equation}
    n_s(k) = 1-\frac{2}{N(k)},
    \label{eq:AttractorEQ1}
\end{equation}
\begin{equation}
    n_t(k) = -\frac{3\alpha}{2N(k)^2}
      \label{eq:AttractorEQ2}
\end{equation}
and
\begin{equation}
    r(k)=\frac{12\alpha}{N(k)^2}.
      \label{eq:AttractorEQ3}
\end{equation}
The cosmological parameters in Eq. (\ref{eq:AttractorEQ1}) - (\ref{eq:AttractorEQ3}) match with that of $\alpha$-attractors, although the form of the potential (\ref{eq:Eq7}) is different. It is actually a modified form of $\alpha$-attractor by the new element - quintessence. We shall see the new results coming from this upgrading over the older one in the next section.
\subsubsection{Running of spectral index}
Just like $n_s(k)$ provides the logarithmic scale dependence of $\Delta_s(k)$, the running of scalar spectral index $\alpha_s(k)$ gives the same over $n_s(k)$ according to
\begin{equation}
    \alpha_s(k) =\frac{dn_s(k)}{d\ln{k}}=\frac{d^2\ln{\Delta_s(k)}}{d\ln{k}^2}.
    \label{eq:runningSpectralINDEXMode}
\end{equation}
Planck-2018 sets the value $\alpha_s(k_{*})=-0.0045\pm 0.0067$ at $68\%$ CL, where the negative sign indicates that $n_s(k)$ should decrease with $k$, which we shall show in section \ref{sec: result}.
\subsubsection{Inflationary Hubble parameter}
According to Friedmann equations \cite{Baumann:2009ds}, the inflationary Hubble parameter is $H_{\mathrm{inf}}=\sqrt{\frac{\rho}{3}}$, which measures the expansion of the universe during inflation, sourced by the inflaton energy density $\rho$. In slow-roll regime it can be related to the potential as
\begin{equation}
    H_{\mathrm{inf}}(\psi) = \sqrt{\frac{V(\psi)}{3}}
\end{equation} which, in our case, is
\begin{equation}
   H_{\mathrm{inf}}(k) =\sqrt{\frac{V(\psi(k))}{3}}=\sqrt{\frac{M^4}{3}\left(1-2ne^{-\sqrt{\frac{2}{3\alpha}}\psi(k)}\right)}.
   \label{eq:DefHubbleINF}
\end{equation}
Following Eq. (\ref{eq:modeEFOLDS}) it can be written in terms of number of e-folds as
\begin{equation}
    H_{\mathrm{inf}}(k)=\frac{M^2}{\sqrt{3}}\left(\frac{\frac{4nN(k)}{3\alpha}+\frac{2n}{\sqrt{3\alpha}}}{\frac{4nN(k)}{3\alpha}+\frac{2n}{\sqrt{3\alpha}}+2n}\right)^{1/2}=\frac{M^2}{\sqrt{3}}\left(\frac{\frac{2N(k)}{3\alpha}+\frac{1}{\sqrt{3\alpha}}}{\frac{2N(k)}{3\alpha}+\frac{1}{\sqrt{3\alpha}}+1}\right)^{1/2}.
    \label{eq:INFHubbleMode}
\end{equation}
According to Planck data $H_{\mathrm{inf}}(k_{*})<2.5\times 10^{-5}$ in the unit of reduced Planck mass or $H_{\mathrm{inf}}(k_{*})<6.07\times 10^{13}$ GeV at $95\%$ CL.\par Now we shall analyse the mode dependencies of the cosmological parameters described earlier using Eqs. (\ref{eq:modeEFOLDS}), (\ref{eq:modeScalarPower}), (\ref{eq:modeTensorPOWER}), (\ref{eq:AttractorEQ1}) - (\ref{eq:runningSpectralINDEXMode}) and (\ref{eq:INFHubbleMode}) satisfying COBE/Planck normalization constraint (\ref{eq:COBEnormalization}) from the self-consistent solutions of $k$-space evolution equations (\ref{eq:modeKeqfinal1}) - (\ref{eq:modeKeqfinal3})  for the inflatinary regime described by the Eq. (\ref{eq:Eq7}) of the quintessential $\alpha$-attractor model of Eq. (\ref{eq:Eq6}). Also, here, we shall compare the obtained values with those from the ordinary $\alpha$-attractor models and Planck-2018 results.
\section{Results and discussion}
\label{sec: result}
We solve numerically the $k$-space evolution equations in Wolfram Mathematica 12 for the following considerations:
\begin{enumerate}
    \item We use the value of $\psi^{(0)}(k)$ corresponding to $N=63.49$ from Eq. (\ref{eq:modeEFOLDS}), $\psi'^{(0)}(k)=0$, $\delta\psi(k)=0.001$, $\delta\psi'(k)=0$, $\Phi_B(k)=0.001$ and $\Phi'_B(k)=0$ at $k=0.001$ as initial conditions, and the set of equations is allowed to plot the solutions for $k$ ranging from $0.001 - 0.009$. This set of initial conditions and range of $k$ are so chosen that the equations converge to a consistent solution where we assume that all the necessary criteria regarding the selection of vacuum over curved space-time (what is simply called the BD vacuum) are already satisfied through the mode selection during DHE.  
    \item Some contemporary studies have already put constraints on the values of the model parameters $\alpha$ and $n$. Ref. \cite{Dimopoulos:2017zvq} chooses $\alpha\lesssim \frac{1}{6}$ and $25\lesssim n\leq 92$ to avoid the super-Planckian  issues of the non-canonical scalar field. These ranges have been made more stringent in \cite{Dimopoulos:2017tud} by selecting $\alpha=1.5$ for $118\leq n\leq 124$ and $\alpha=4.2$ for $121\leq n\leq 125$ incorporating the effects of instant preheating, perturbative reheating, over-production of gravitinos and backreaction.\par Based on these surveys in our work we choose $n=122$ and $\alpha=\frac{1}{10}, \frac{1}{6}, 1, 4.3$. We find that below $\alpha=\frac{1}{10}$ the cosmological parameters are almost insensitive to variation of $\alpha$ and above $\alpha=4.3$ the system does not converge to any solution. Actually in our formalism, constraining $n$ is very difficult because the cosmological parameters do not depend upon $n$ explicitly and also, its precise value depends on detailed processes involving preheating and reheating, which is not considered in the present framework. Our aim is to constrain $\alpha$ through mode analysis of cosmological parameters. Therefore we keep the value of $n$ fixed at $n=122$ throughout the entire study. Also we have used the same choice in comparison with ordinary $\alpha$-attractors and Planck-2018 data.
    \item The inflationary scale-determining factor $M$ is calculated with the above choices of $n$ and $\alpha$ based on COBE/Planck normalization from the Eq. (\ref{eq:COBEnormalization}) and the obtained values are enlisted in the table below.
    \begin{table}[H]
    \captionsetup{justification=centering,width=0.7\textwidth}
    \caption{Constraints for $M$ by COBE/Planck normalisation for a given value of $\alpha$. The conversion factor is $M_P=2.43\times 10^{18}$ GeV.}
    \begin{center}
        \begin{adjustbox}{width=0.45\textwidth}
        \begin{tabular}{|c|c|c|c|}
    \hline
    $n$ & $\alpha$ & $M (M_P)$ & $M$ (GeV)\\
    \hline\hline
    $122$ & $\frac{1}{10}$ & $1.76\times 10^{-3}$ & $4.28\times 10^{15}$\\
    $122$ & $\frac{1}{6}$ & $2.00\times 10^{-3}$ & $4.86\times 10^{15}$\\
    $122$ & $1$ & $3.14\times 10^{-3}$ & $7.63\times 10^{15}$\\
    $122$ & $4.3$ & $4.57\times 10^{-3}$ & $1.11\times 10^{16}$\\
    \hline
    \end{tabular}
    \end{adjustbox}
    \end{center}
         \label{tab:Table1}
    \end{table}
    All values of $M$ are within the given limit set by Planck-2018 \textit{i.e.} ${V_{*}}^{1/4} = M<1.7\times 10^{16}$ GeV (fouth column) or ${V_{*}}^{1/4} = M<6.99\times 10^{-3}$ (third column) in reduced Planck unit. We shall use these $M$ values in the subsequent presentations of the graphical and numerical results. 
\end{enumerate}
\subsection{Validity of the linear perturbative framework}
\label{subsec:Validity}
We plot the self-consistent solutions of the $k$-space evolution equations (\ref{eq:modeKeqfinal1})-(\ref{eq:modeKeqfinal3}) subjected to the considerations described above in figures \ref{fig:unperturbedPOT} through \ref{fig:perturbedINF} for $\alpha=1/10, 1/6, 1, 4.3$ and $n=122$ within $k=0.001 - 0.009$. Figures \ref{fig:unperturbedPOT} and \ref{fig:perturbedPOT} describe the mode dependencies of unperturbed part $V^{(0)}(k)$ and its first order perturbed part $\delta V(k)$ according to
    \begin{equation}
    \begin{split}
        V(\psi(k))&=V(\psi^{(0)}(k)+\delta\psi(k))=V(\psi^{(0)}(k))+\frac{\partial V(\psi^{(0)}(k))}{\partial \psi^{(0)}(k)}\delta\psi(k)\\
        &=V^{(0)}(k)+2V^{(0)}(k)F_1\delta\psi(k)=V^{(0)}(k)+\delta V(k).
    \end{split}
    \label{eq:modeINFpot}
    \end{equation}
    This equation suggests that the effect of the inflaton perturbation $\delta\psi (k)$ is induced microscopically to the perturbation in the potential $\delta V(k)$.
    Figures \ref{fig:unperturbedINF} and \ref{fig:perturbedINF} describe the same for unperturbed inflaton $\psi^{(0)}(k)$ and the inflaton perturbation $\delta\psi(k)$. From these figures its very clear that the perturbed parts $\delta V(k)$ ($\sim 10^{-17} - 10^{-14}$) and $\delta\psi (k)$ ($\sim 10^{-4} - 10^{-3}$) are quite smaller than $V^{(0)}(k)$ ($\sim 10^{-12} - 10^{-10}$) and $\psi^{(0)}(k)$ ($\mathcal{O}(1) - \mathcal{O}(10)$) respectively for all values of $\alpha$ signifying the fact that linear cosmological perturbative framework in sub-Planckian limit is working satisfactorily in $k$-space just like in field space. The order of the $V^{(0)}(k)$ values indicate that the results are in perfect agreement with the $M$ values of table \ref{tab:Table1} \textit{i.e.} the potentials in momentum space also satisfy the COBE/Planck normalization condition. We observe an interesting fact from these figures - for $\alpha< 1$ the amount of perturbation both in potential and inflaton field is larger for $k\leq 0.005$ than that of the values for $k>0.005$. And the reverse is observed for $\alpha\geq 1$. This indicates that the effect of perturbations is shifted from the bands $0.001\leq k \leq 0.005$ to $0.005< k \leq 0.009$ as we increase $\alpha$ from low to high values. This is a new finding regarding the effect of tuning of $\alpha$ on the cosmological perturbations.
    \subsection{Mode behaviours of cosmological parameters}
\label{subsec:ModeBehaviours}
Figures \ref{fig:scalarPowerSpectrum} and \ref{fig:tensorPowerSpectrum} illustrate the COBE/Planck normalized scalar and tensor power spectra of the inflaton perturbation obtained from Eqs. (\ref{eq:modeScalarPower}) and (\ref{eq:modeTensorPOWER}). The orders of the scalar power spectra ($\Delta_s(k)$) are $10^{-9}$ while that of the tensor power spectra ($\Delta_t(k)$) are roughly $10^{-12}$ which indicate that the primordial gravitational waves are very weak which is in fact verified in recent cosmological surveys like Planck-2018 \cite{Planck:2018jri,Planck:2018vyg}. One of the important consequences of this fact is that the tensor to scalar ratio is too small (see figure \ref{fig:tensorToScalarRatio}), which is relevant in the study of $B$ modes of cosmic microwave background radiation. Now as $\alpha$ increases $\Delta_s(k)$ decreases and $\Delta_t(k)$ increases such that their ratio $r$ increases (see Eq. (\ref{eq:LogR})), which we can verify in figure \ref{fig:tensorToScalarRatio}.\par In figure \ref{fig:numberOfEFolds} we have plotted the mode-dependent number of e-folds $N(k)$ from Eq. (\ref{eq:modeEFOLDS}) for the specified $k$ range. The values of $N(k)$ remain almost same for $\alpha=1/10,1/6$ and for $\alpha=1,4.3$. Although its variation is very little with the increase in $\alpha$, it has a decreasing mode behaviour with $k$. This signifies that the higher modes (having high $k$ or low wavelengths) exit the horizon with smaller number of e-folds in comparison to the small modes (having low $k$ or large wavelengths). This scale-dependency is a direct proof of DHE of the inflationary modes over quasi de-Sitter background which is the basis of our formalism. These modes re-enter the Hubble horizon after inflation and evolve as classical density perturbations during hot big bang (HBB), producing the primary and secondary anisotropies, observed in the CMB angular power spectra and polarization maps. The statistical scalar ($\Delta_s(k)$) and tensor ($\Delta_t(k)$) correlations of quantum fluctuations are implanted in the post-inflationary matter perturbations as initial conditions which result in large scale structures and its distribution in the universe.\par The mode dependencies of power spectra and number of e-folds can also be visualised in the spectral indices and tensor to scalar ratio, which we plot in figures \ref{fig:scalarSpectralIndex} - \ref{fig:tensorToScalarRatio} according to Eqs. (\ref{eq:AttractorEQ1}) - (\ref{eq:AttractorEQ3}). In figure \ref{fig:scalarSpectralIndex} we show that the scalar spectral index $n_s$ is found to be around $0.9676 - 0.9684$ for entire range of $k$ values, which is almost same for all $\alpha$ values. One of the reason for this is, in this $k$ range the number of e-folds remains around $61.5 - 63.5$ (see figure \ref{fig:numberOfEFolds}) and according to Eq. (\ref{eq:AttractorEQ1}), $n_s$ depends only on $N(k)$. So, $n_s$ also traces the same variation as that of $N(k)$. Another reason can be found by observing the variation of $\Delta_s(k)$ with $\alpha$ which is very small. Its order remains at $10^{-9}$, only the change happens in the magnitude part which shifts from $2.16 - 1.66$. For that reason its log-dependence over $k$ is quite small which we observe in $n_s(k)$ according to (\ref{eq:LOGspectralIndex}). All $n_s(k)$ values remain very close to the Planck quoted values: $n_s=0.9649\pm 0.0042$ at $68\%$ CL with $TT+TE+EE+\mathrm{low}l+\mathrm{low}E+\mathrm{lensing}+BAO$ effects. The values of $n_s(k)\approx 0.96$ signify that the cosmological perturbations are almost scale invariant.\par In figures \ref{fig:tensorSpectralIndex} and \ref{fig:tensorToScalarRatio}, we plot tensor spectral index $n_t(k)$ and tensor to scalar ratio $r(k)$ respectively. Both the parameters are proportional to $\alpha$ according to (\ref{eq:AttractorEQ2}) and (\ref{eq:AttractorEQ3}), for which $|n_t(k)|$ and $r(k)$ increase with the increase in $\alpha$ for a given value of $k$. Specifically in case of $r$, the monotonic increment with $k$ is caused due to fast decrease of $\Delta_s(k)$ (see figure \ref{fig:scalarPowerSpectrum}) and slow decrease of $\Delta_t(k)$ (see figure \ref{fig:tensorPowerSpectrum}) with $k$ for a specific $\alpha$ according to (\ref{eq:LogR}). All values of $r$ satisfy the Planck constraint: $r_{0.002}<0.064$ at $95\%$ CL with the same polarization effects as that of $n_s$.\par The log-scale dependence of $n_s(k)$ is interpreted by the running of spectral index $\alpha_s(k)$, which we show in figure \ref{fig:RunningSpectralIndex} according to (\ref{eq:runningSpectralINDEXMode}). The negative sign indicates that $n_s(k)$ decreases as $k$ increases. Its values are found to be nearly independent of $\alpha$ just like $n_s$ and are in good agreement with Planck data: $\alpha_s(k_{*})=-0.0045\pm 0.0067$ at $68\%$ CL. These values also satisfy the constraint given in Ref. \cite{Dimopoulos:2017zvq}.\par
\begin{figure}[H]
\begin{subfigure}{0.52\linewidth}
  \centering
   \includegraphics[width=70mm,height=65mm]{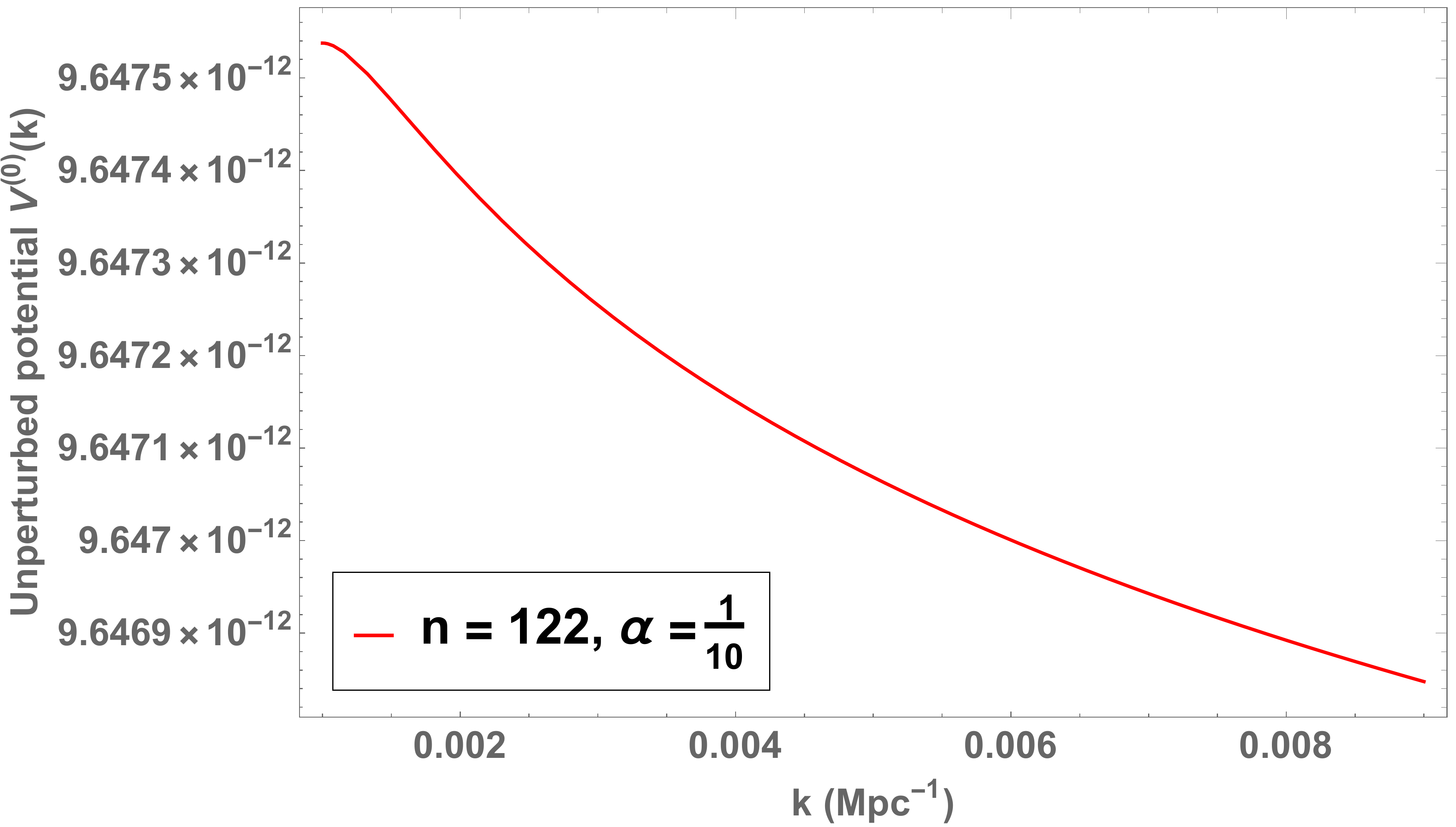} 
   \subcaption{}
   \label{fig:unperturbedPOT_1}
\end{subfigure}%
\begin{subfigure}{0.52\linewidth}
  \centering
   \includegraphics[width=70mm,height=65mm]{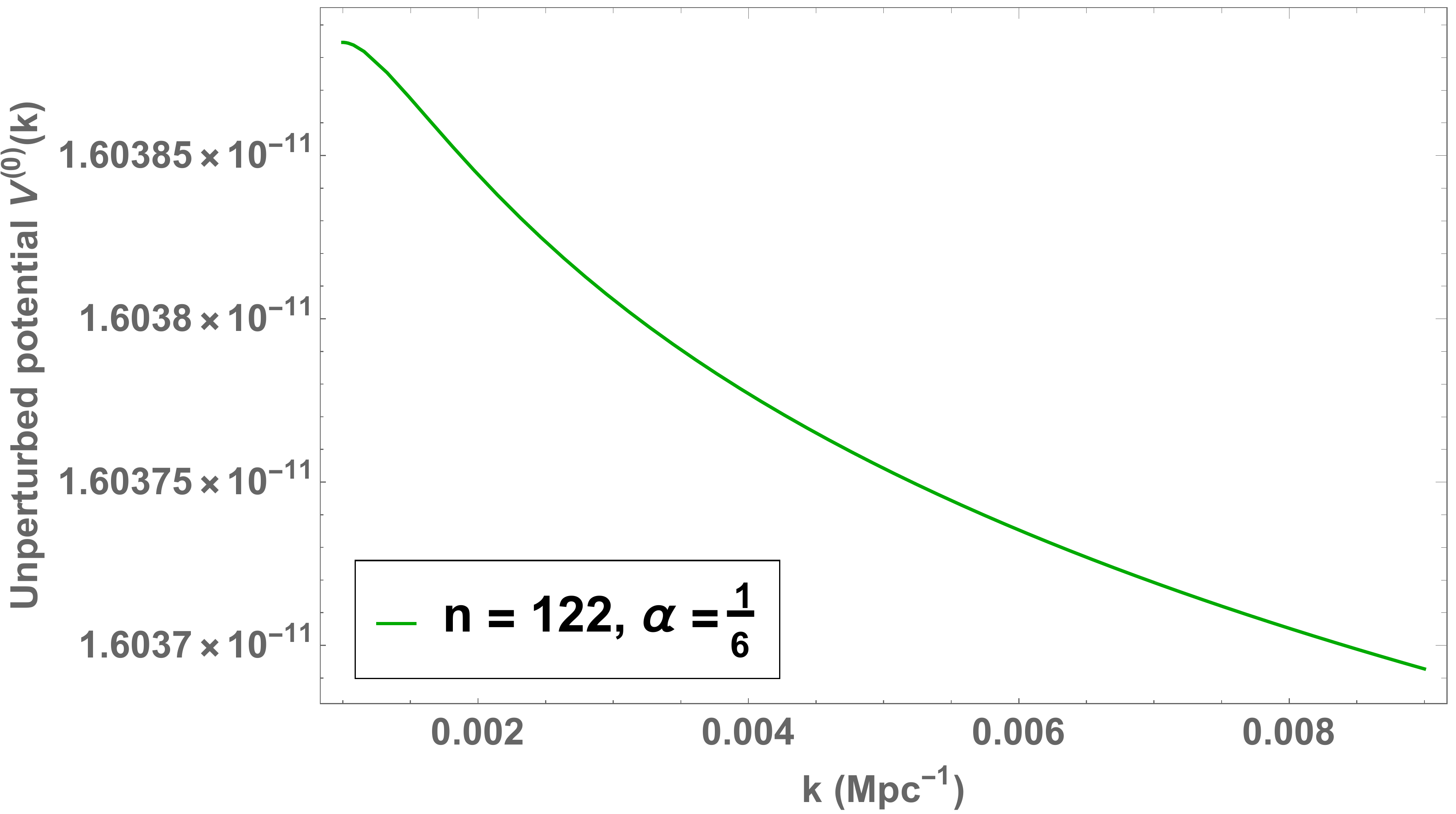}
   \subcaption{}
   \label{fig:unperturbedPOT_2}
\end{subfigure}%
\vspace{0.1\linewidth}
\begin{subfigure}{0.52\linewidth}
  \centering
   \includegraphics[width=70mm,height=65mm]{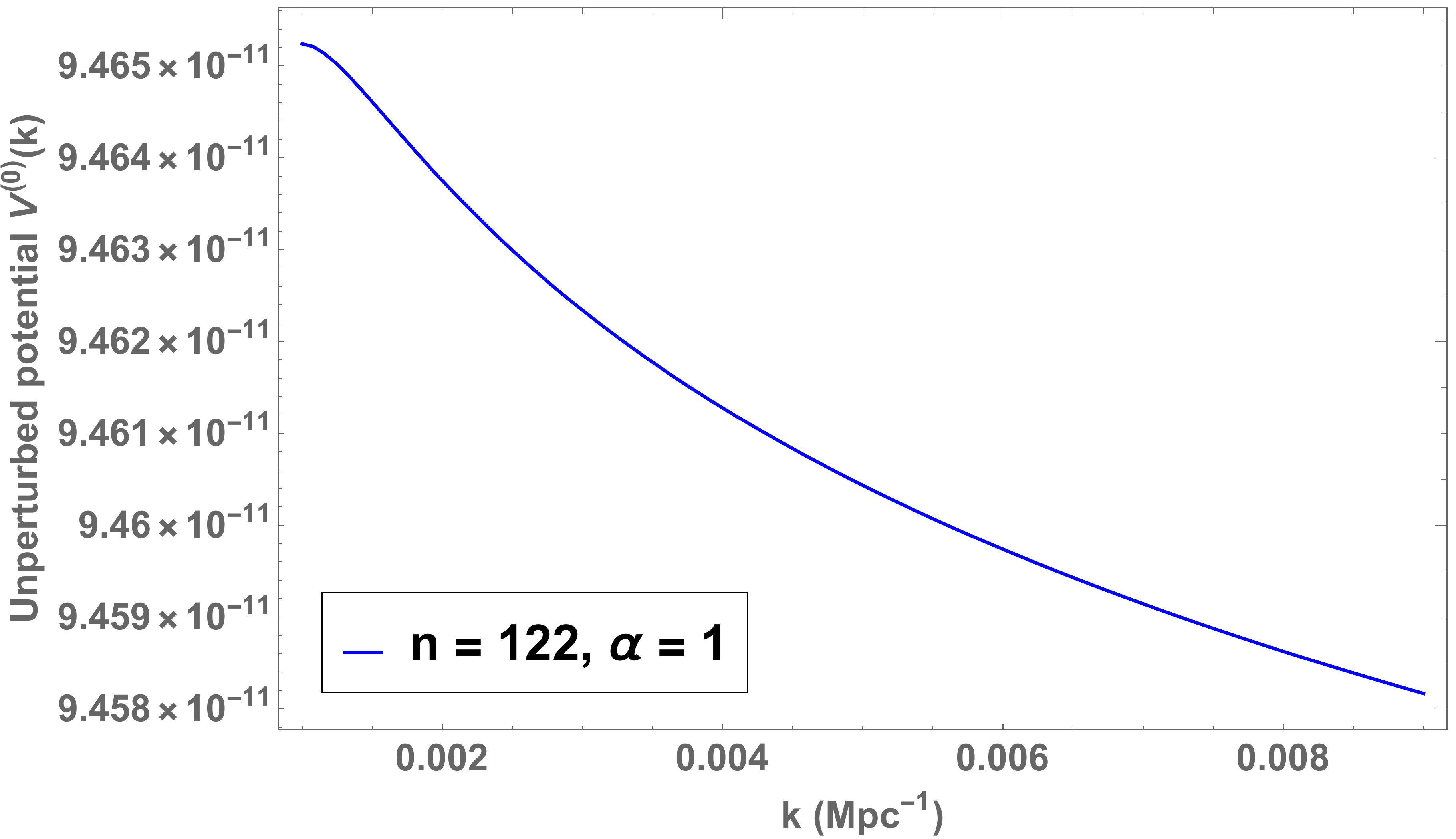}
   \subcaption{}
    \label{fig:unperturbedPOT_3}
\end{subfigure}%
\begin{subfigure}{0.52\linewidth}
  \centering
   \includegraphics[width=70mm,height=65mm]{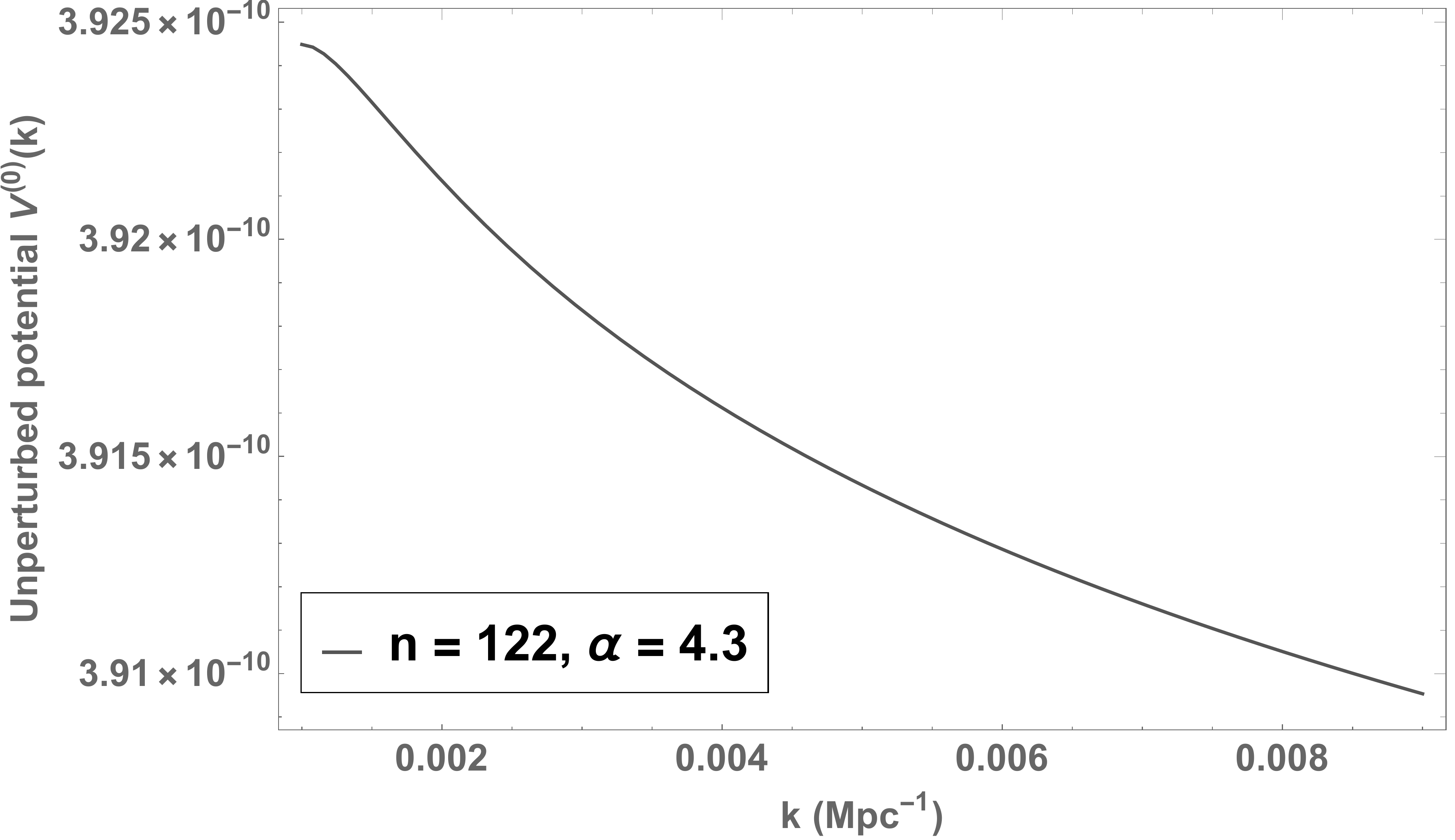}
   \subcaption{}
    \label{fig:unperturbedPOT_4}
\end{subfigure}
\caption{Unperturbed parts of the potential (\ref{eq:Eq7}), where $\psi\equiv \psi(k)$ is the mode function for four values of $\alpha$ for a given value of $n$. The values of $V^{(0)}(k)$ increase on increase in $\alpha$ for a particular value of $k$.}
\label{fig:unperturbedPOT}
\end{figure}

\begin{figure}[H]
\begin{subfigure}{0.52\linewidth}
  \centering
   \includegraphics[width=70mm,height=65mm]{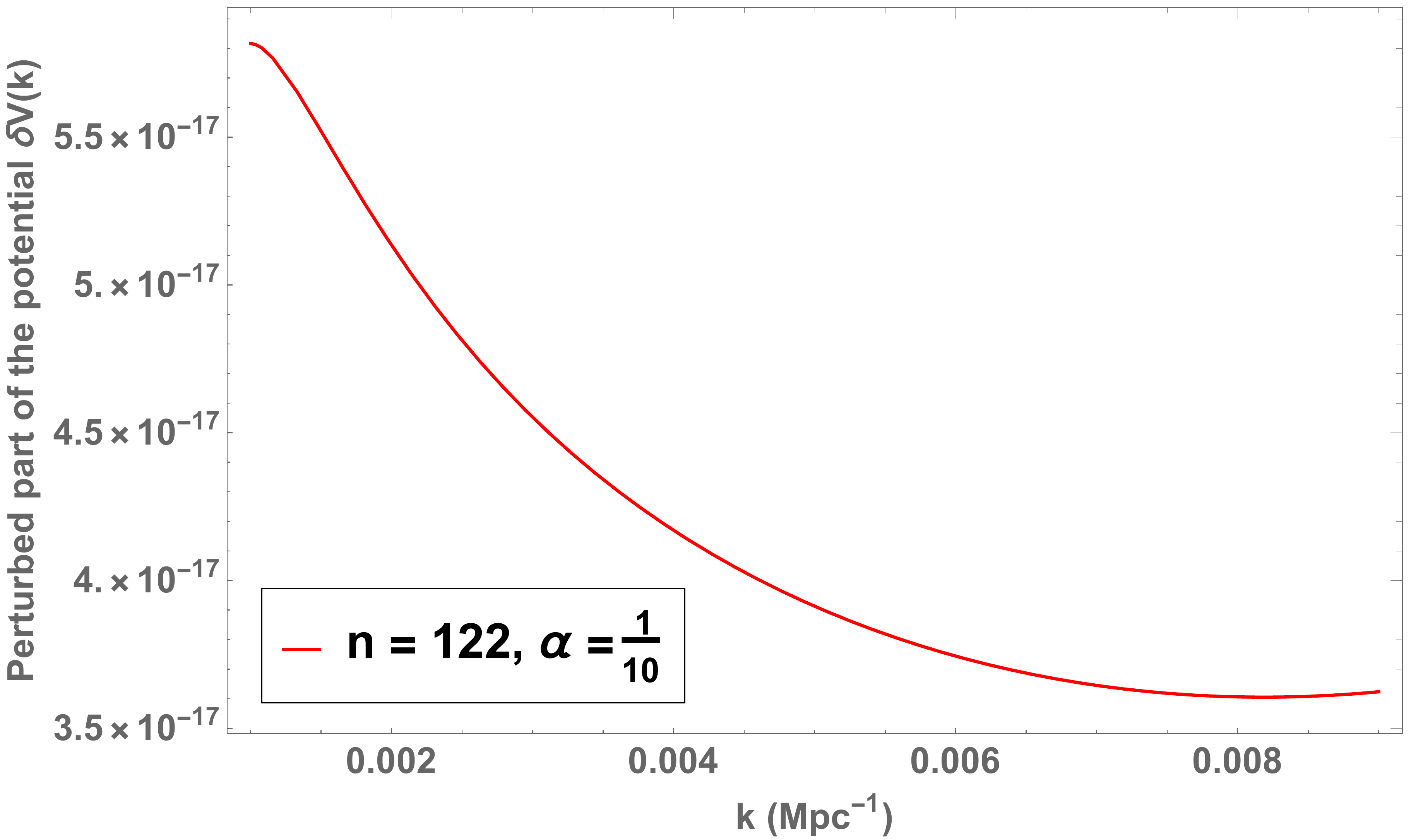} 
   \subcaption{}
   \label{fig:perturbedPOT_1}
\end{subfigure}%
\begin{subfigure}{0.52\linewidth}
  \centering
   \includegraphics[width=70mm,height=65mm]{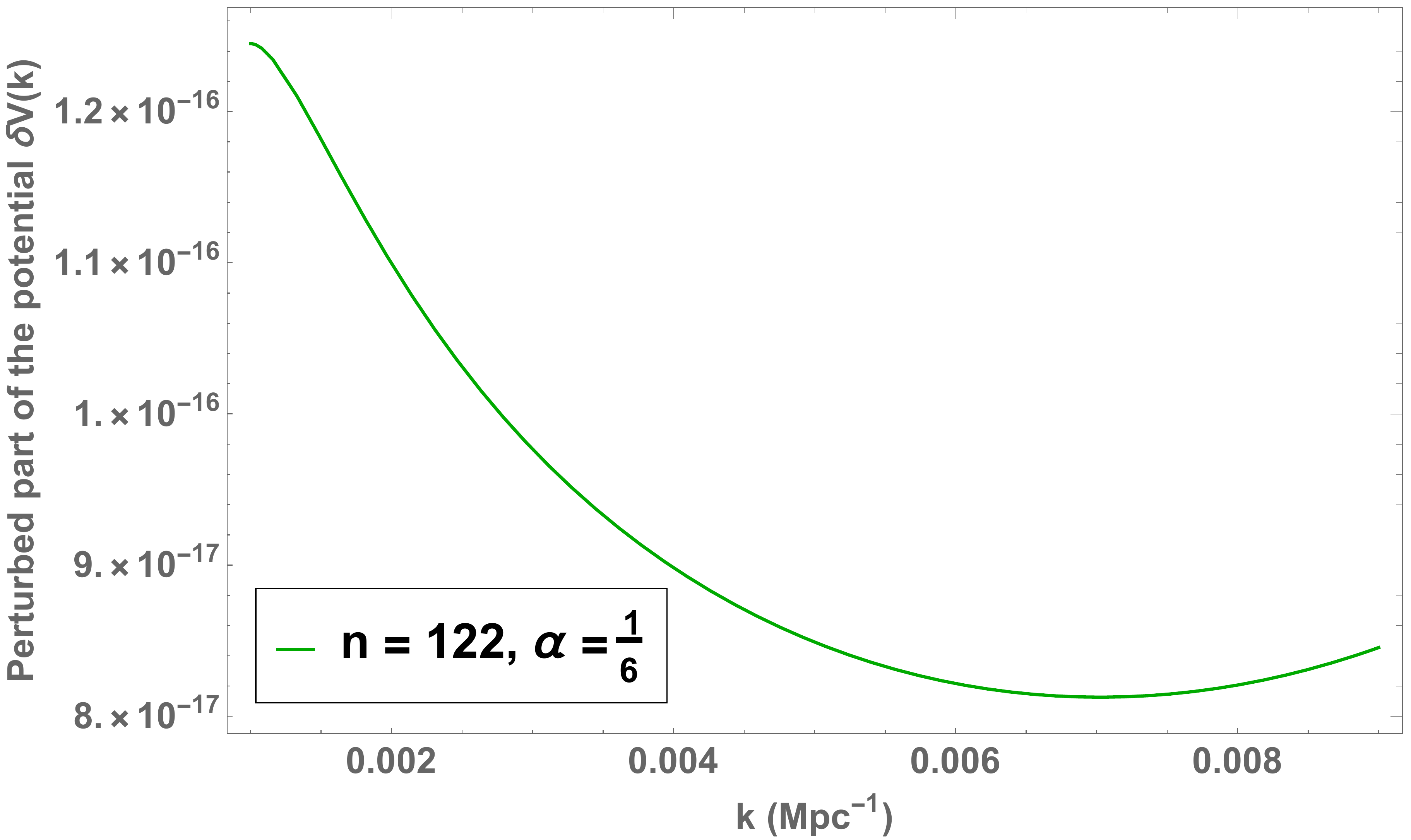}
   \subcaption{}
   \label{fig:perturbedPOT_2}
\end{subfigure}%
\vspace{0.1\linewidth}
\begin{subfigure}{0.52\linewidth}
  \centering
   \includegraphics[width=70mm,height=65mm]{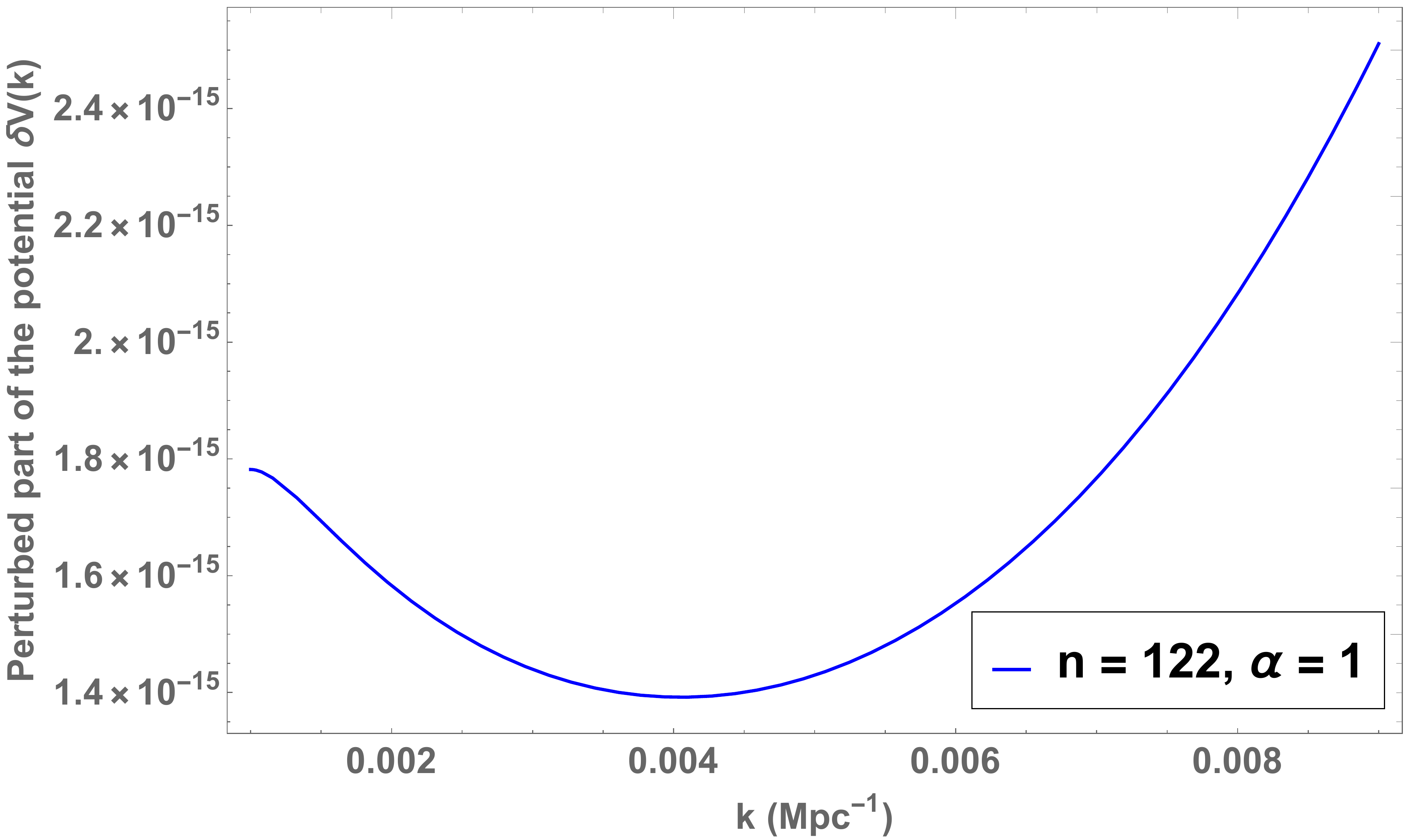}
   \subcaption{}
    \label{fig:perturbedPOT_3}
\end{subfigure}%
\begin{subfigure}{0.52\linewidth}
  \centering
   \includegraphics[width=70mm,height=65mm]{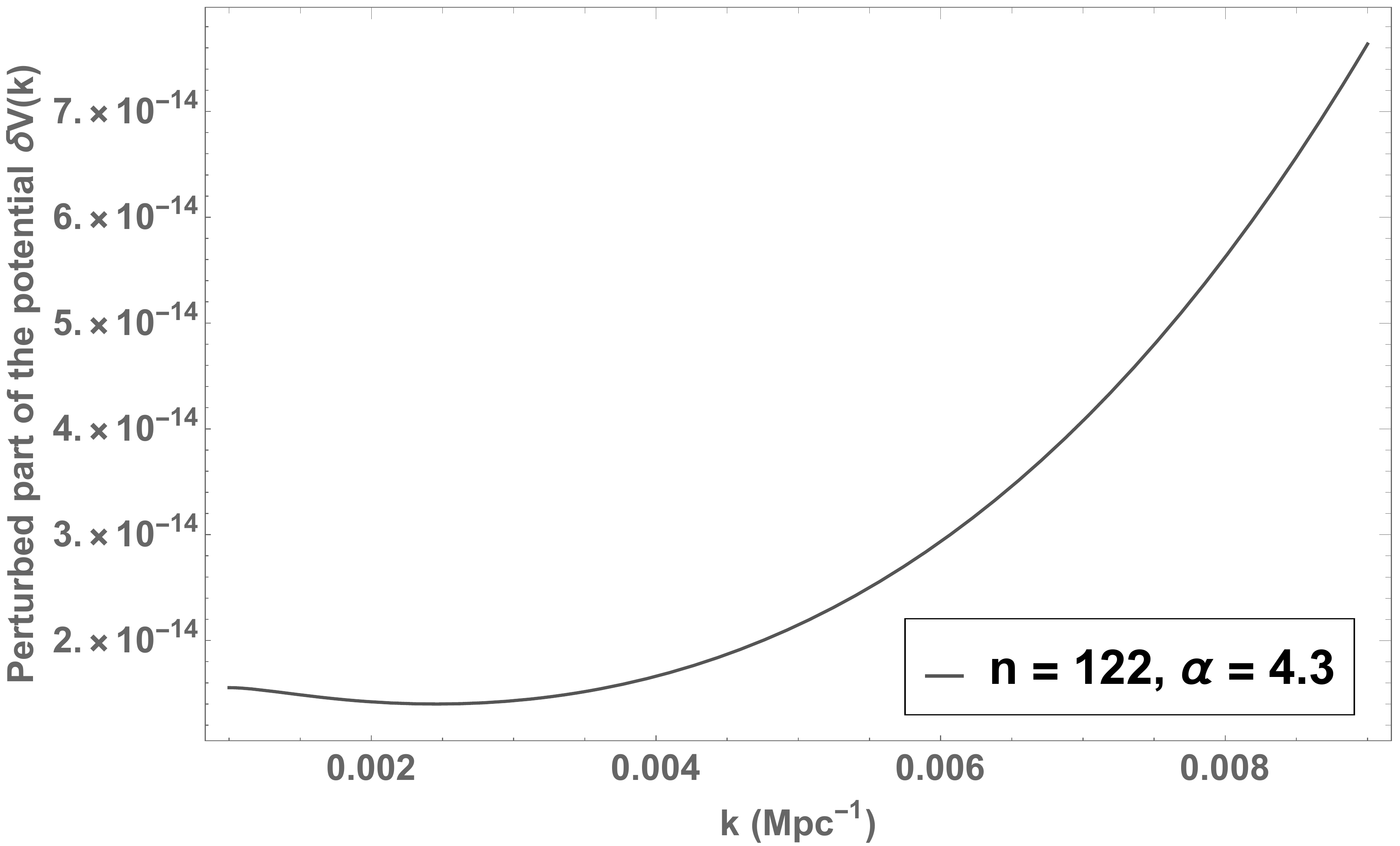}
   \subcaption{}
    \label{fig:perturbedPOT_4}
\end{subfigure}
\caption{First order perturbed parts of the potential (\ref{eq:Eq7}) for four values of $\alpha$ for a given value of $n$. The values of $\delta V(k)$ increase on increase in $\alpha$ for a particular value of $k$. The amount of perturbation is larger for small $k$'s than that of high $k$'s for $\alpha=1/10,1/6$ and likewise for $\alpha=1,4.3$ the corresponding values for small $k$'s are smaller than that of high $k$'s. }
\label{fig:perturbedPOT}
\end{figure}

\begin{figure}[H]
\begin{subfigure}{0.52\linewidth}
  \centering
   \includegraphics[width=70mm,height=65mm]{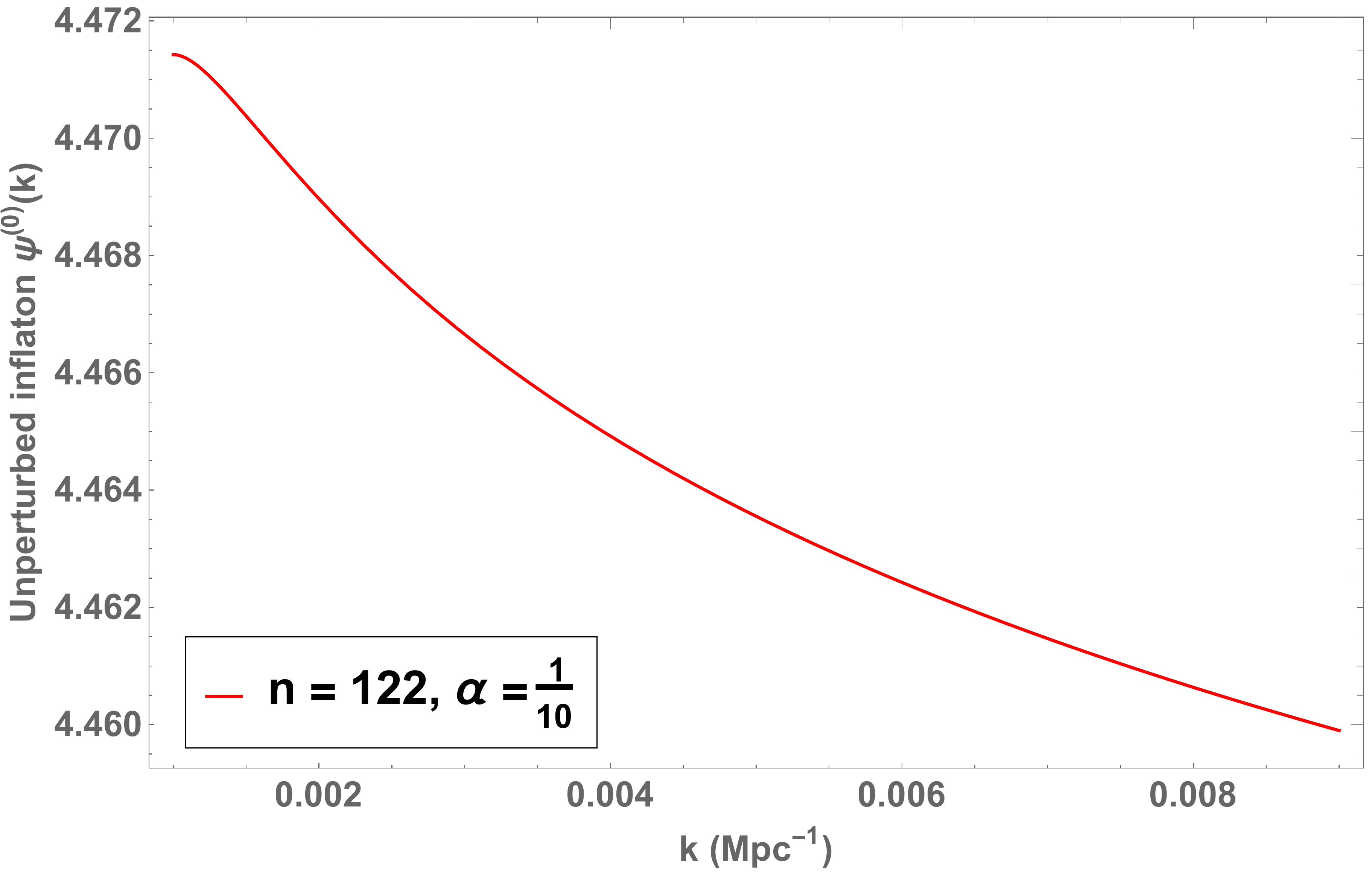} 
   \subcaption{}
   \label{fig:unperturbedINF_1}
\end{subfigure}%
\begin{subfigure}{0.52\linewidth}
  \centering
   \includegraphics[width=70mm,height=65mm]{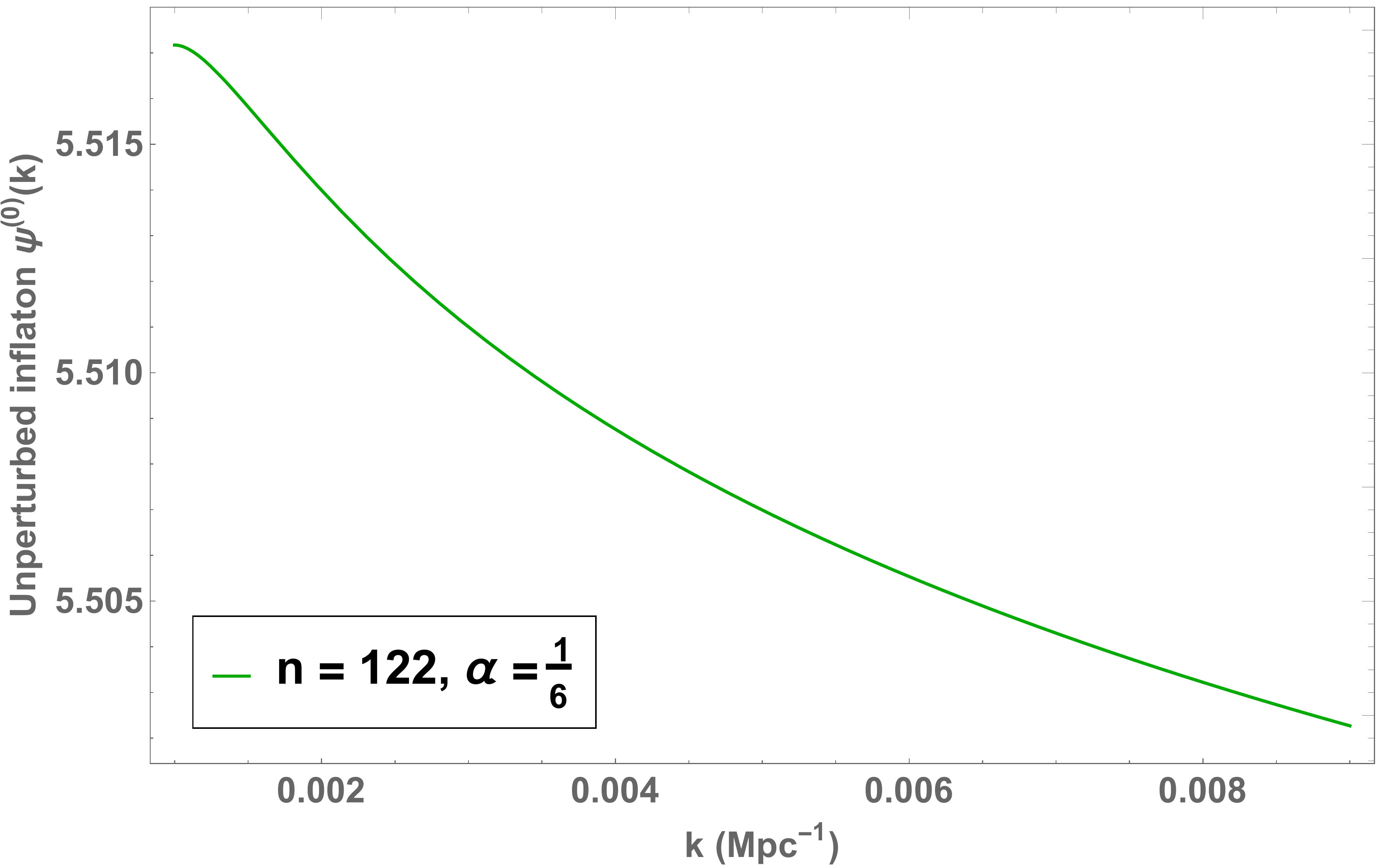}
   \subcaption{}
   \label{fig:unperturbedINF_2}
\end{subfigure}%
\vspace{0.1\linewidth}
\begin{subfigure}{0.52\linewidth}
  \centering
   \includegraphics[width=70mm,height=65mm]{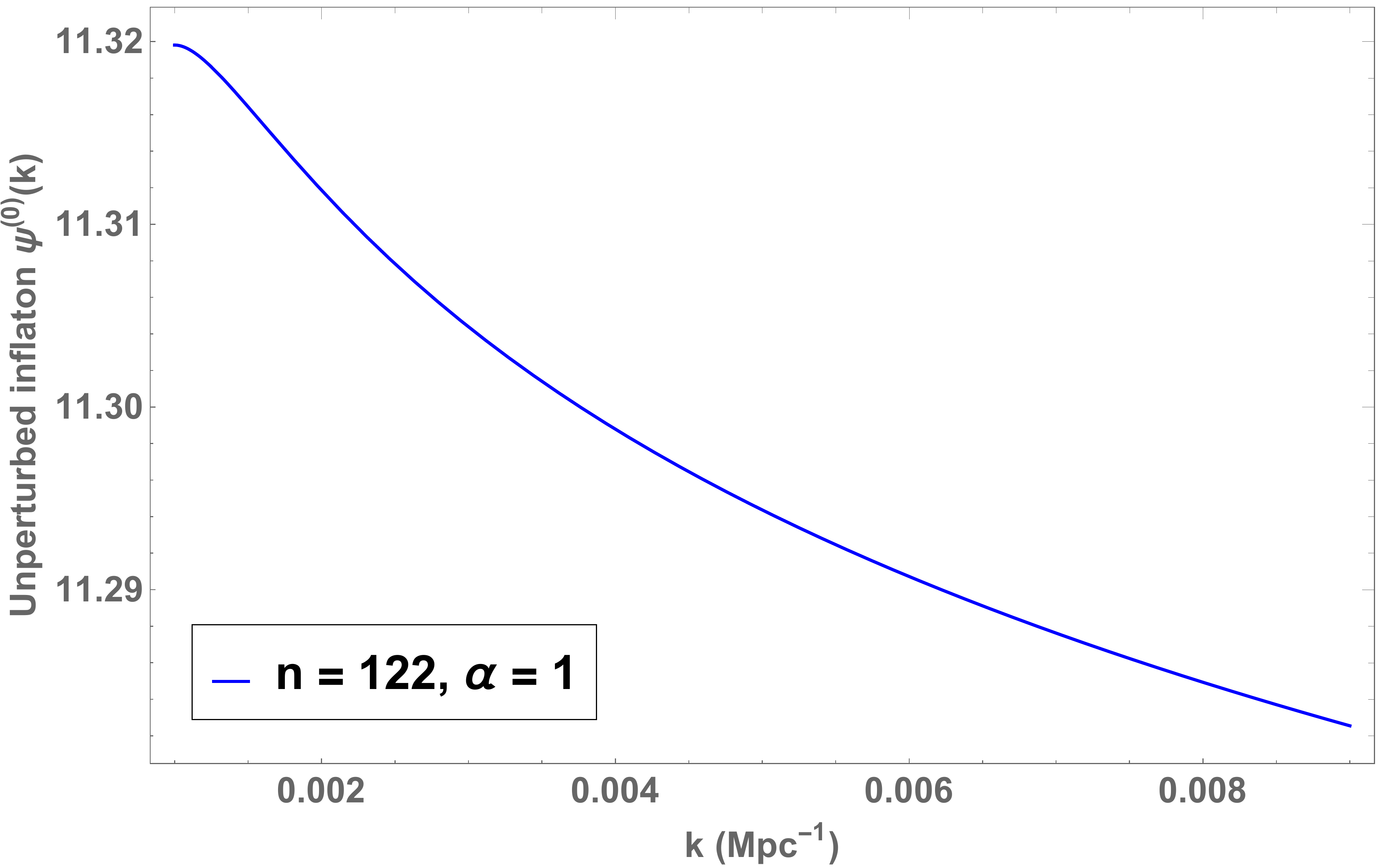}
   \subcaption{}
    \label{fig:unperturbedINF_3}
\end{subfigure}%
\begin{subfigure}{0.52\linewidth}
  \centering
   \includegraphics[width=70mm,height=65mm]{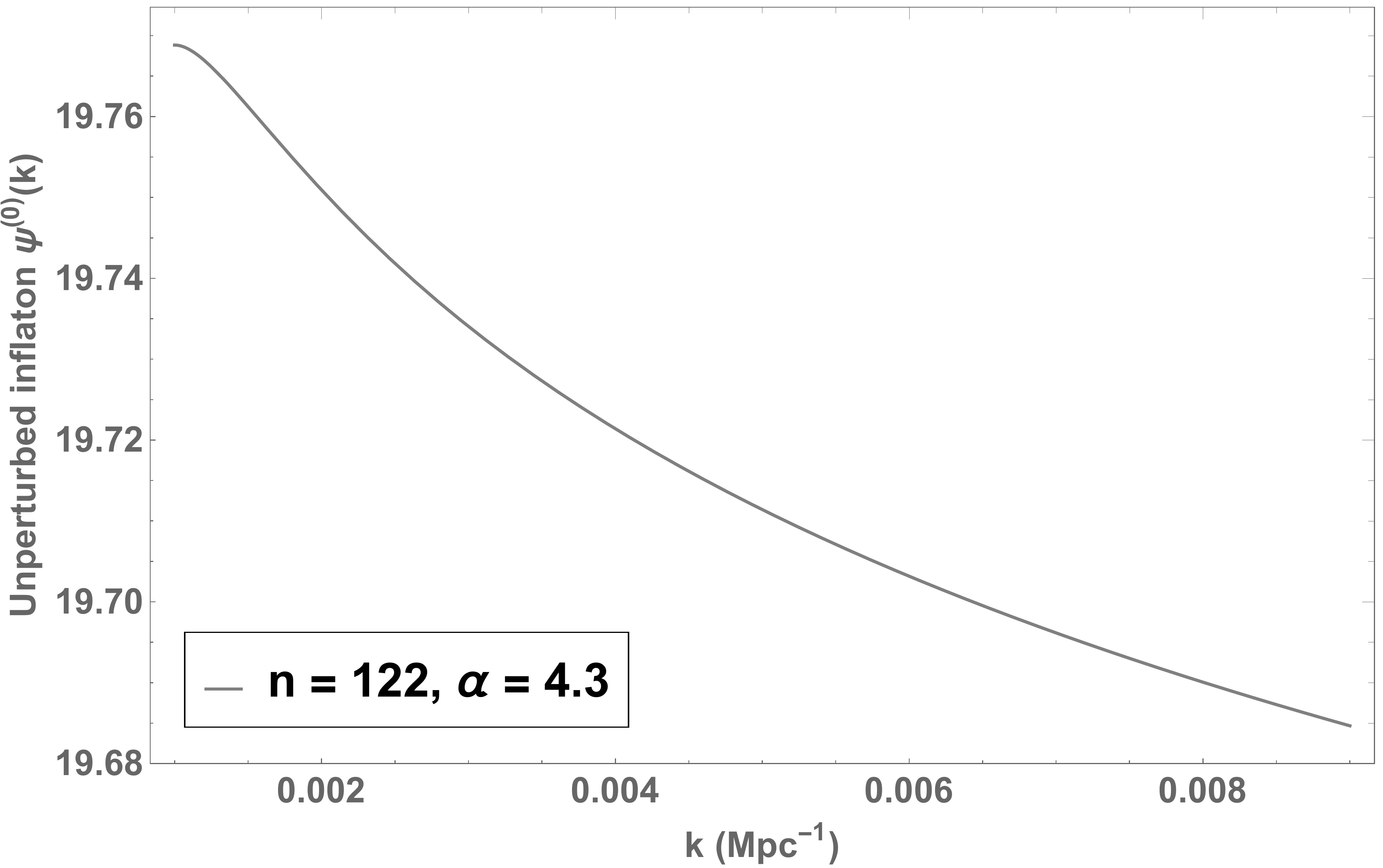}
   \subcaption{}
    \label{fig:unperturbedINF_4}
\end{subfigure}
\caption{Unperturbed parts of the inflaton field $\psi(k)$ for four values of $\alpha$ for a given value of $n$. The values of $\psi^{(0)}(k)$ increase on increase in $\alpha$ for a particular value of $k$.}
\label{fig:unperturbedINF}
\end{figure}

\begin{figure}[H]
\begin{subfigure}{0.52\linewidth}
  \centering
   \includegraphics[width=70mm,height=65mm]{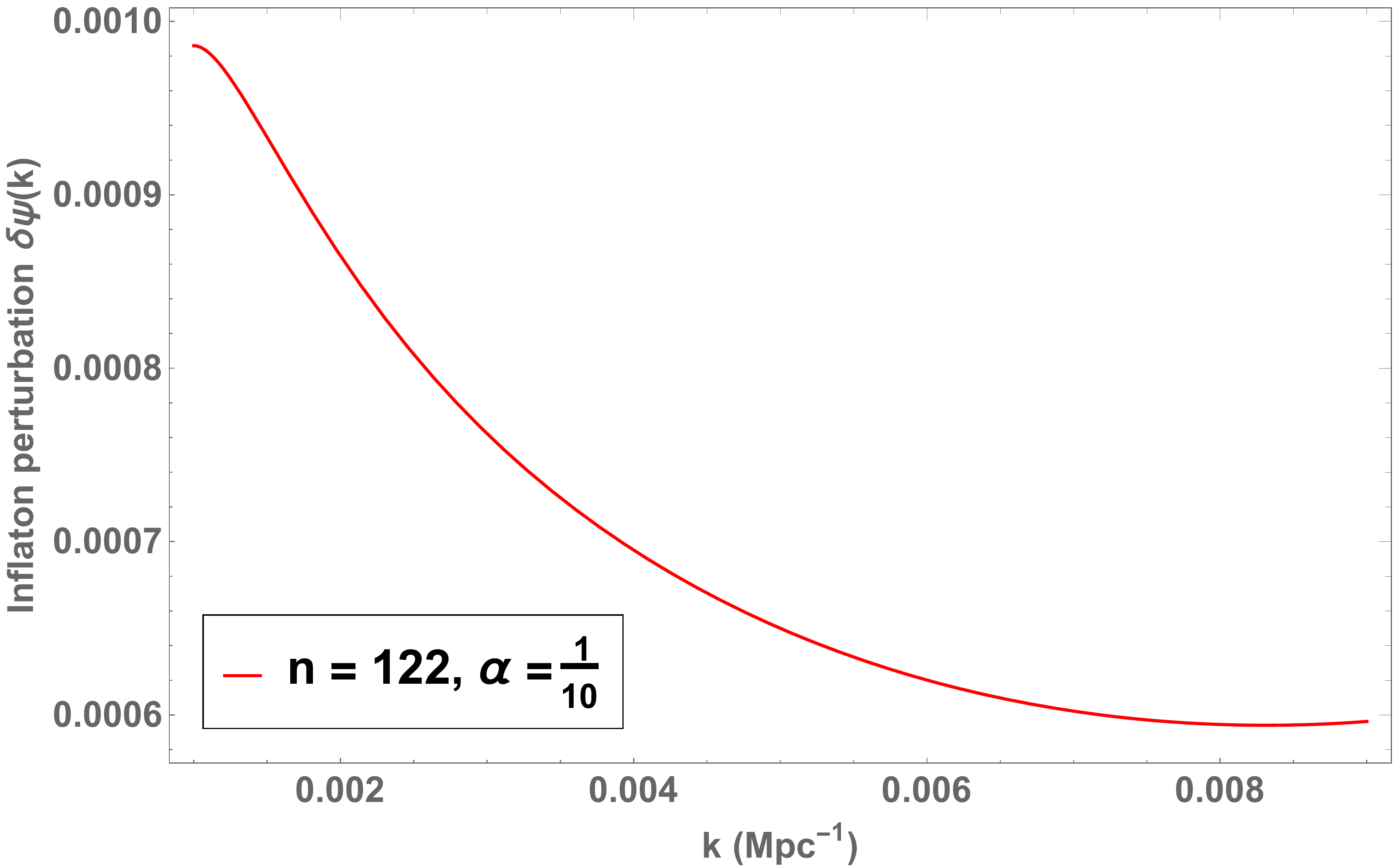} 
   \subcaption{}
   \label{fig:perturbedINF_1}
\end{subfigure}%
\begin{subfigure}{0.52\linewidth}
  \centering
   \includegraphics[width=70mm,height=65mm]{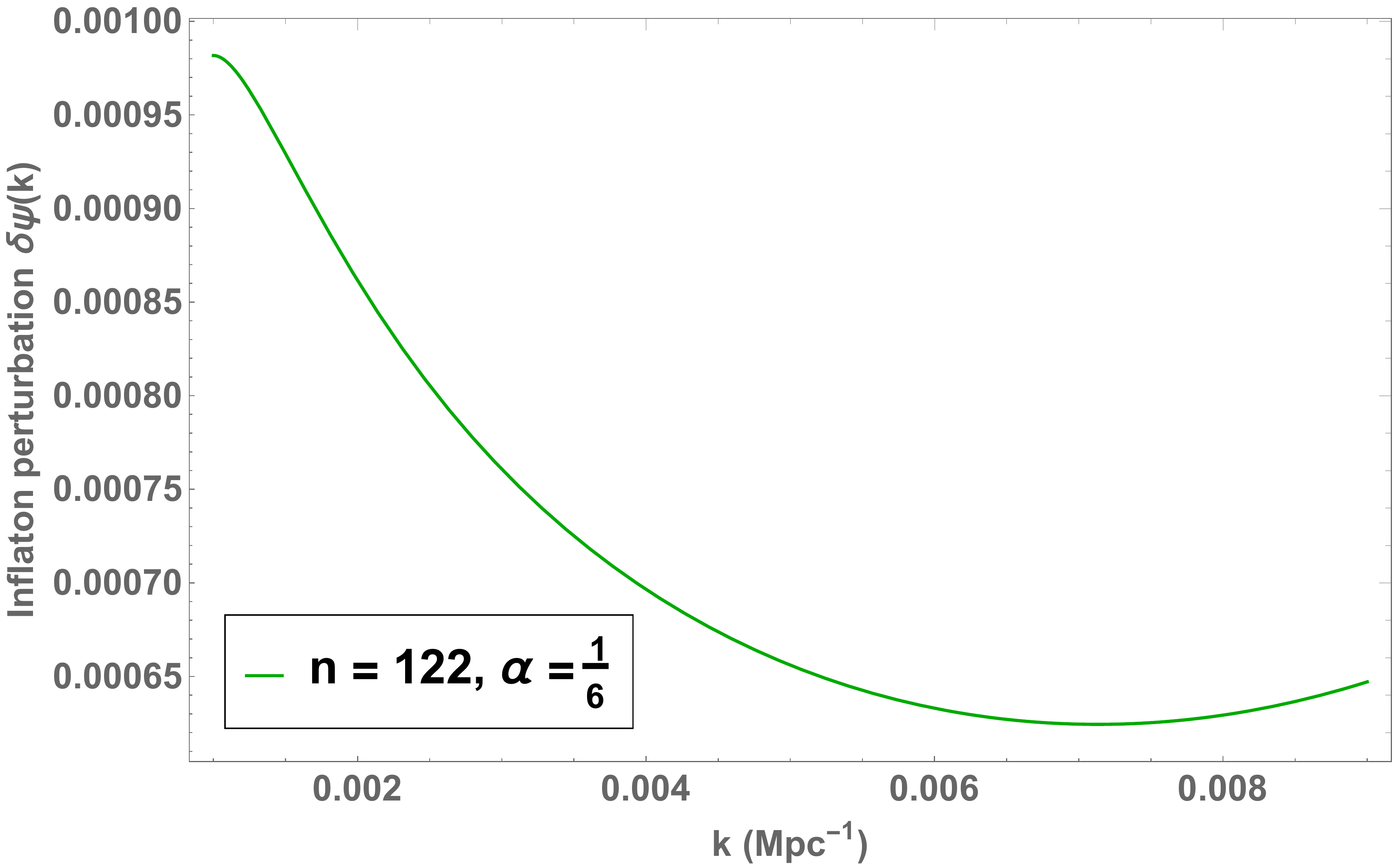}
   \subcaption{}
   \label{fig:perturbedINF_2}
\end{subfigure}%
\vspace{0.1\linewidth}
\begin{subfigure}{0.52\linewidth}
  \centering
   \includegraphics[width=70mm,height=65mm]{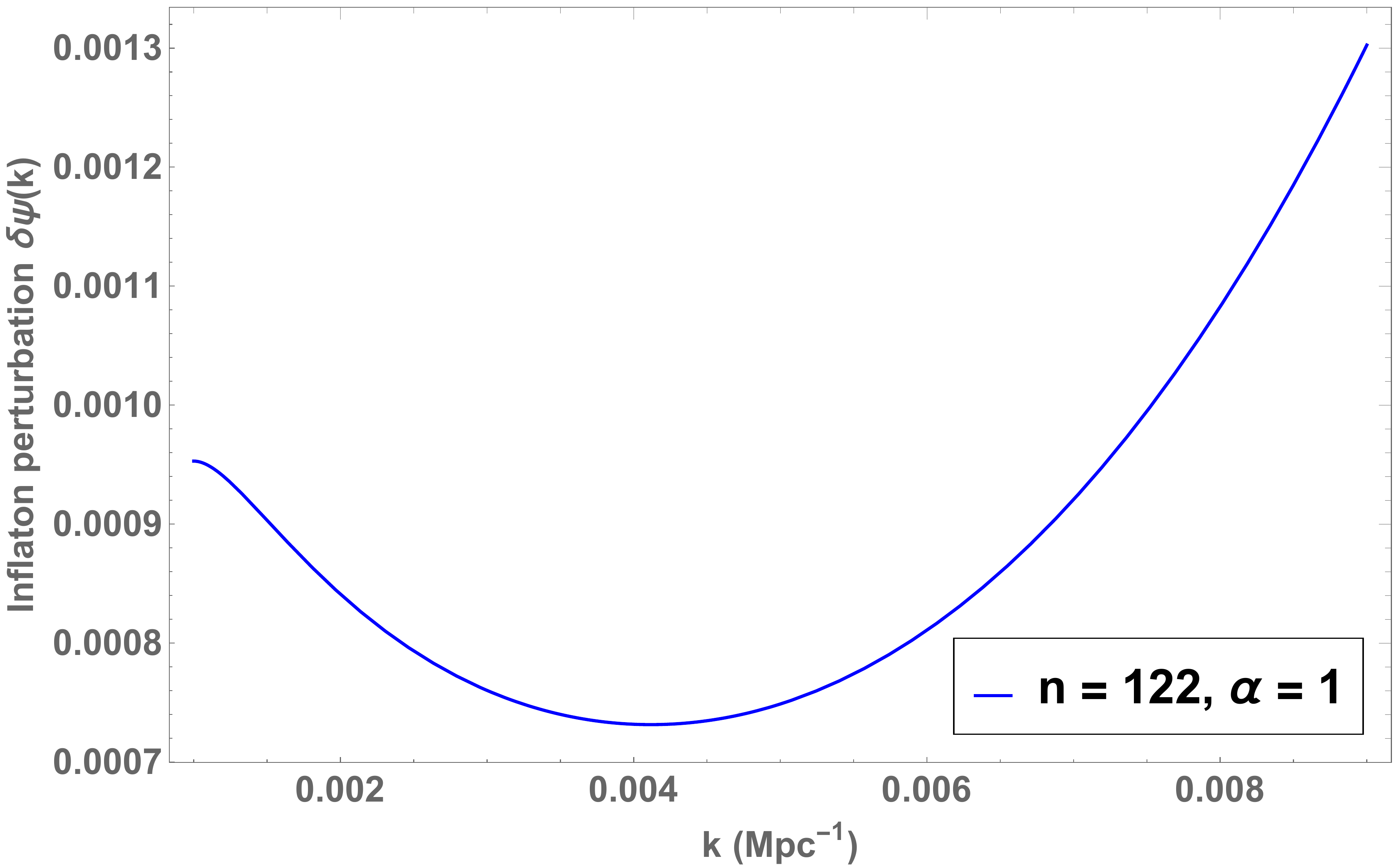}
   \subcaption{}
    \label{fig:perturbedINF_3}
\end{subfigure}%
\begin{subfigure}{0.52\linewidth}
  \centering
   \includegraphics[width=70mm,height=65mm]{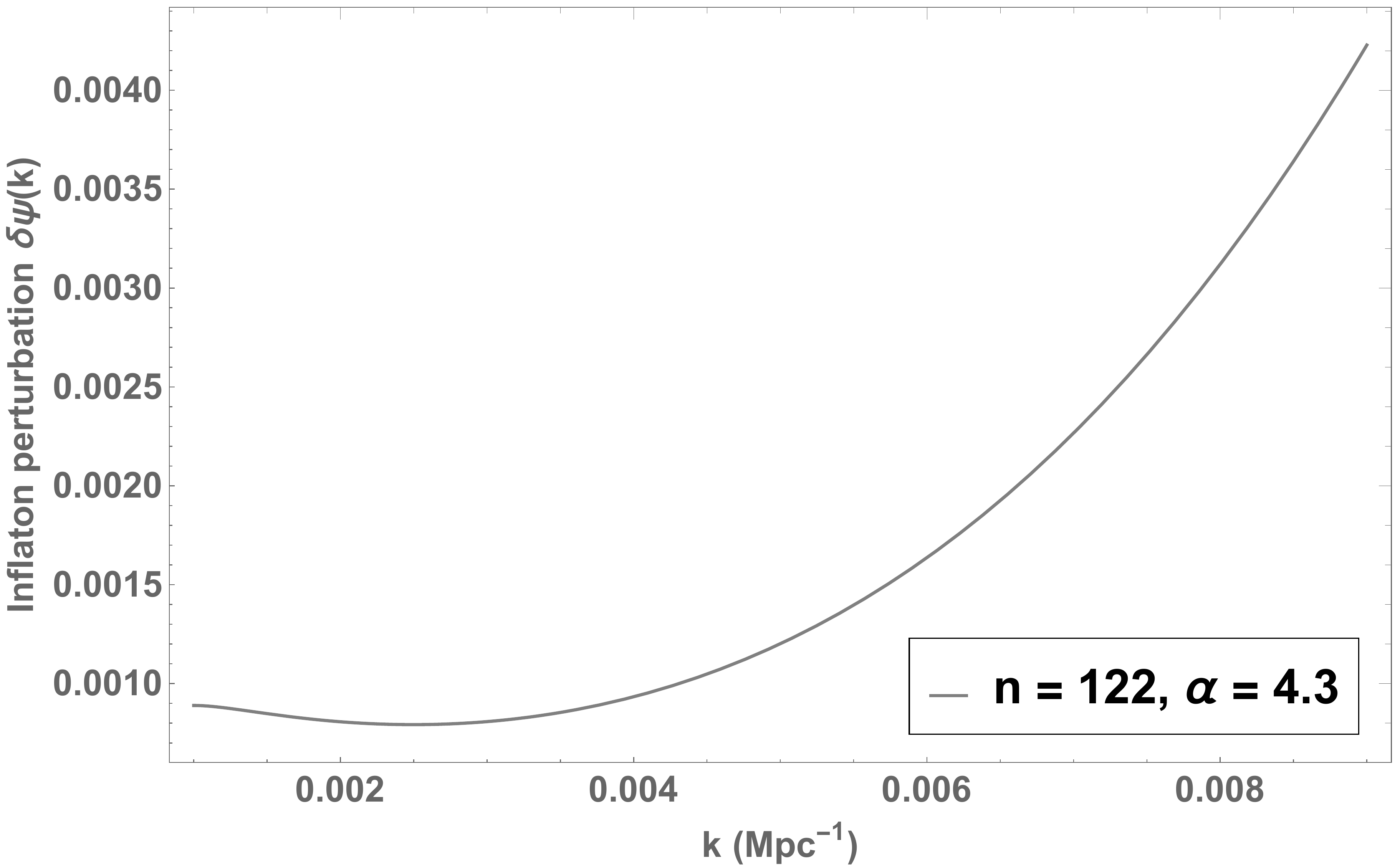}
   \subcaption{}
    \label{fig:perturbedINF_4}
\end{subfigure}
\caption{First order perturbed parts of the inflaton field $\psi(k)$ for four values of $\alpha$ for a given value of $n$. The values of $\delta \psi(k)$ increase on increase in $\alpha$ for a particular value of $k$. Just like $\delta V(k)$ here also the amount of inflaton perturbation is larger for small $k$'s than that of high $k$'s for $\alpha=1/10,1/6$ and for $\alpha=1,4.3$ the corresponding values for small $k$'s are smaller than that of high $k$'s.}
\label{fig:perturbedINF}
\end{figure}

\begin{figure}[H]
\begin{subfigure}{0.52\linewidth}
  \centering
   \includegraphics[width=70mm,height=65mm]{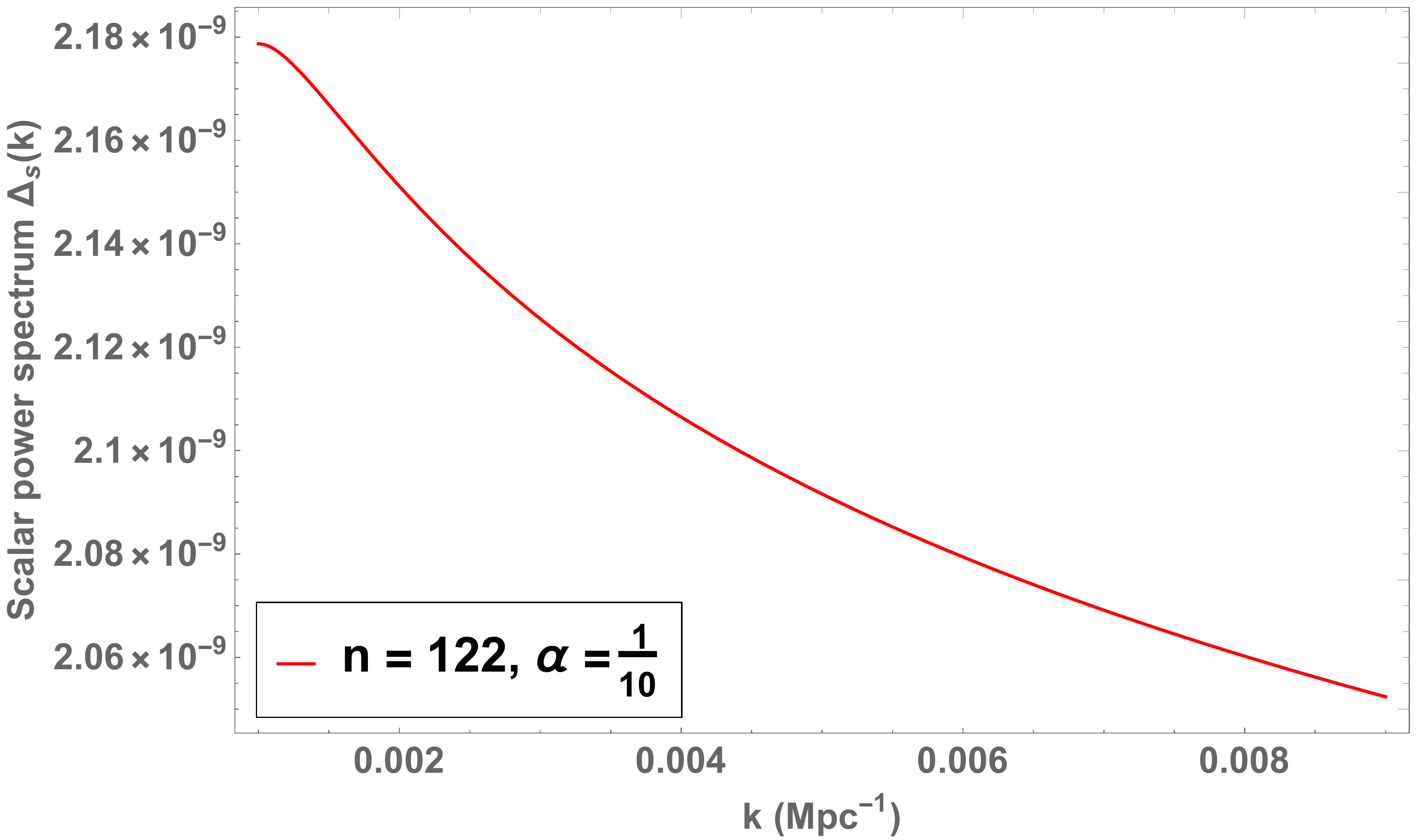} 
   \subcaption{}
   \label{fig:scalarPowerSpectrum_1}
\end{subfigure}%
\begin{subfigure}{0.52\linewidth}
  \centering
   \includegraphics[width=70mm,height=65mm]{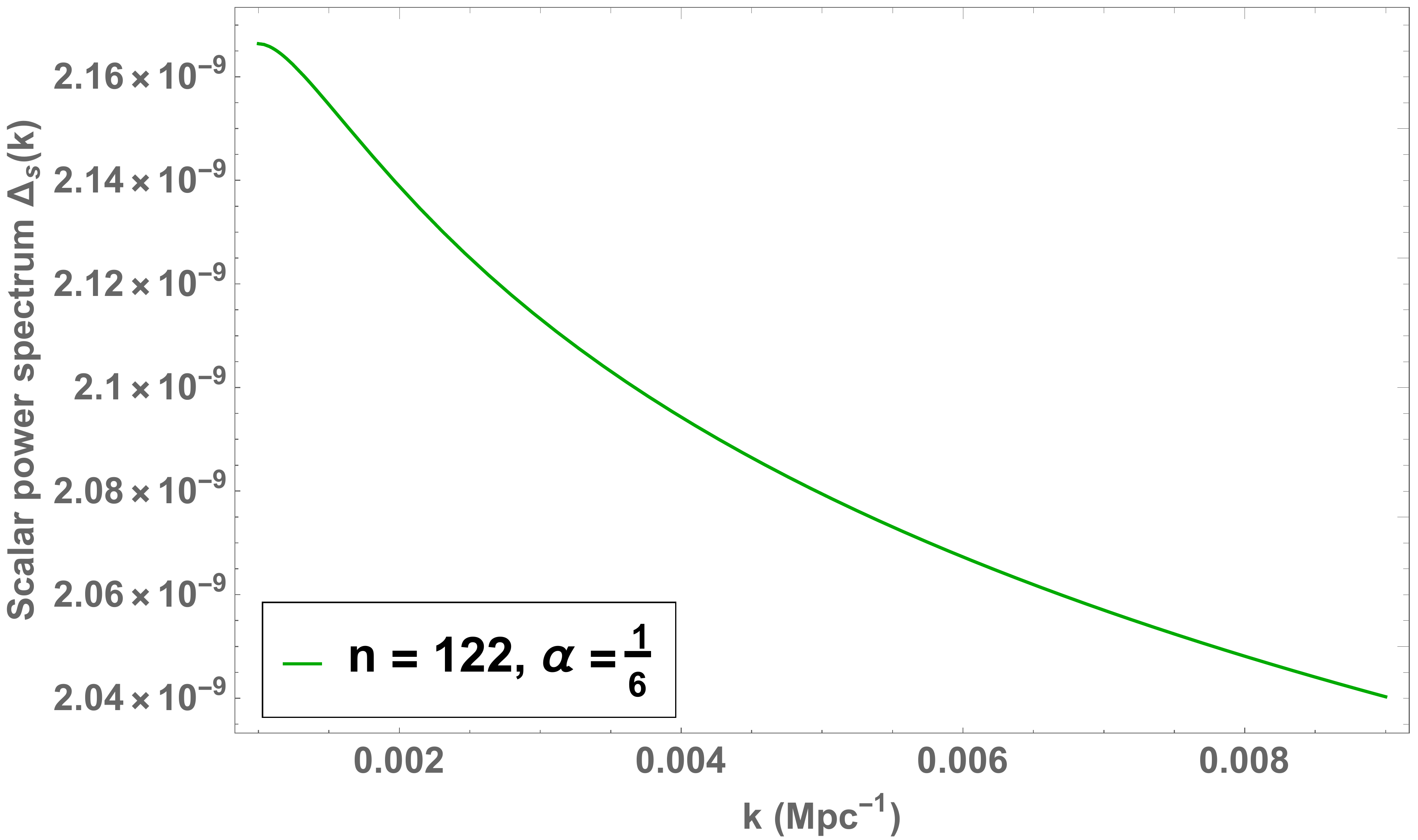}
   \subcaption{}
   \label{fig:scalarPowerSpectrum_2}
\end{subfigure}%
\vspace{0.1\linewidth}
\begin{subfigure}{0.52\linewidth}
  \centering
   \includegraphics[width=70mm,height=65mm]{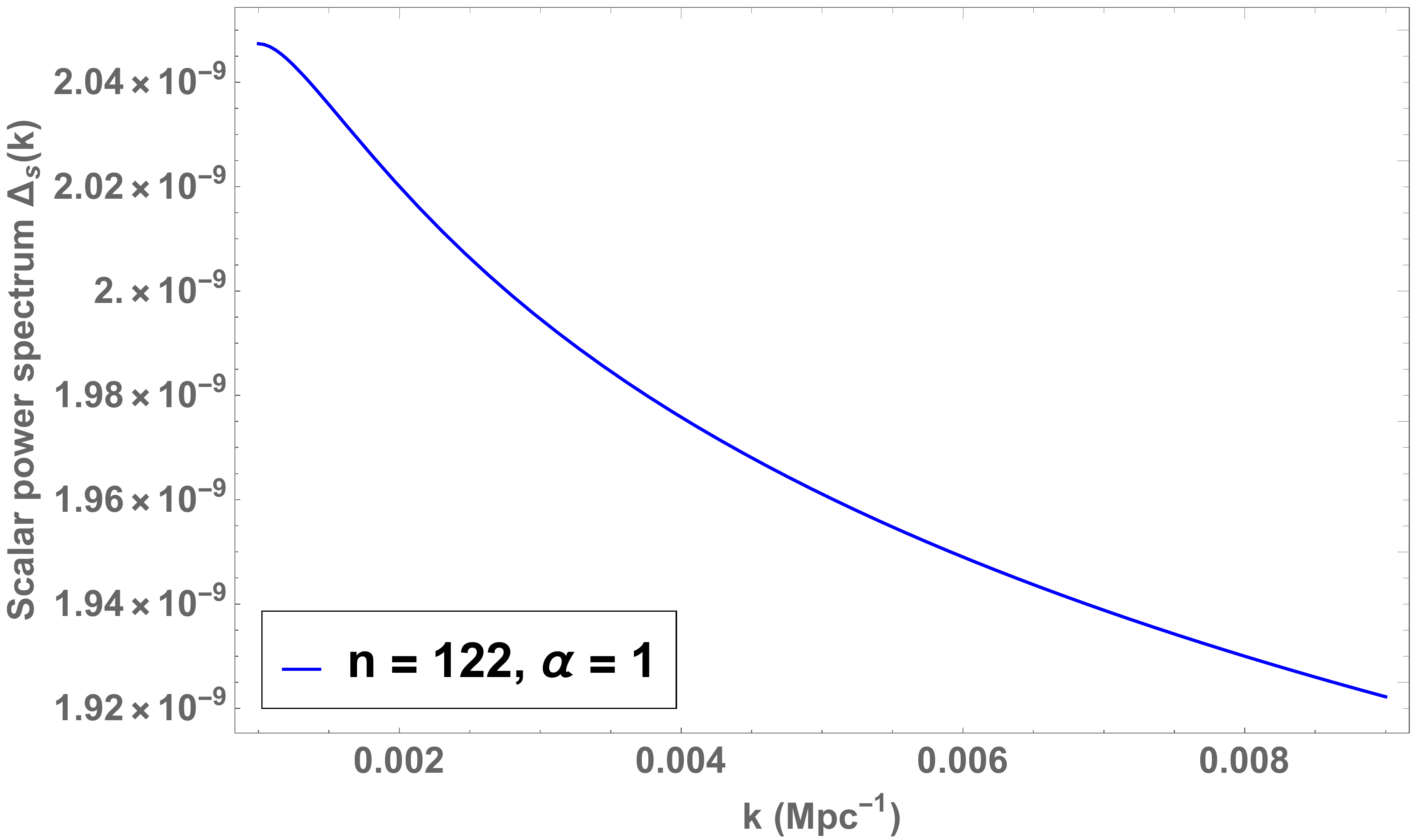}
   \subcaption{}
    \label{fig:scalarPowerSpectrum_3}
\end{subfigure}%
\begin{subfigure}{0.52\linewidth}
  \centering
   \includegraphics[width=70mm,height=65mm]{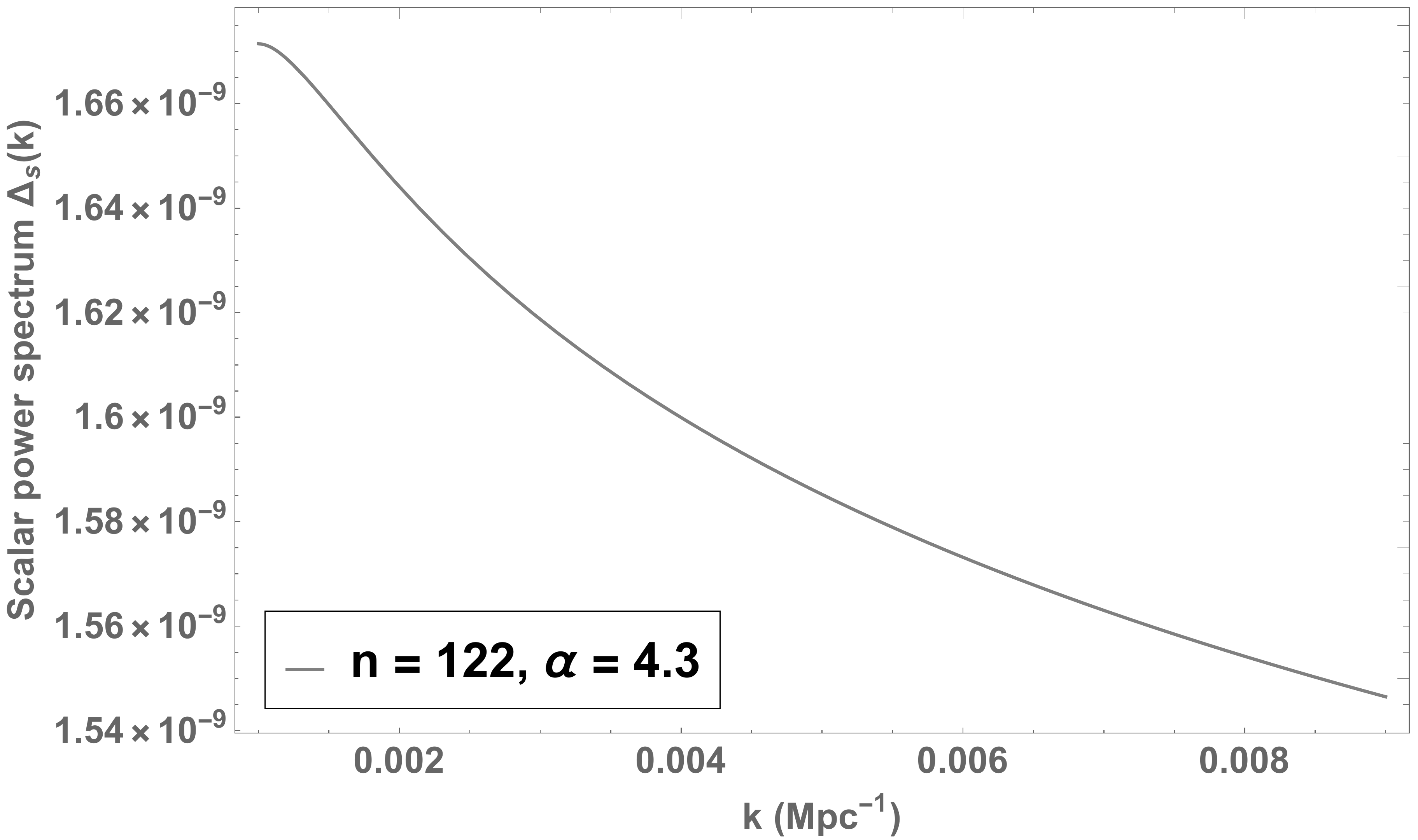}
   \subcaption{}
    \label{fig:scalarPowerSpectrum_4}
\end{subfigure}
\caption{Scalar power spectra for four values of $\alpha$ for a given value of $n$. The values of $\Delta_s(k)$ decrease on increase in $\alpha$ for a particular value of $k$.}
\label{fig:scalarPowerSpectrum}
\end{figure}

\begin{figure}[H]
\begin{subfigure}{0.52\linewidth}
  \centering
   \includegraphics[width=70mm,height=65mm]{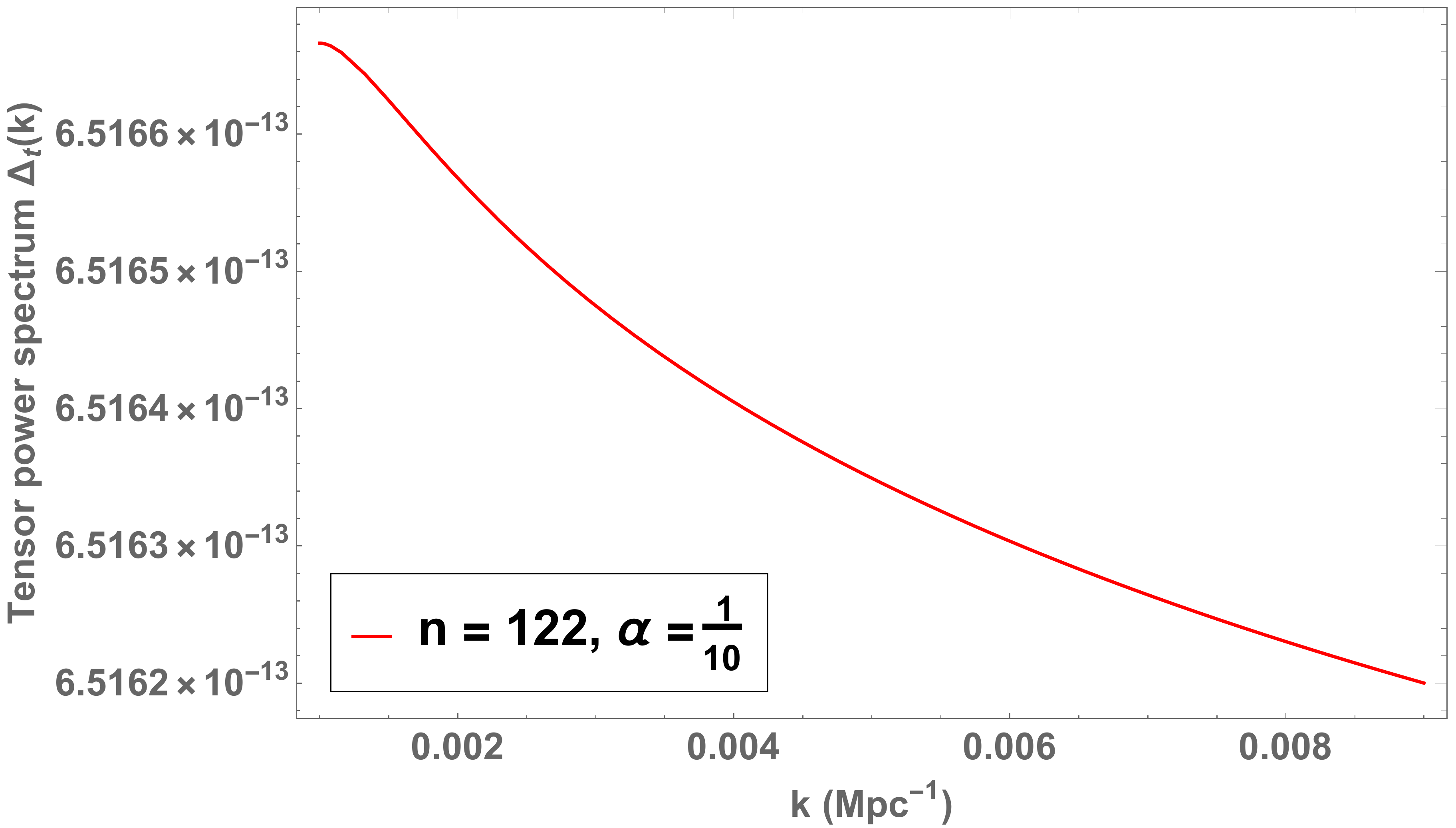} 
   \subcaption{}
   \label{fig:tensorPowerSpectrum_1}
\end{subfigure}%
\begin{subfigure}{0.52\linewidth}
  \centering
   \includegraphics[width=70mm,height=65mm]{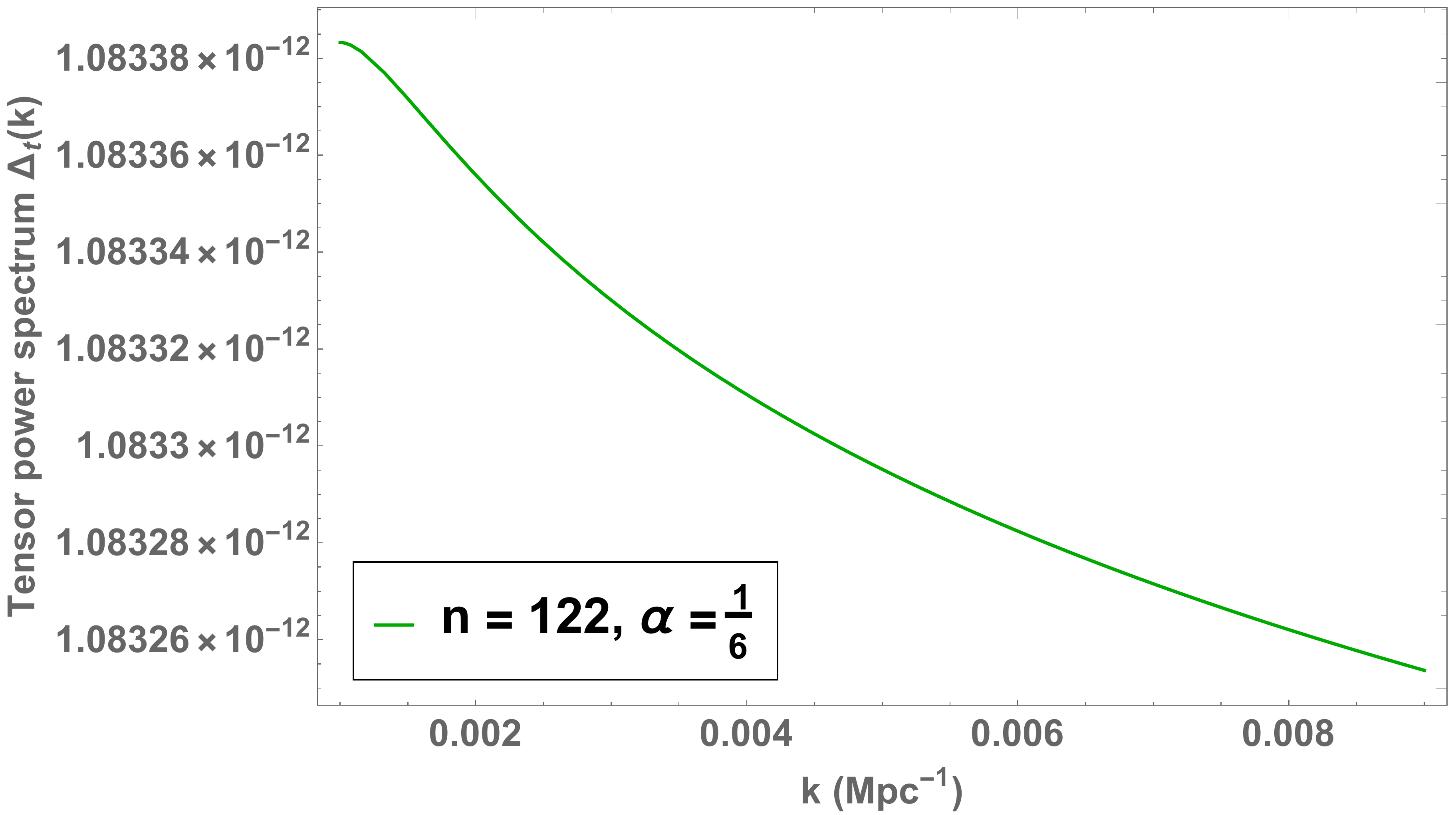}
   \subcaption{}
   \label{fig:tensorPowerSpectrum_2}
\end{subfigure}%
\vspace{0.1\linewidth}
\begin{subfigure}{0.52\linewidth}
  \centering
   \includegraphics[width=70mm,height=65mm]{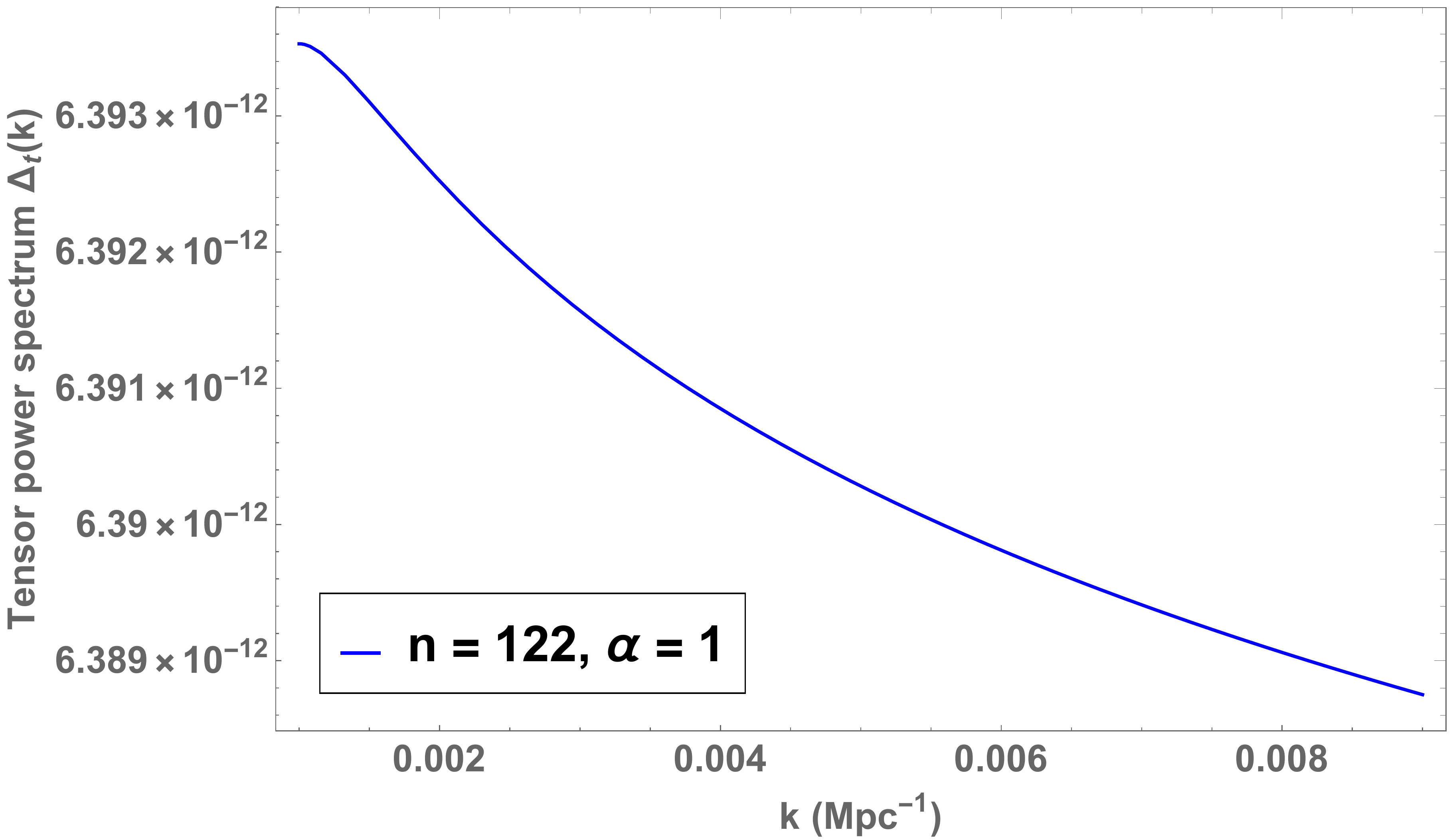}
   \subcaption{}
    \label{fig:tensorPowerSpectrum_3}
\end{subfigure}%
\begin{subfigure}{0.52\linewidth}
  \centering
   \includegraphics[width=70mm,height=65mm]{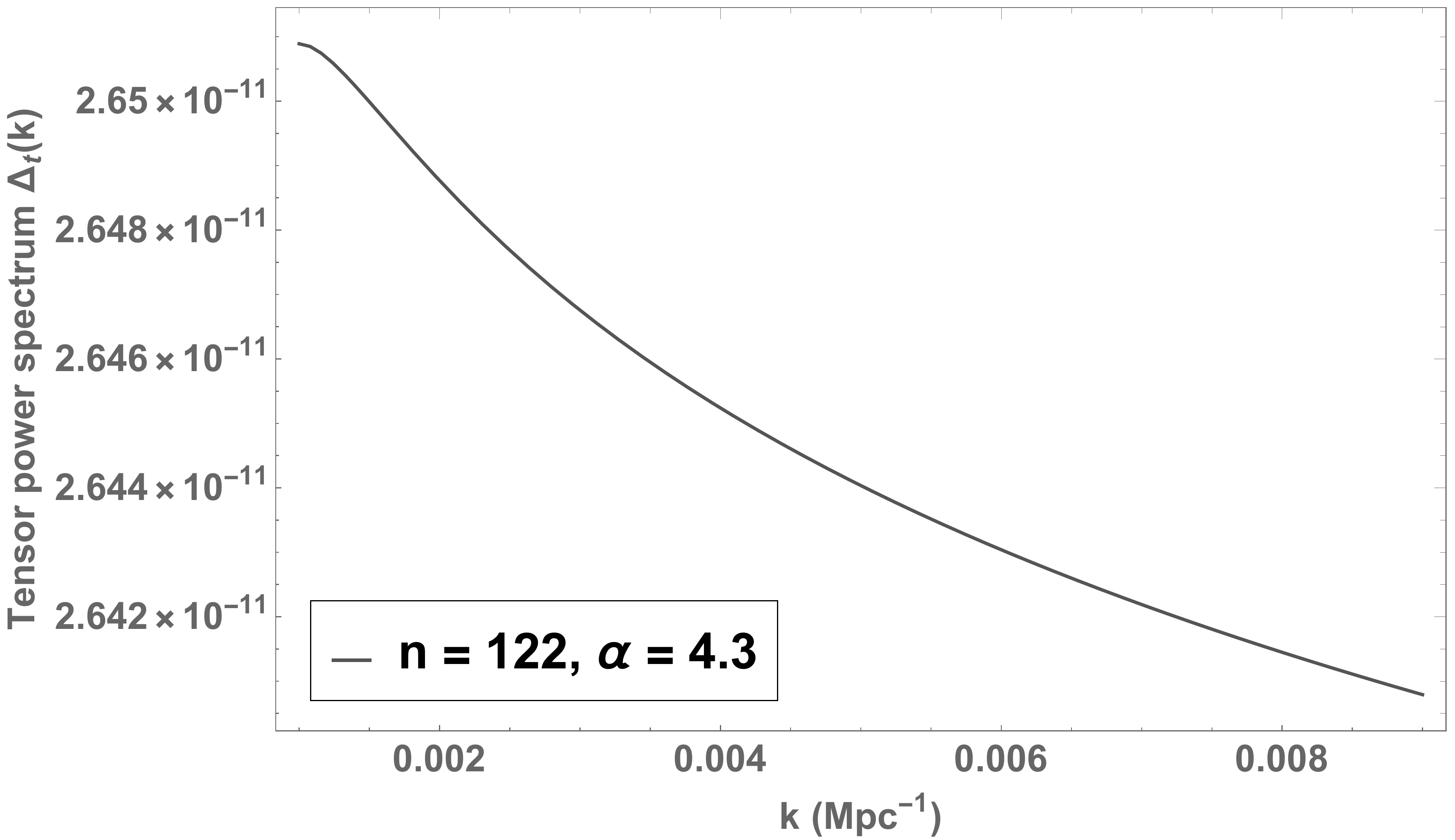}
   \subcaption{}
    \label{fig:tensorPowerSpectrum_4}
\end{subfigure}
\caption{Tensor power spectra for four values of $\alpha$ for a given value of $n$. The values of $\Delta_t(k)$ increase on increase in $\alpha$ for a particular value of $k$.}
\label{fig:tensorPowerSpectrum}
\end{figure}

\begin{figure}[H]
\begin{subfigure}{0.52\linewidth}
  \centering
   \includegraphics[width=70mm,height=65mm]{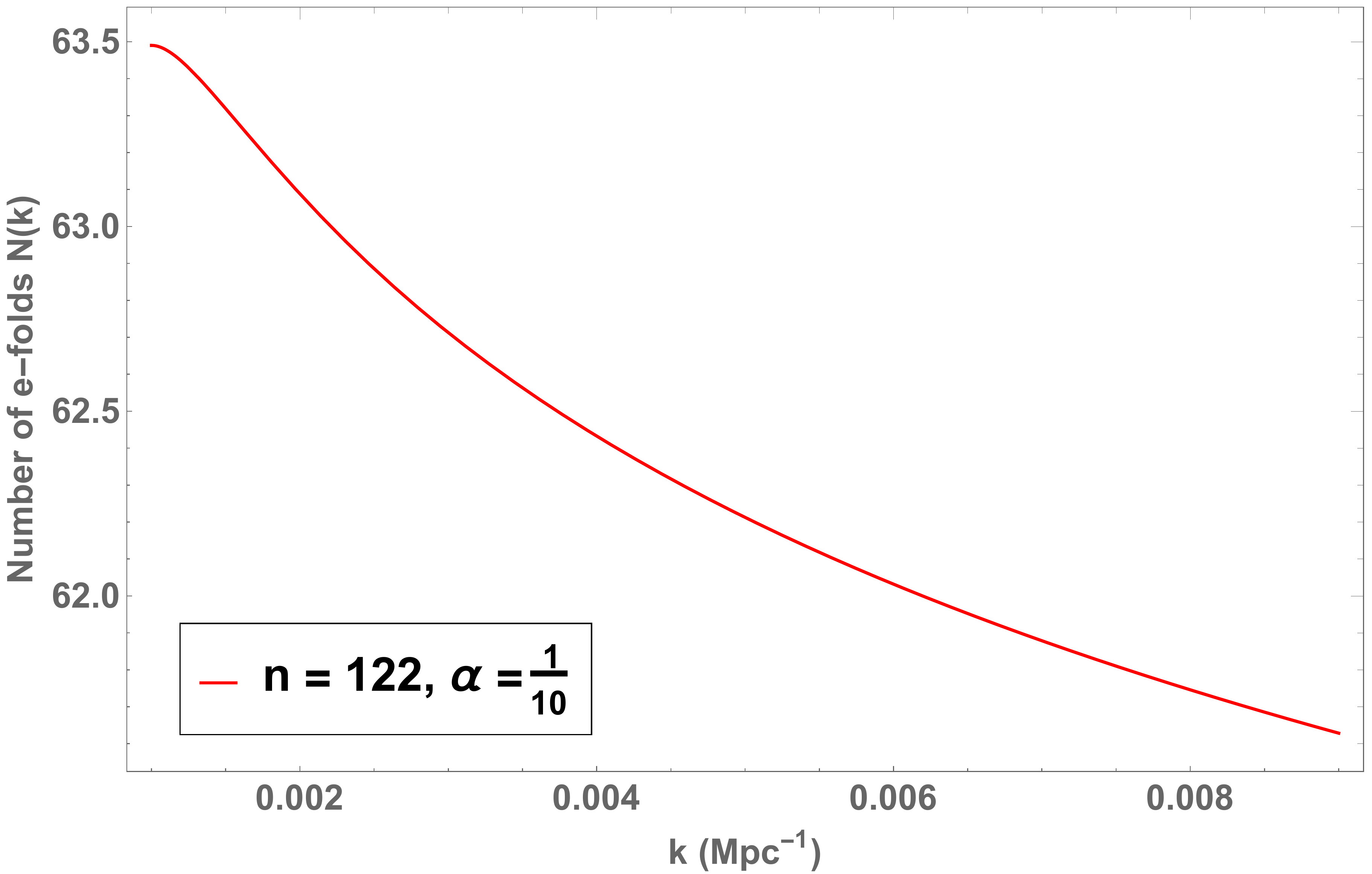} 
   \subcaption{}
   \label{fig:numberOfEFolds_1}
\end{subfigure}%
\begin{subfigure}{0.52\linewidth}
  \centering
   \includegraphics[width=70mm,height=65mm]{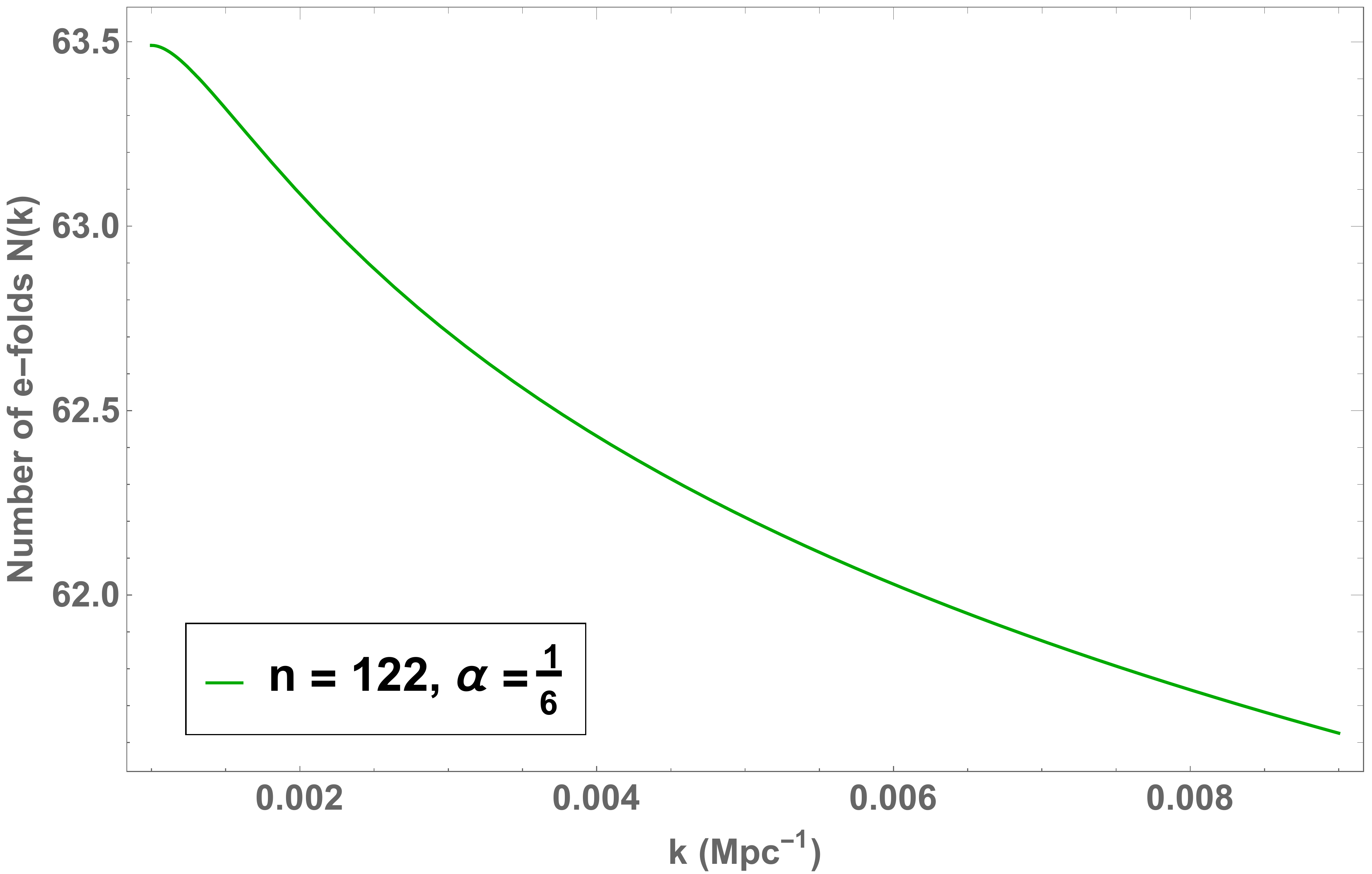}
   \subcaption{}
   \label{fig:numberOfEFolds_2}
\end{subfigure}%
\vspace{0.1\linewidth}
\begin{subfigure}{0.52\linewidth}
  \centering
   \includegraphics[width=70mm,height=65mm]{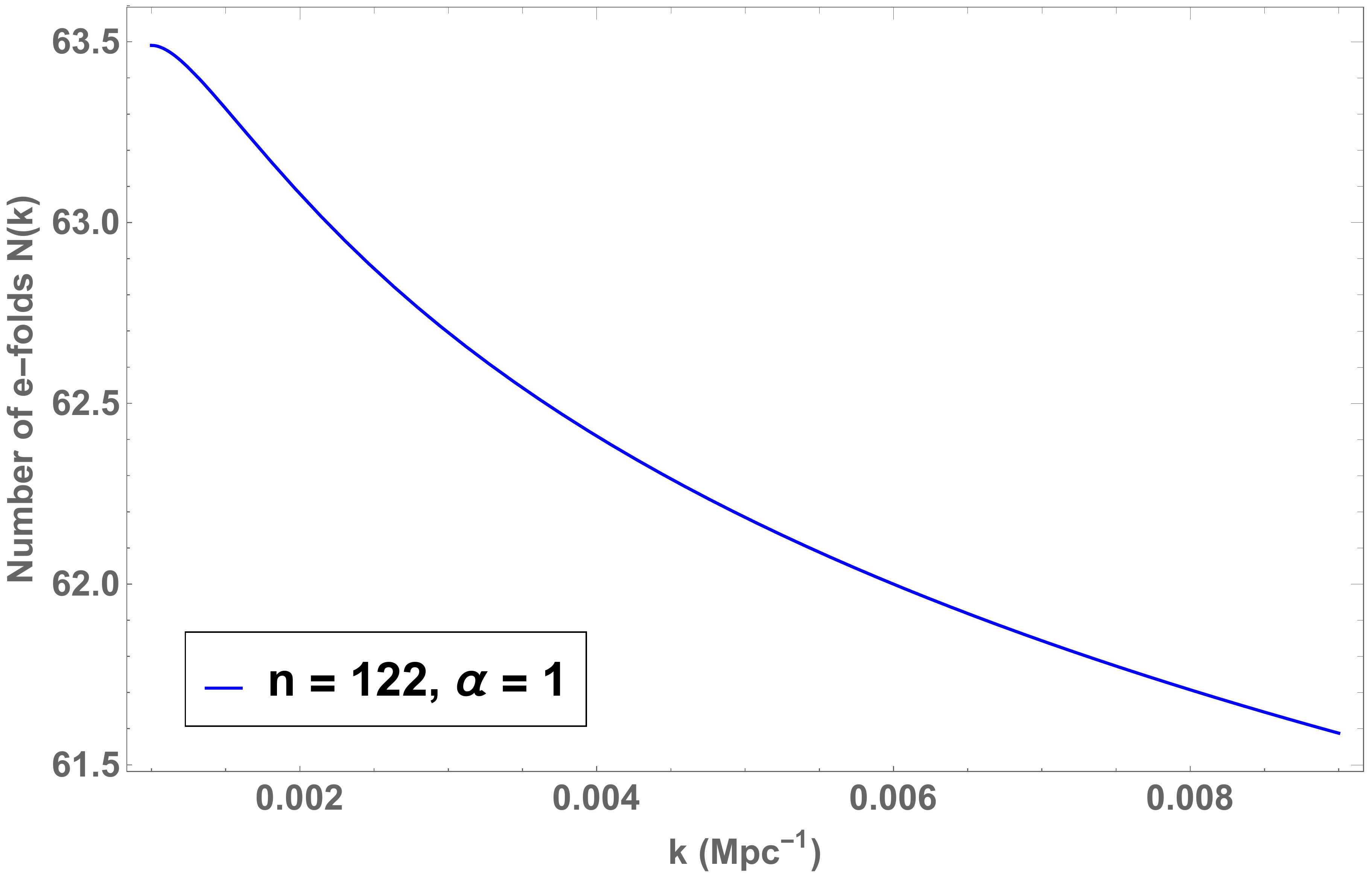}
   \subcaption{}
    \label{fig:numberOfEFolds_3}
\end{subfigure}%
\begin{subfigure}{0.52\linewidth}
  \centering
   \includegraphics[width=70mm,height=65mm]{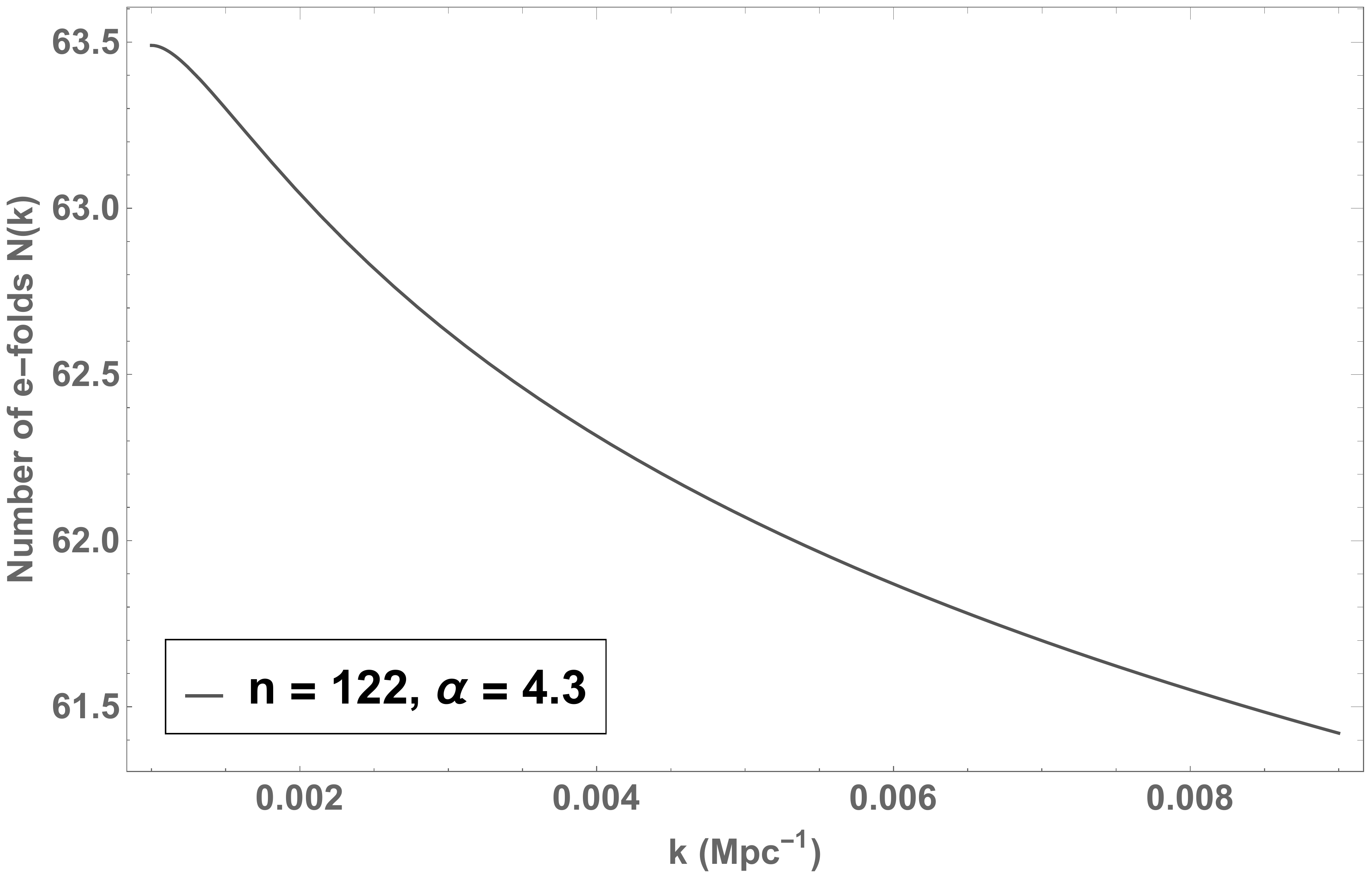}
   \subcaption{}
    \label{fig:numberOfEFolds_4}
\end{subfigure}
\caption{Number of e-folds for four values of $\alpha$ for a given value of $n$. The values of $N(k)$ remain almost same for $\alpha=\frac{1}{10},\frac{1}{6}$ and $\alpha=1,4.3$ for a particular value of $k$.}
\label{fig:numberOfEFolds}
\end{figure}

\begin{figure}[H]
\begin{subfigure}{0.52\linewidth}
  \centering
   \includegraphics[width=70mm,height=65mm]{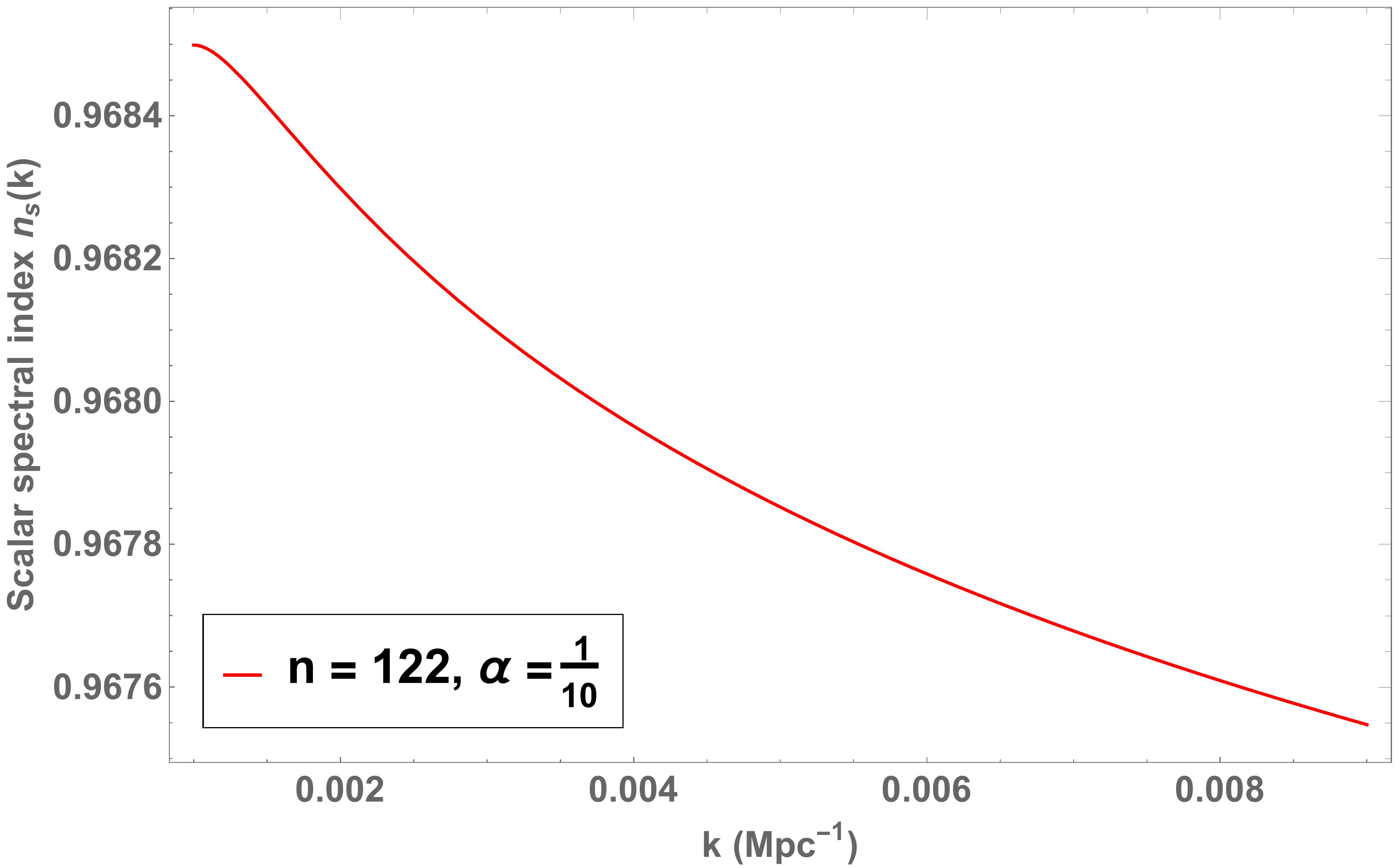} 
   \subcaption{}
   \label{fig:scalarSpectralIndex_1}
\end{subfigure}%
\begin{subfigure}{0.52\linewidth}
  \centering
   \includegraphics[width=70mm,height=65mm]{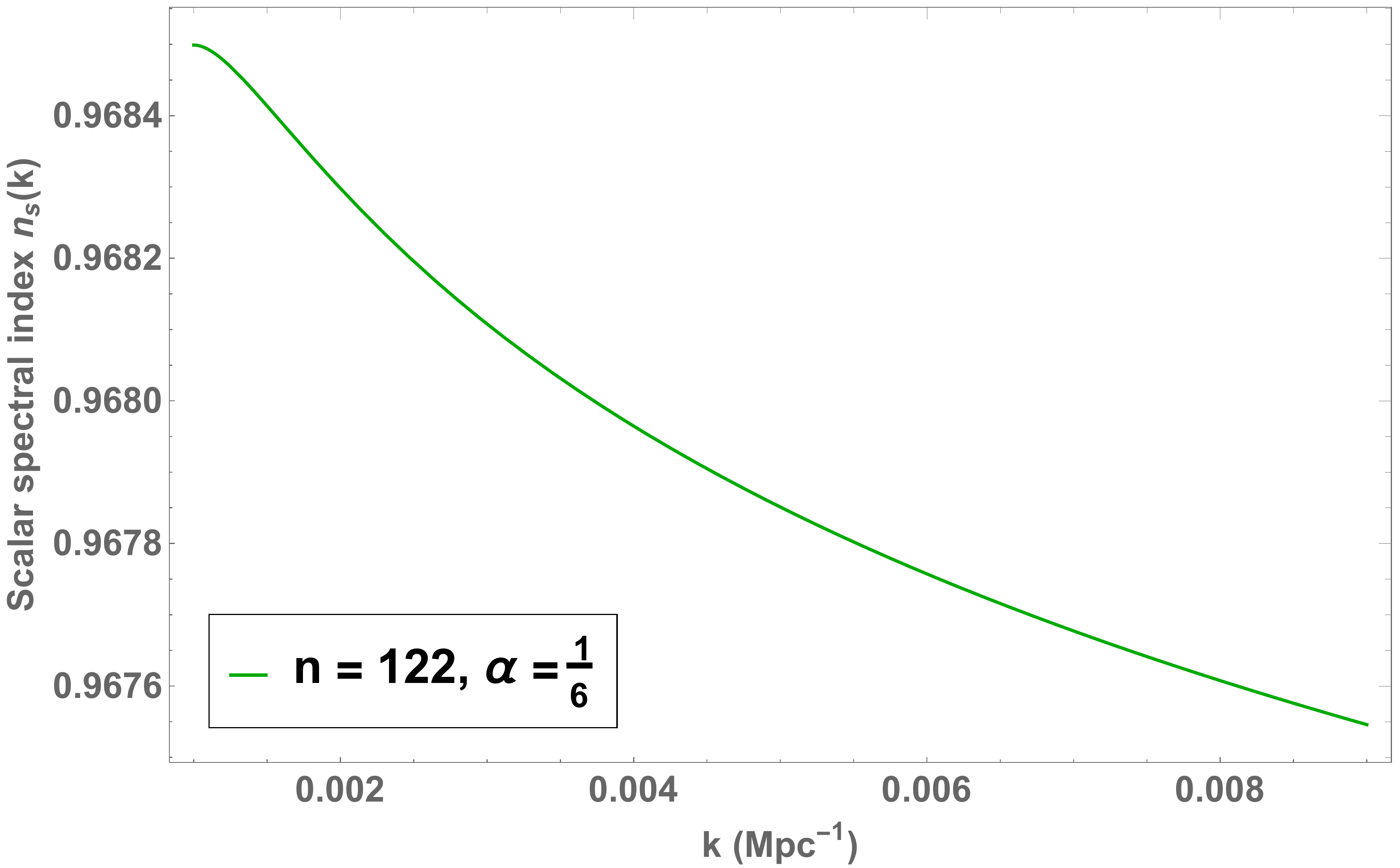}
   \subcaption{}
   \label{fig:scalarSpectralIndex_2}
\end{subfigure}%
\vspace{0.1\linewidth}
\begin{subfigure}{0.52\linewidth}
  \centering
   \includegraphics[width=70mm,height=65mm]{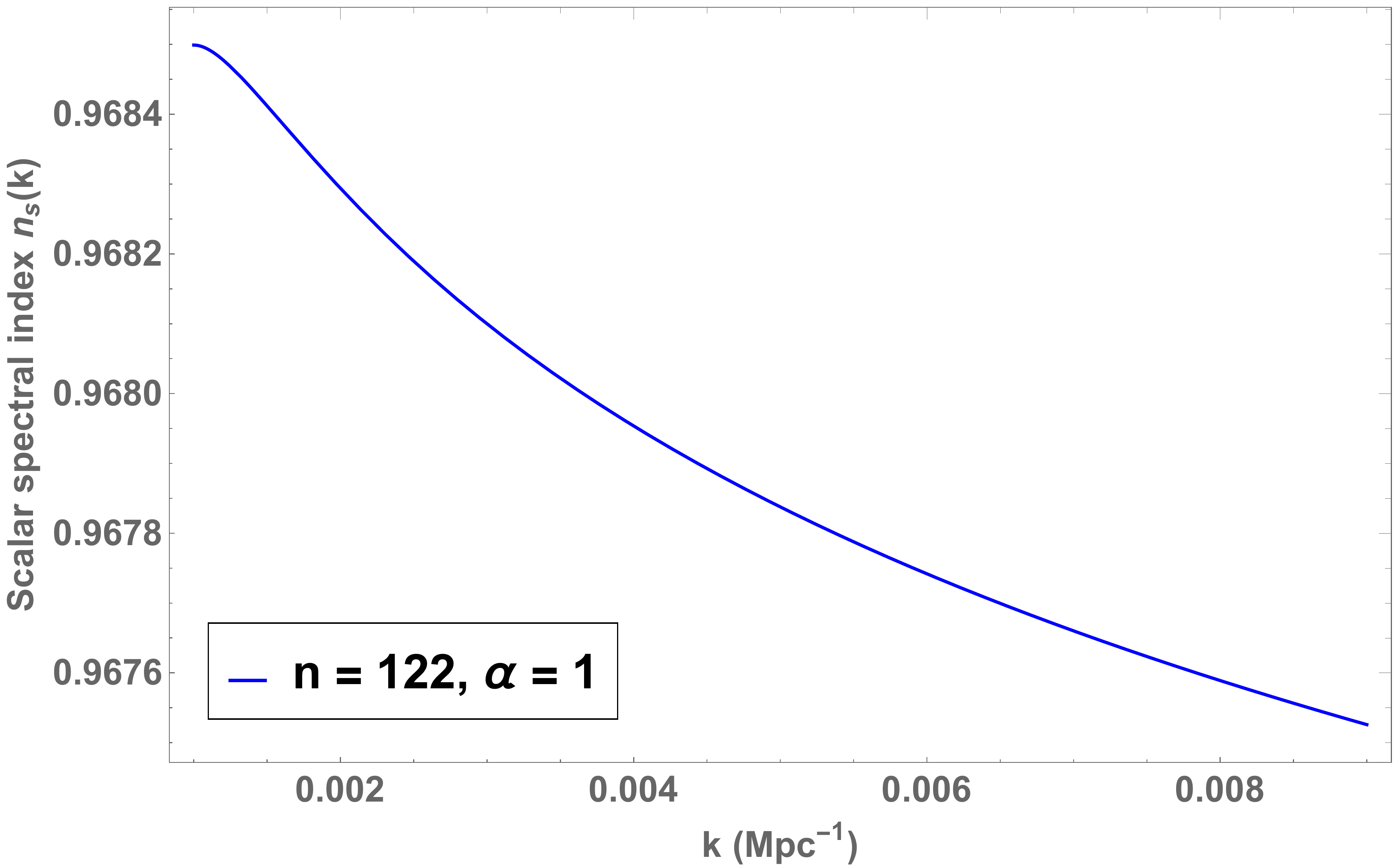}
   \subcaption{}
    \label{fig:scalarSpectralIndex_3}
\end{subfigure}%
\begin{subfigure}{0.52\linewidth}
  \centering
   \includegraphics[width=70mm,height=65mm]{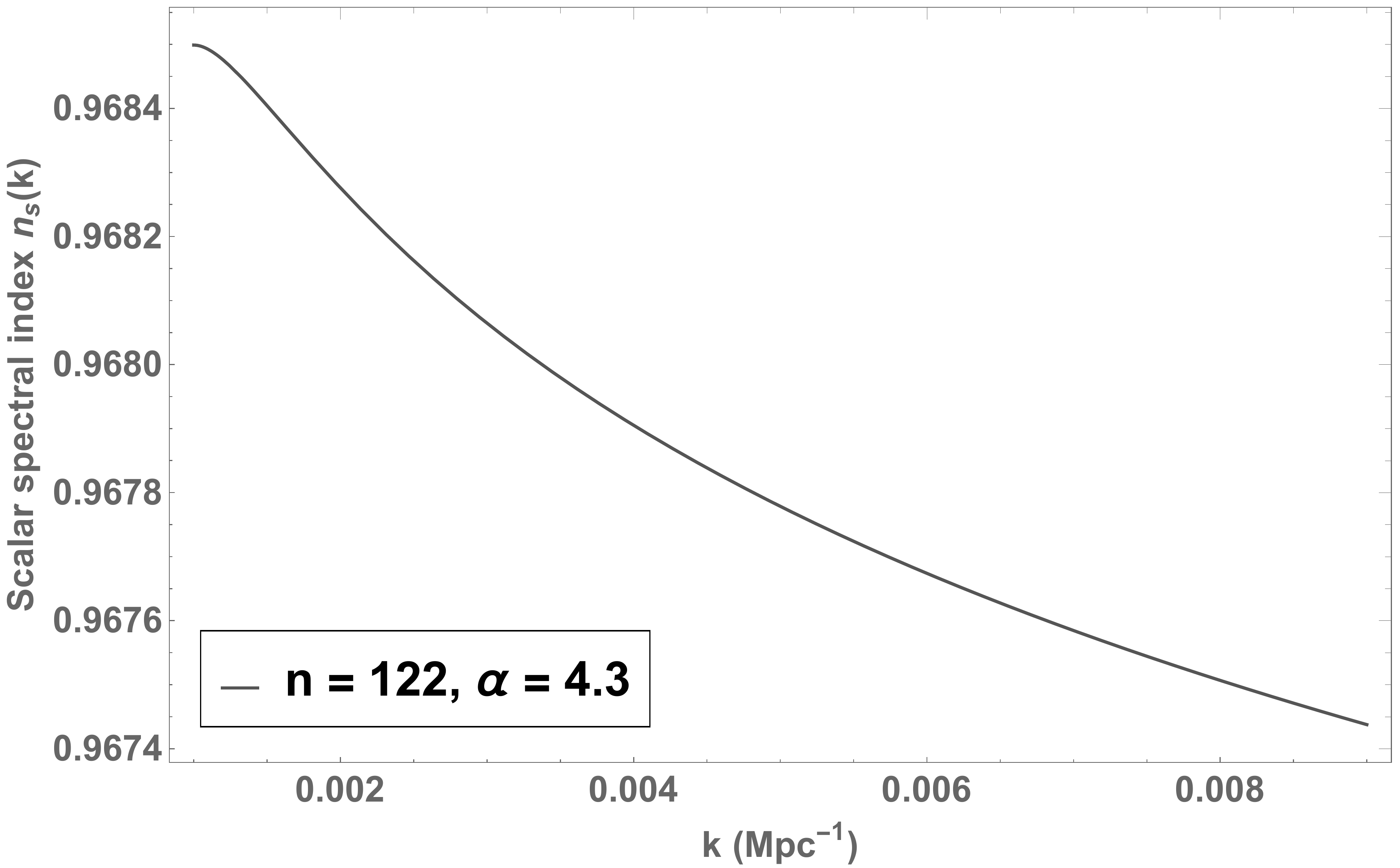}
   \subcaption{}
    \label{fig:scalarSpectralIndex_4}
\end{subfigure}
\caption{Scalar spectral indices for four values of $\alpha$ for a given value of $n$. The values of $n_s(k)$ remain almost fixed for all values of $\alpha$.}
\label{fig:scalarSpectralIndex}
\end{figure}

\begin{figure}[H]
\begin{subfigure}{0.52\linewidth}
  \centering
   \includegraphics[width=70mm,height=65mm]{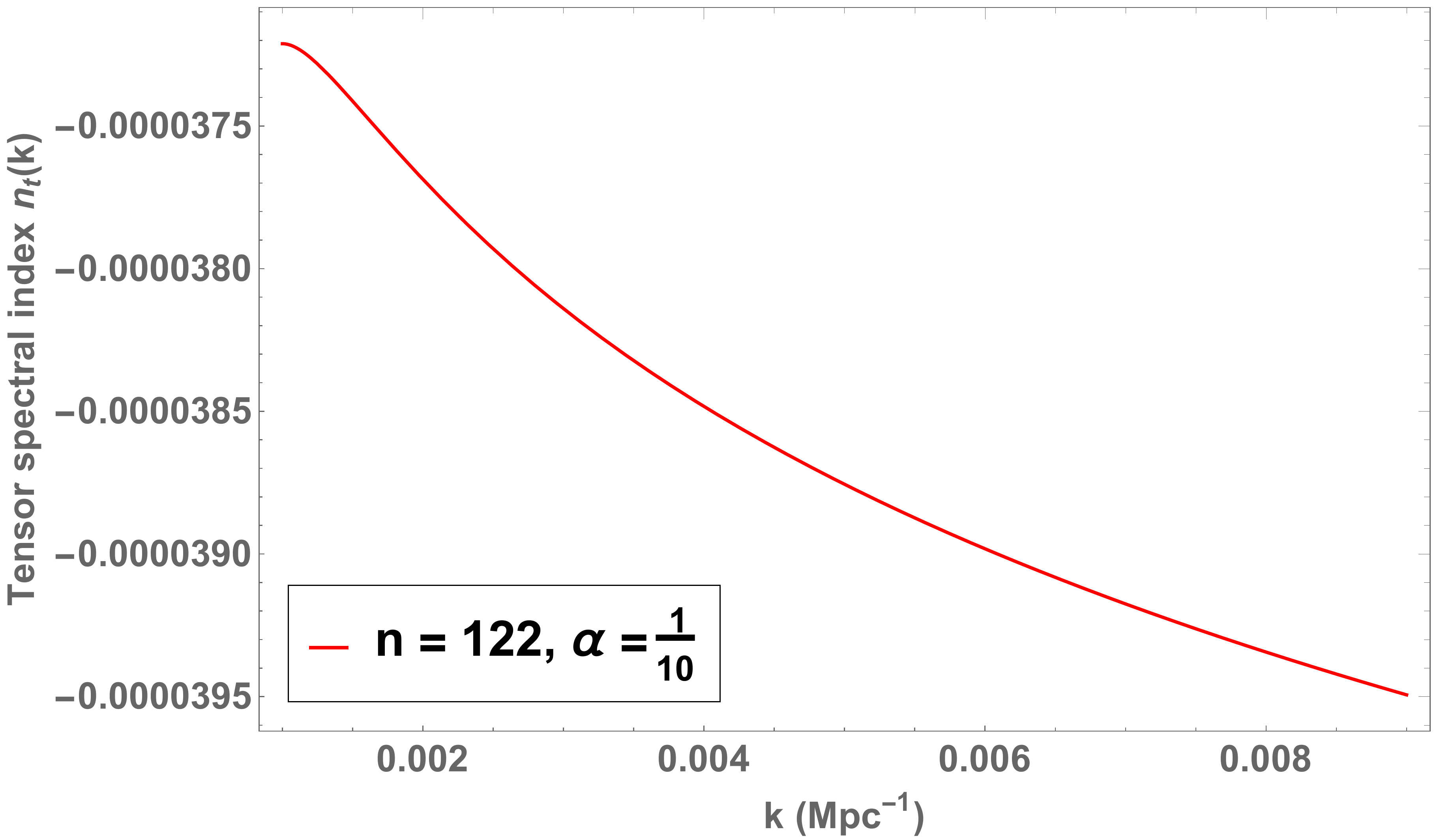} 
   \subcaption{}
   \label{fig:tensorSpectralIndex_1}
\end{subfigure}%
\begin{subfigure}{0.52\linewidth}
  \centering
   \includegraphics[width=70mm,height=65mm]{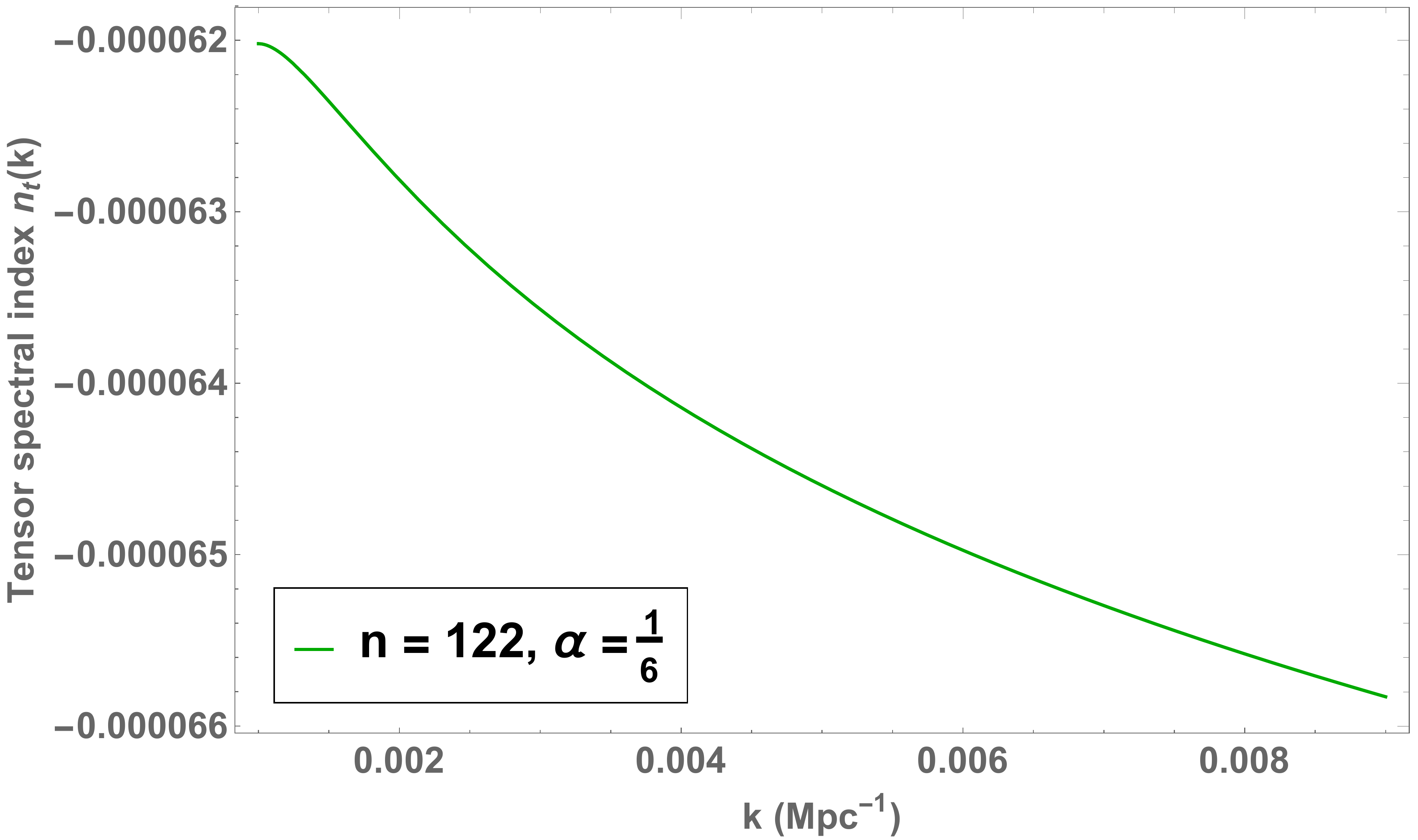}
   \subcaption{}
   \label{fig:tensorSpectralIndex_2}
\end{subfigure}%
\vspace{0.1\linewidth}
\begin{subfigure}{0.52\linewidth}
  \centering
   \includegraphics[width=70mm,height=65mm]{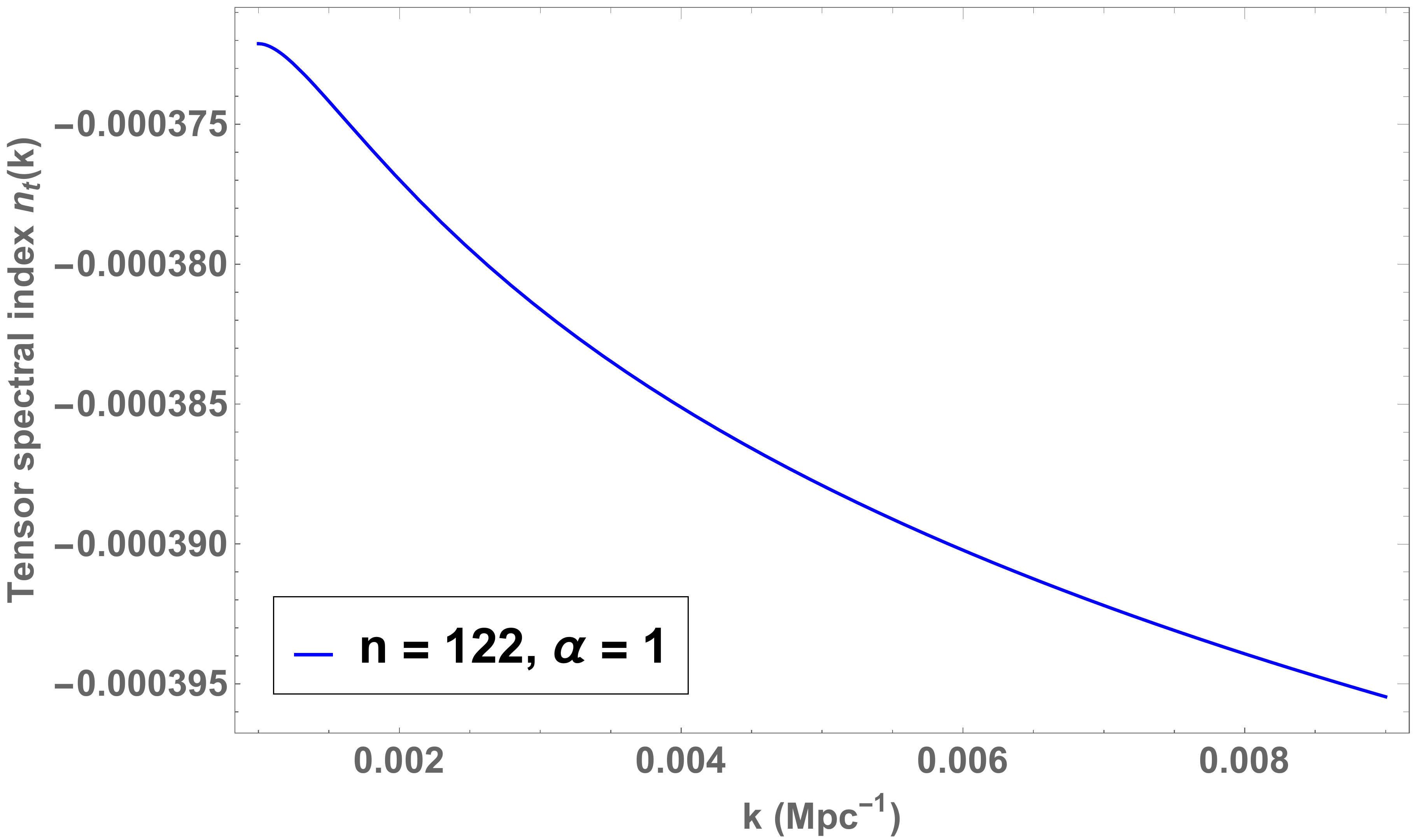}
   \subcaption{}
    \label{fig:tensorSpectralIndex_3}
\end{subfigure}%
\begin{subfigure}{0.52\linewidth}
  \centering
   \includegraphics[width=70mm,height=65mm]{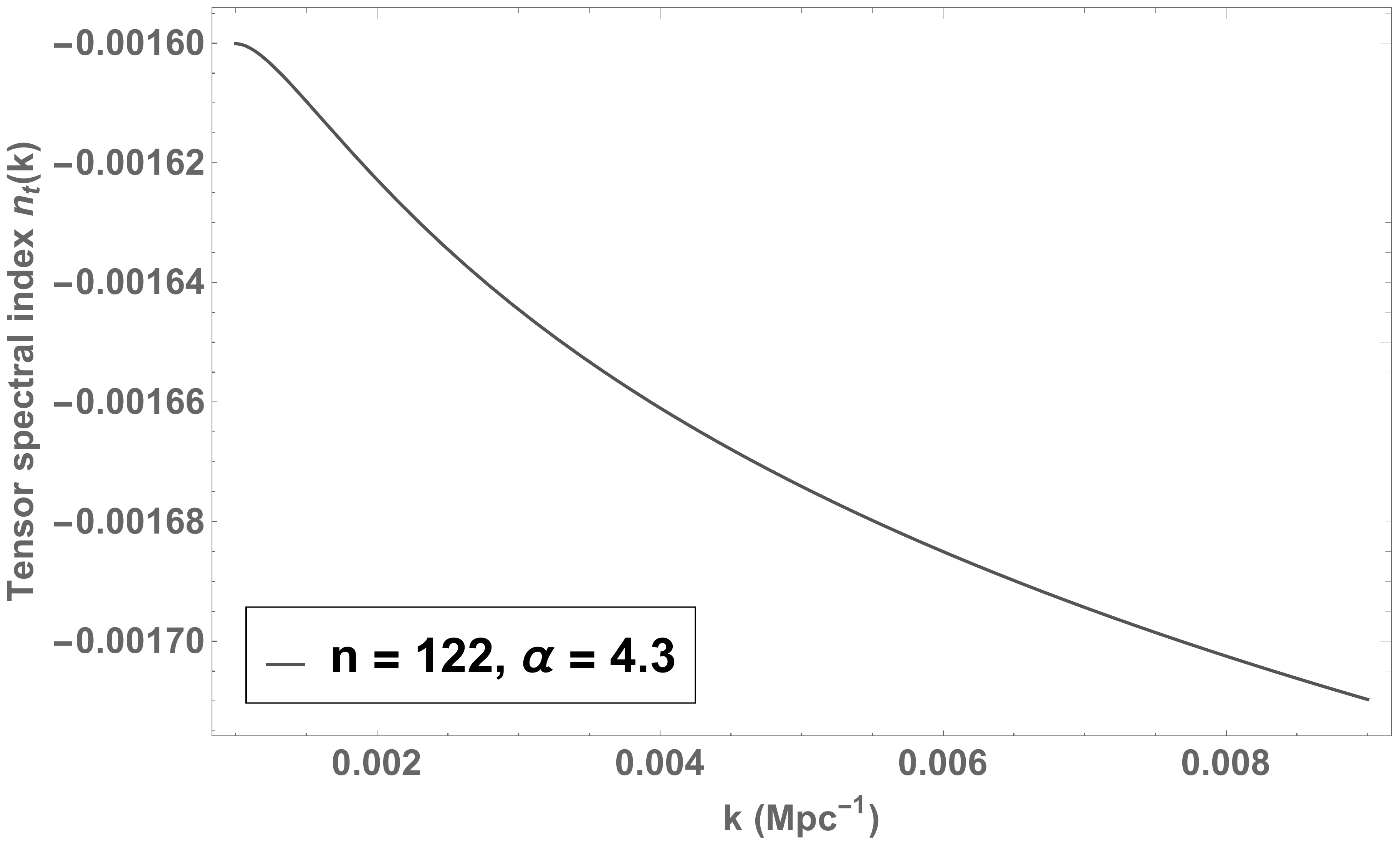}
   \subcaption{}
    \label{fig:tensorSpectralIndex_4}
\end{subfigure}
\caption{Tensor spectral indices for four values of $\alpha$ for a given value of $n$. The values of $|n_t(k)|$ increase on increase in $\alpha$ for a particular value of $k$.}
\label{fig:tensorSpectralIndex}
\end{figure}

\begin{figure}[H]
\begin{subfigure}{0.52\linewidth}
  \centering
   \includegraphics[width=70mm,height=65mm]{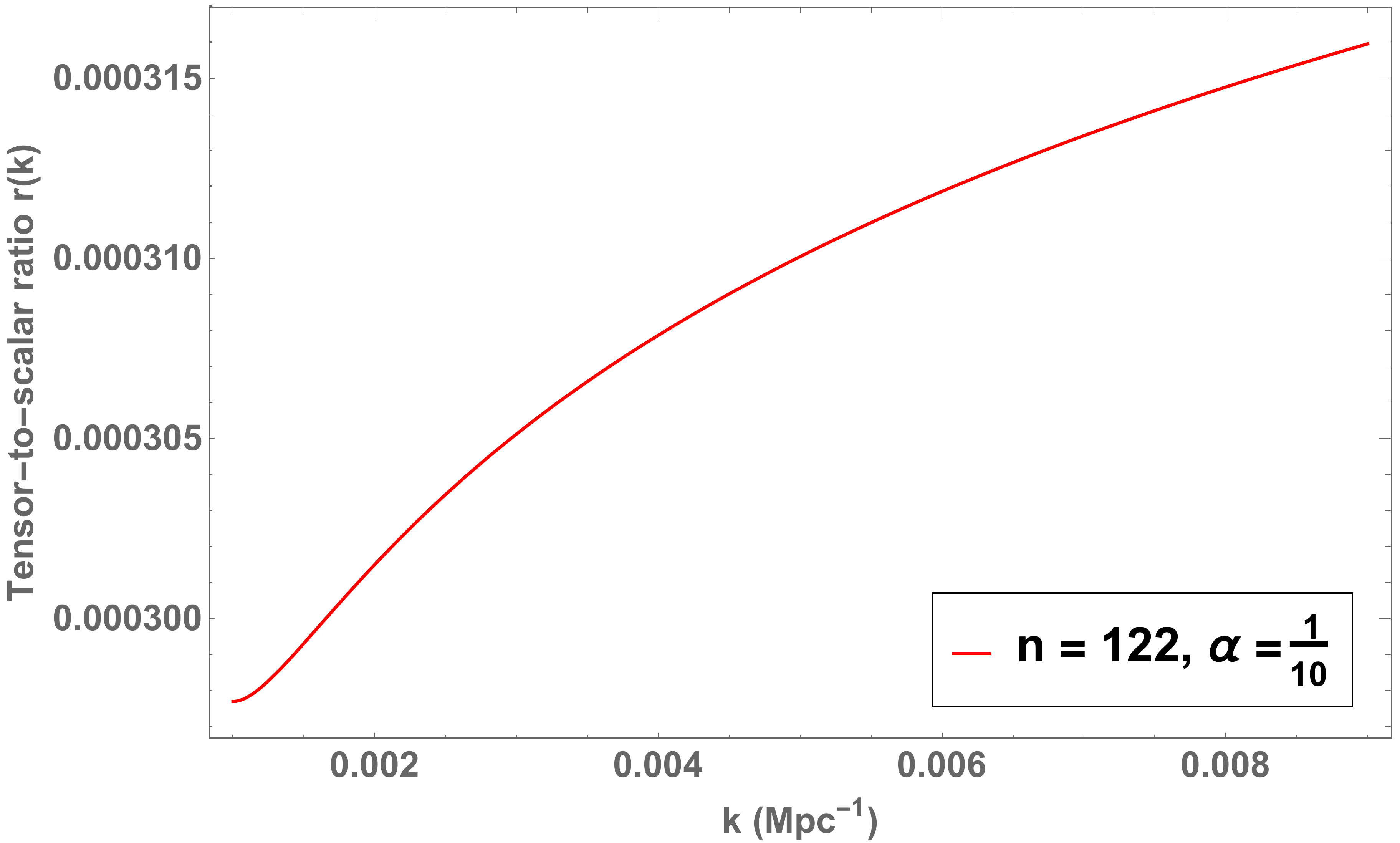} 
   \subcaption{}
   \label{fig:tensorToScalarRatio_1}
\end{subfigure}%
\begin{subfigure}{0.52\linewidth}
  \centering
   \includegraphics[width=70mm,height=65mm]{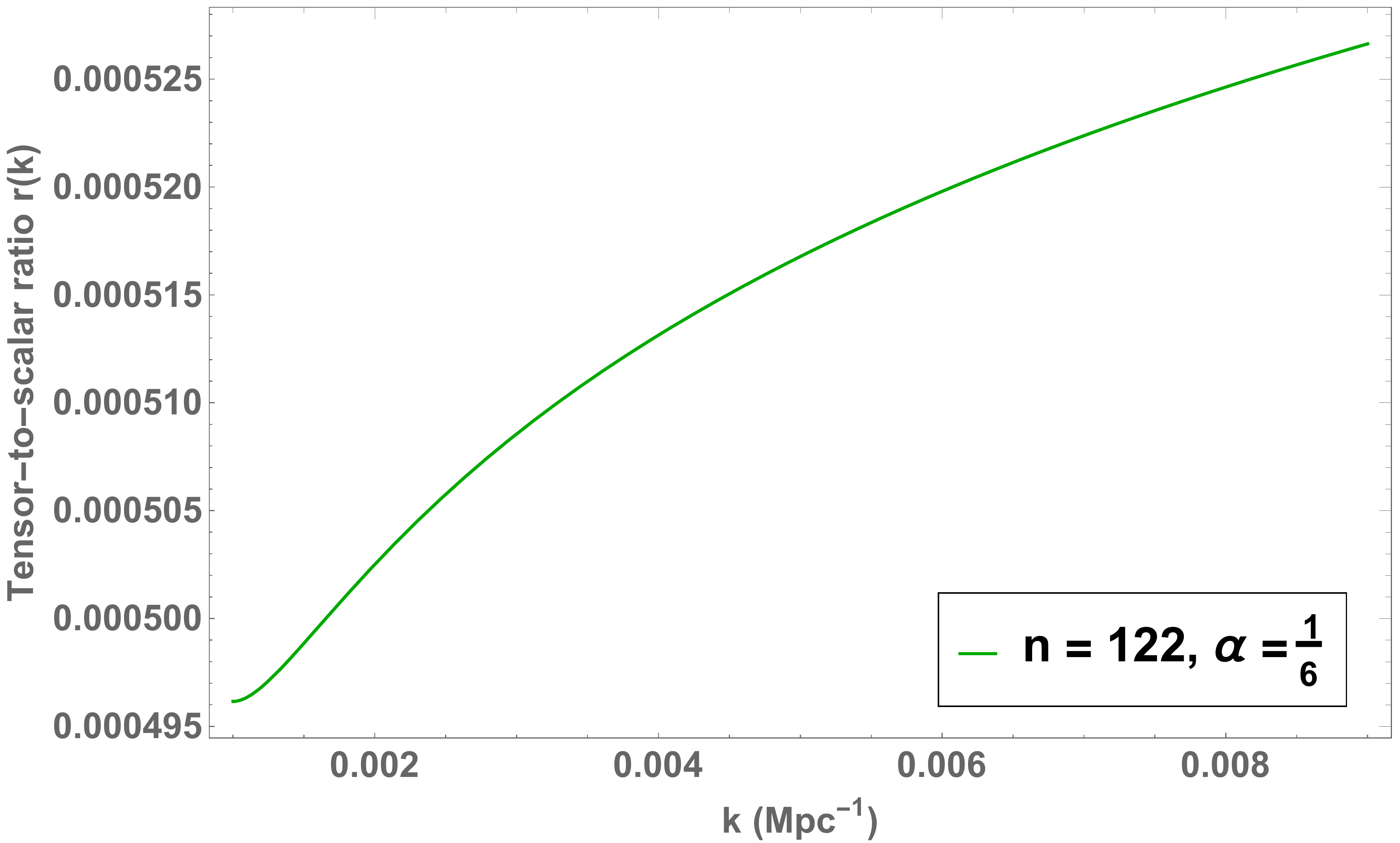}
   \subcaption{}
   \label{fig:tensorToScalarRatio_2}
\end{subfigure}%
\vspace{0.1\linewidth}
\begin{subfigure}{0.52\linewidth}
  \centering
   \includegraphics[width=70mm,height=65mm]{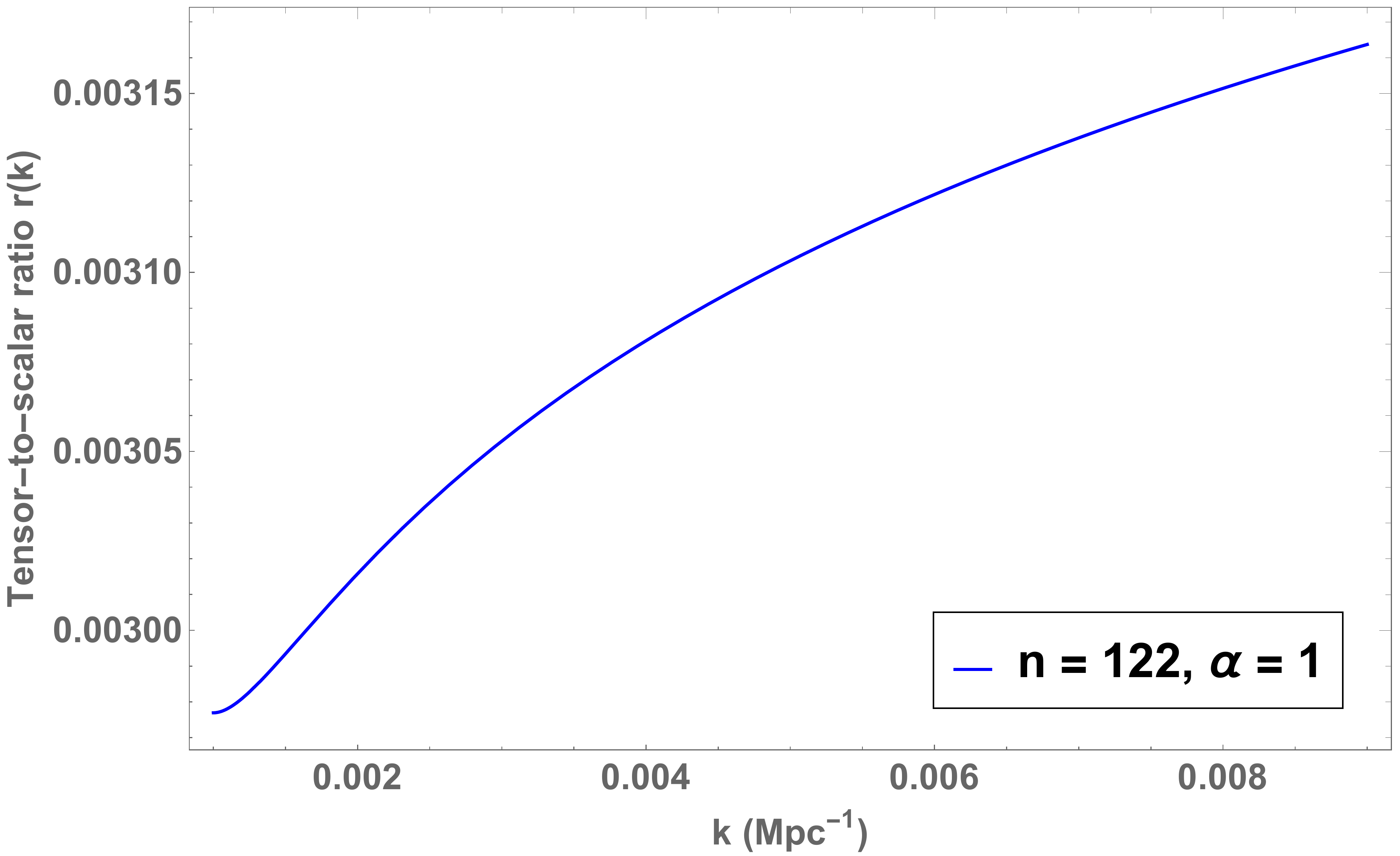}
   \subcaption{}
    \label{fig:tensorToScalarRatio_3}
\end{subfigure}%
\begin{subfigure}{0.52\linewidth}
  \centering
   \includegraphics[width=70mm,height=65mm]{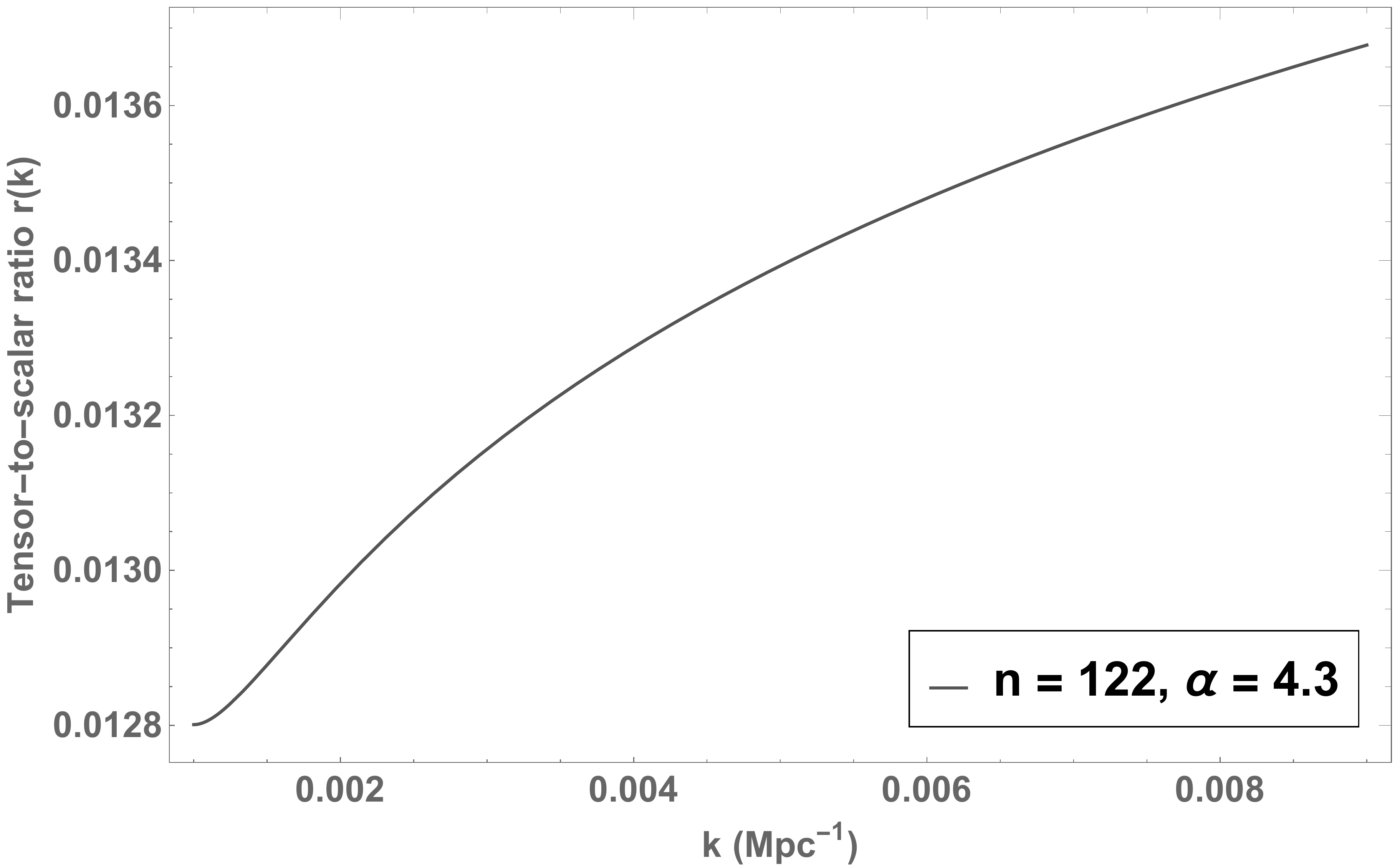}
   \subcaption{}
    \label{fig:tensorToScalarRatio_4}
\end{subfigure}
\caption{Tensor to scalar ratios for four values of $\alpha$ for a given value of $n$. The values of $r(k)$ increase on increase in $\alpha$ for a particular value of $k$.}
\label{fig:tensorToScalarRatio}
\end{figure}

\begin{figure}[H]
\begin{subfigure}{0.52\linewidth}
  \centering
   \includegraphics[width=70mm,height=65mm]{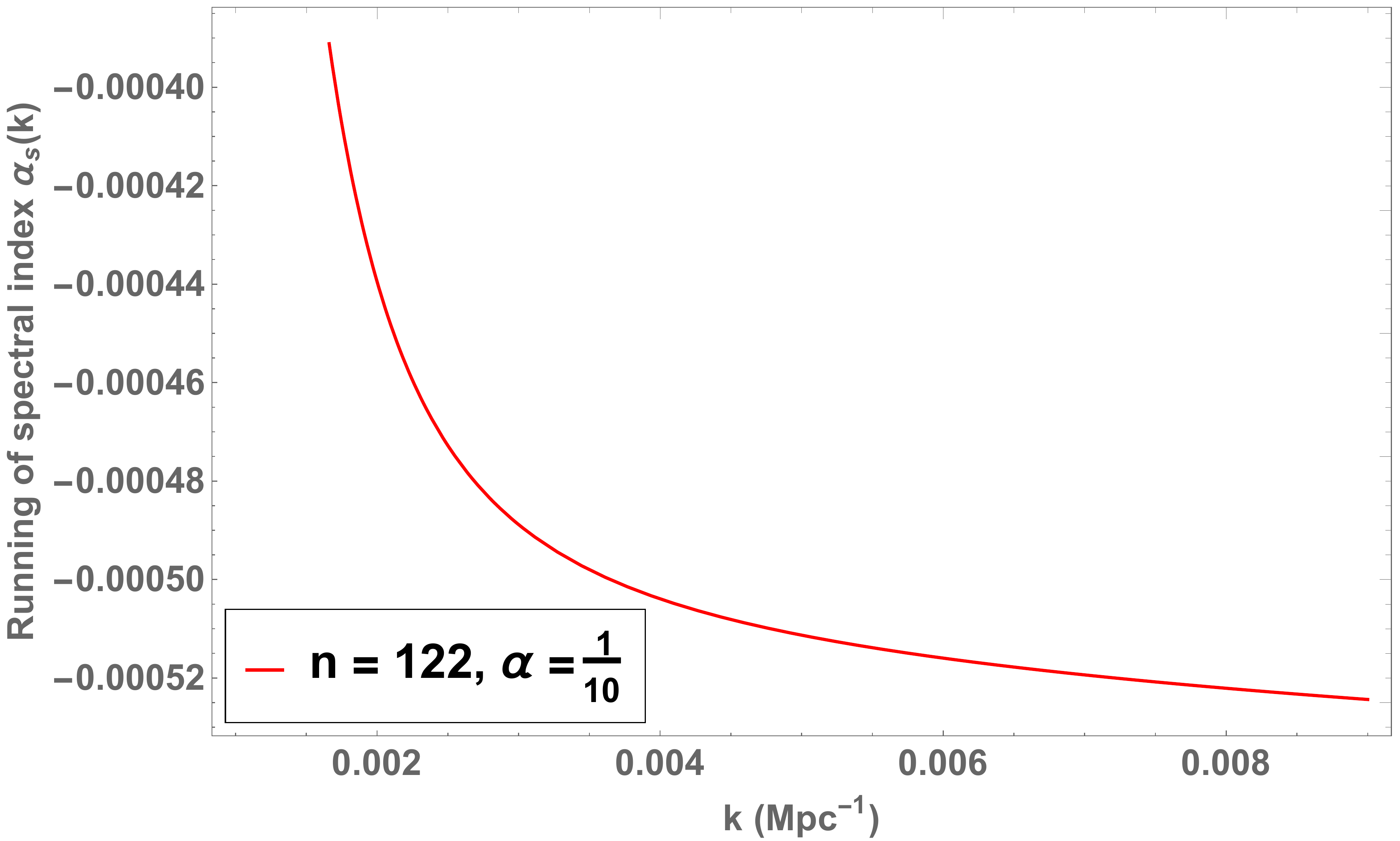} 
   \subcaption{}
   \label{fig:RunningSpectralIndex1}
\end{subfigure}%
\begin{subfigure}{0.52\linewidth}
  \centering
   \includegraphics[width=70mm,height=65mm]{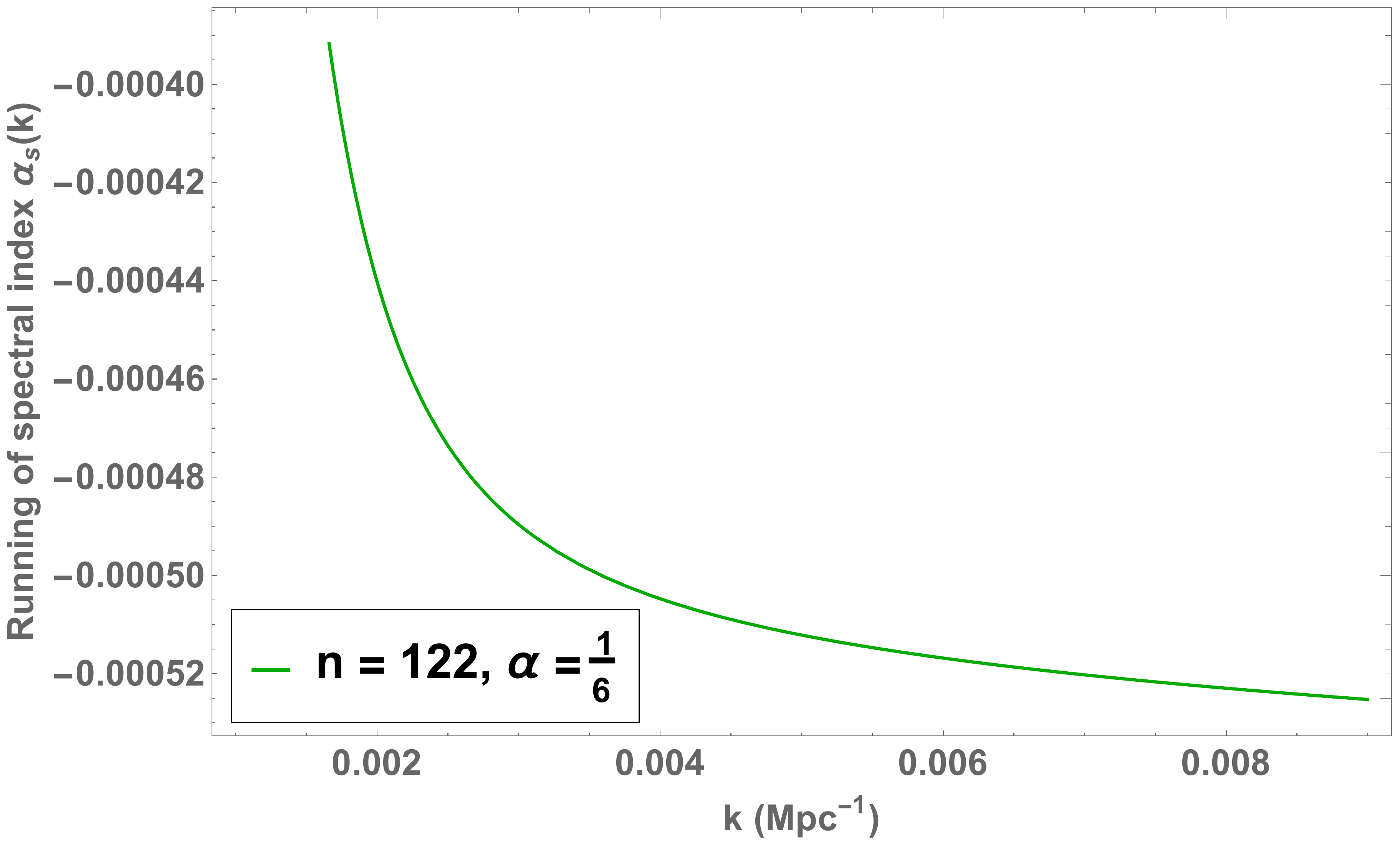}
   \subcaption{}
   \label{fig:RunningSpectralIndex2}
\end{subfigure}%
\vspace{0.1\linewidth}
\begin{subfigure}{0.52\linewidth}
  \centering
   \includegraphics[width=70mm,height=65mm]{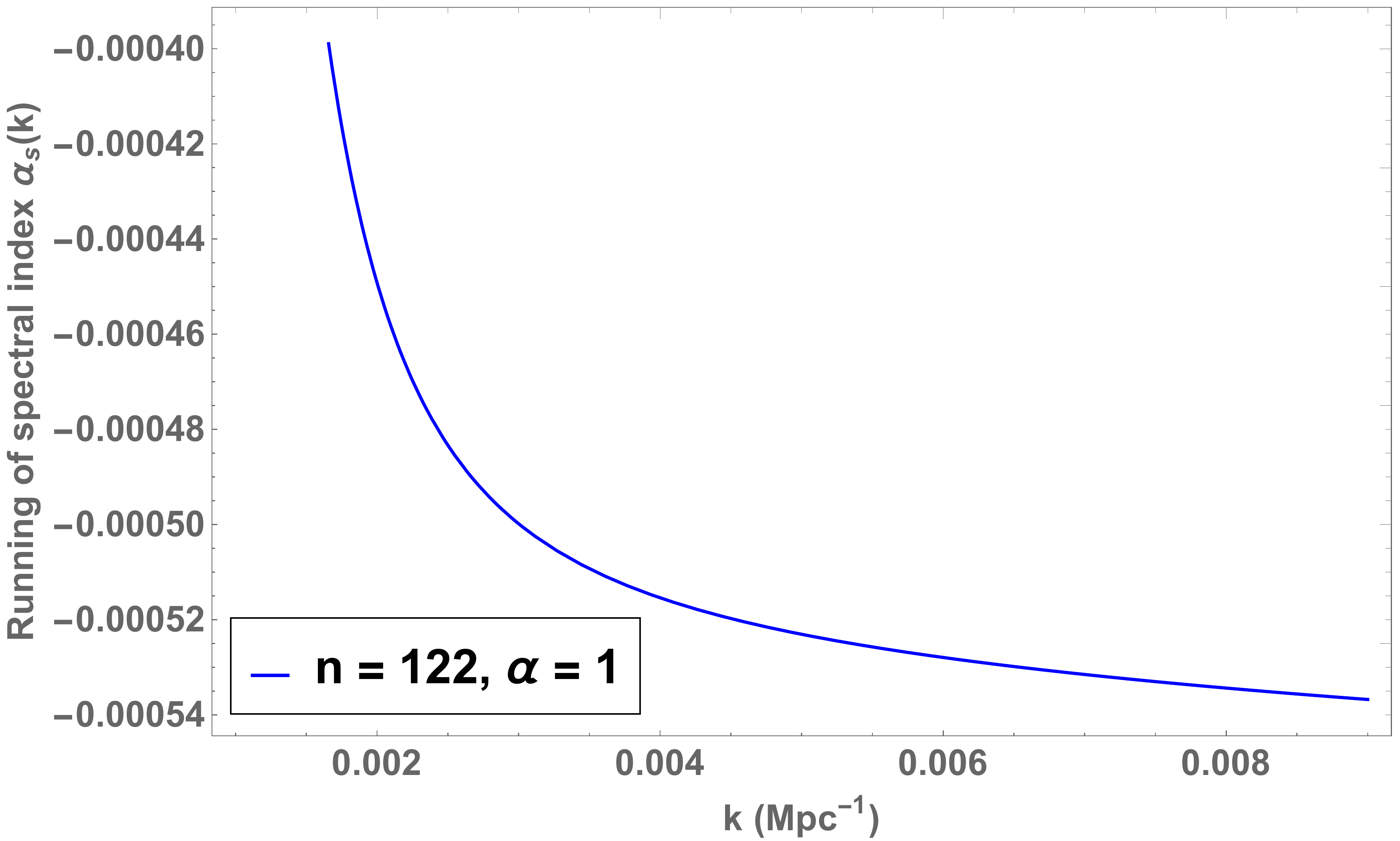}
   \subcaption{}
    \label{fig:RunningSpectralIndex3}
\end{subfigure}%
\begin{subfigure}{0.52\linewidth}
  \centering
   \includegraphics[width=70mm,height=65mm]{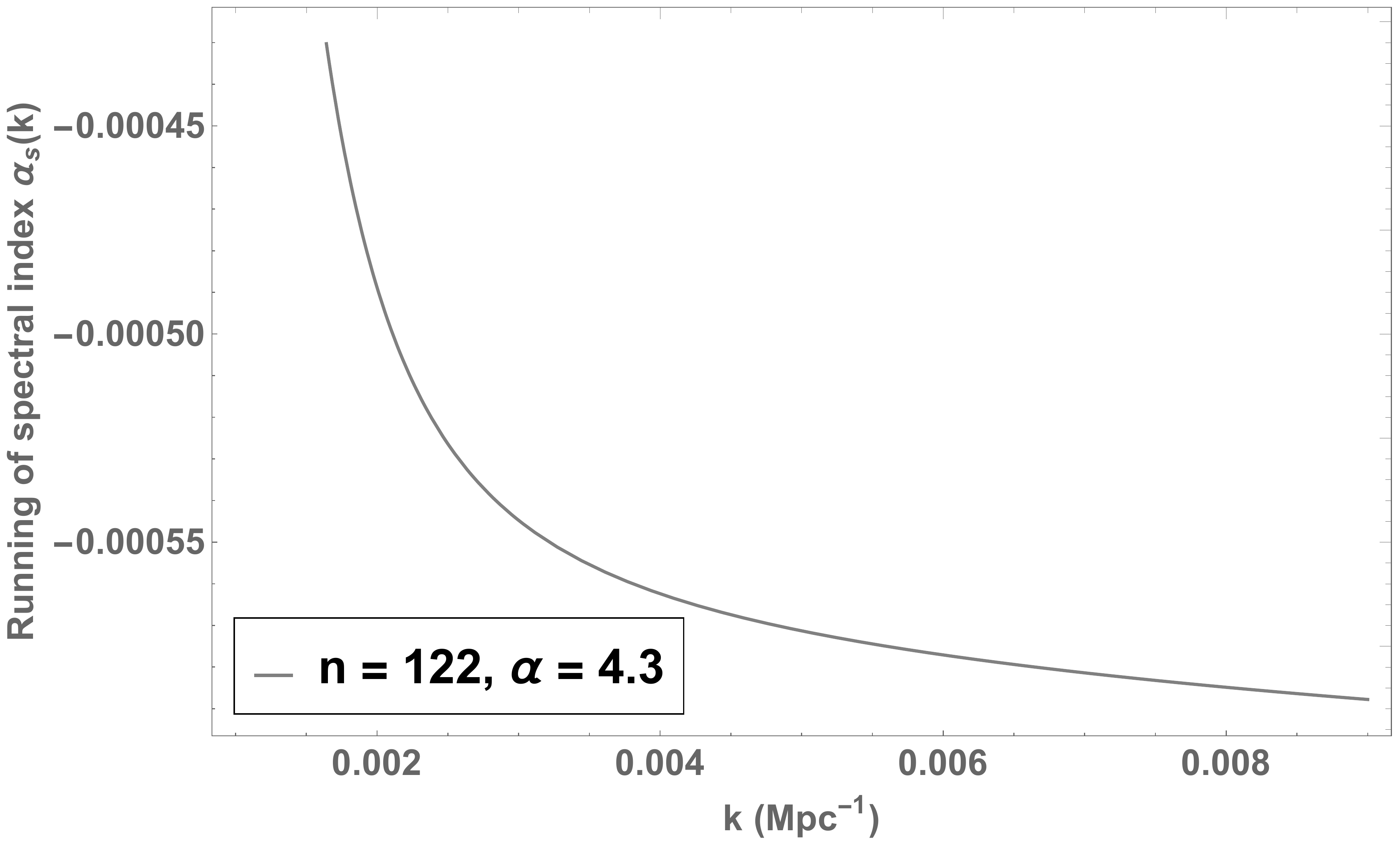}
   \subcaption{}
    \label{fig:RunningSpectralIndex4}
\end{subfigure}
\caption{Running of the spectral index for four values of $\alpha$ for a given value of $n$. Very small variation is observed in $\alpha_s(k)$ on increase in $\alpha$.}
\label{fig:RunningSpectralIndex}
\end{figure}

\begin{figure}[H]
\begin{subfigure}{0.52\linewidth}
  \centering
   \includegraphics[width=70mm,height=65mm]{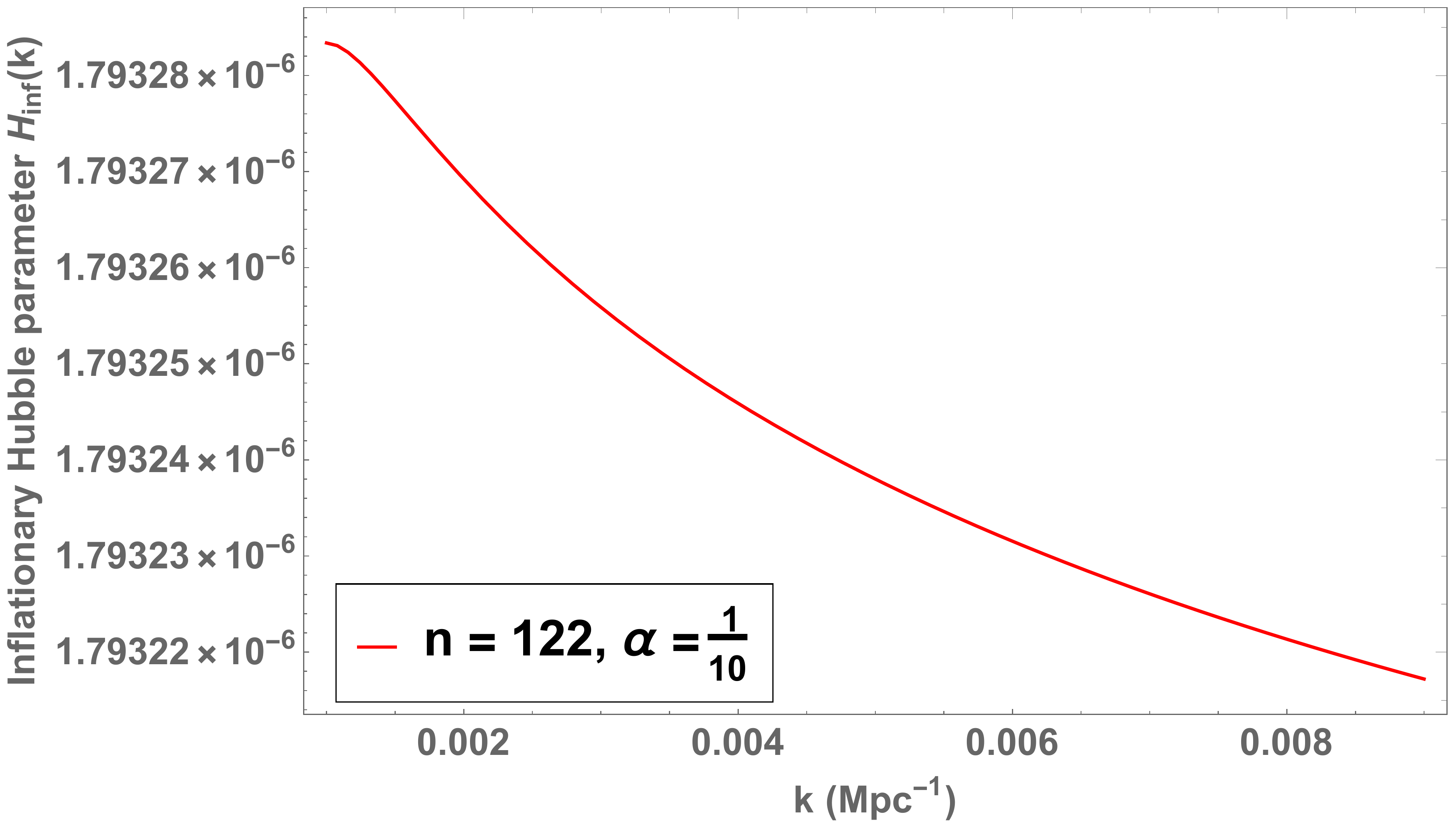} 
   \subcaption{}
   \label{fig:InfHubbleParameter_1}
\end{subfigure}%
\begin{subfigure}{0.52\linewidth}
  \centering
   \includegraphics[width=70mm,height=65mm]{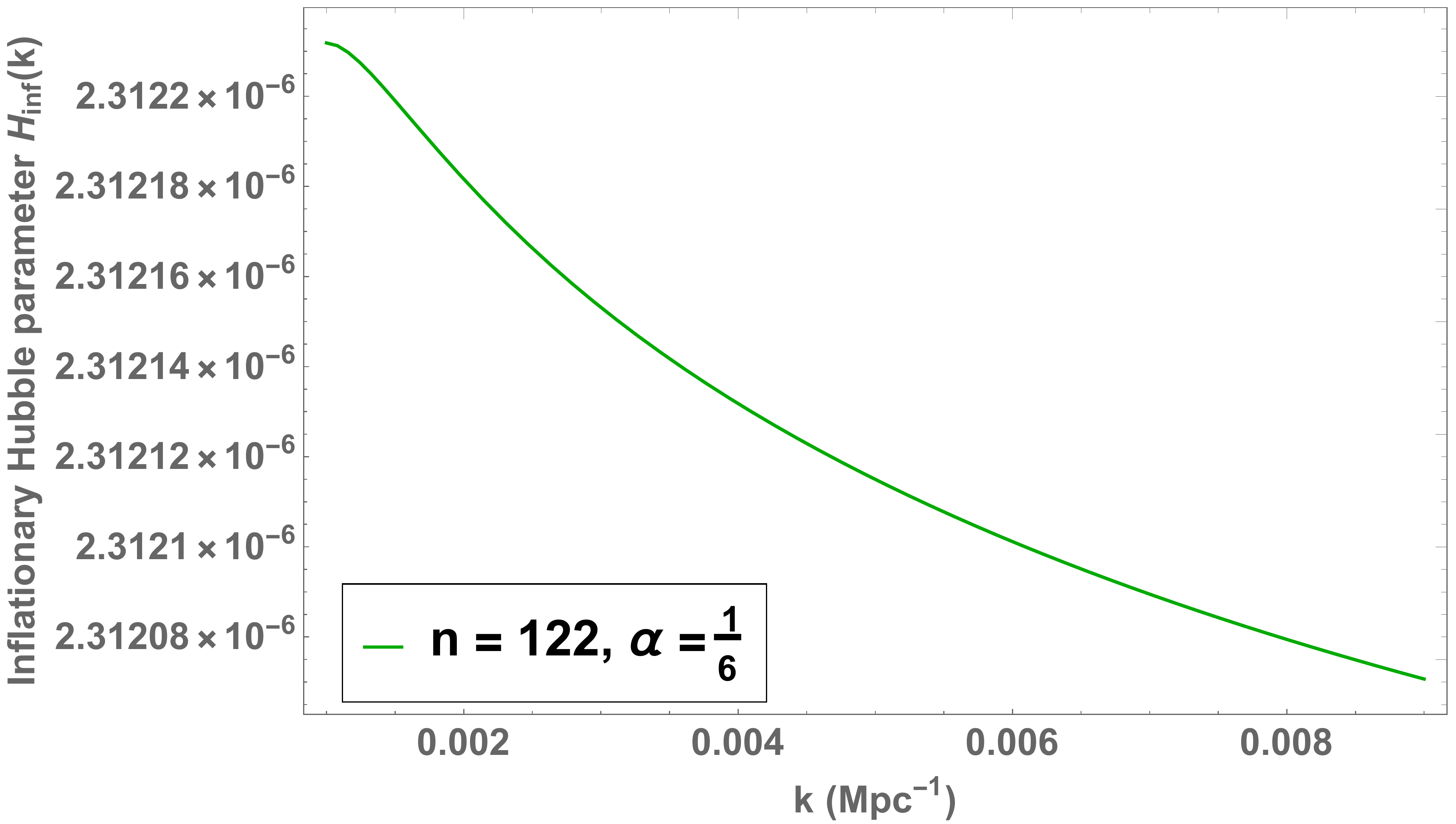}
   \subcaption{}
   \label{fig:InfHubbleParameter_2}
\end{subfigure}%
\vspace{0.1\linewidth}
\begin{subfigure}{0.52\linewidth}
  \centering
   \includegraphics[width=70mm,height=65mm]{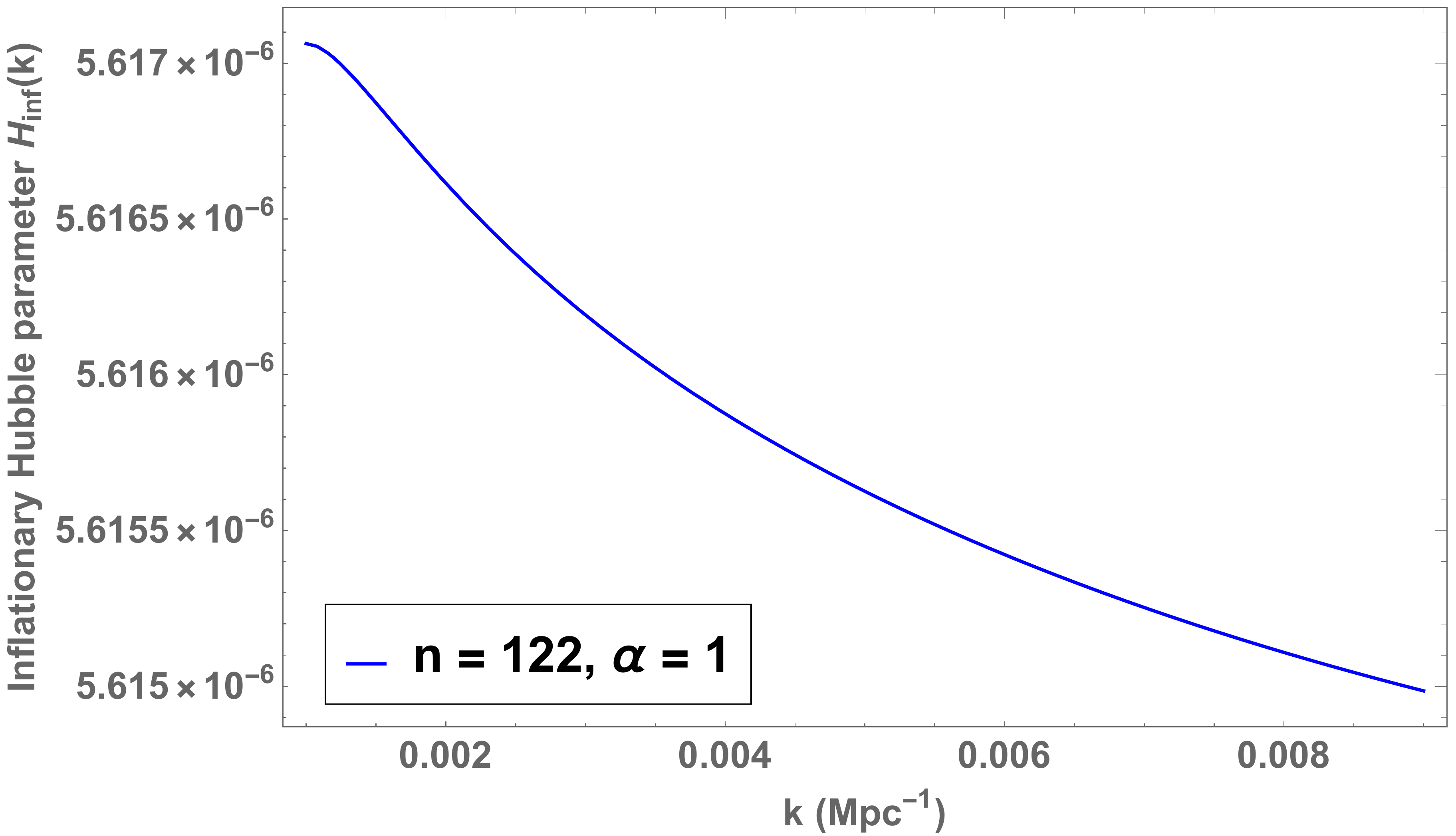}
   \subcaption{}
    \label{fig:InfHubbleParameter_3}
\end{subfigure}%
\begin{subfigure}{0.52\linewidth}
  \centering
   \includegraphics[width=70mm,height=65mm]{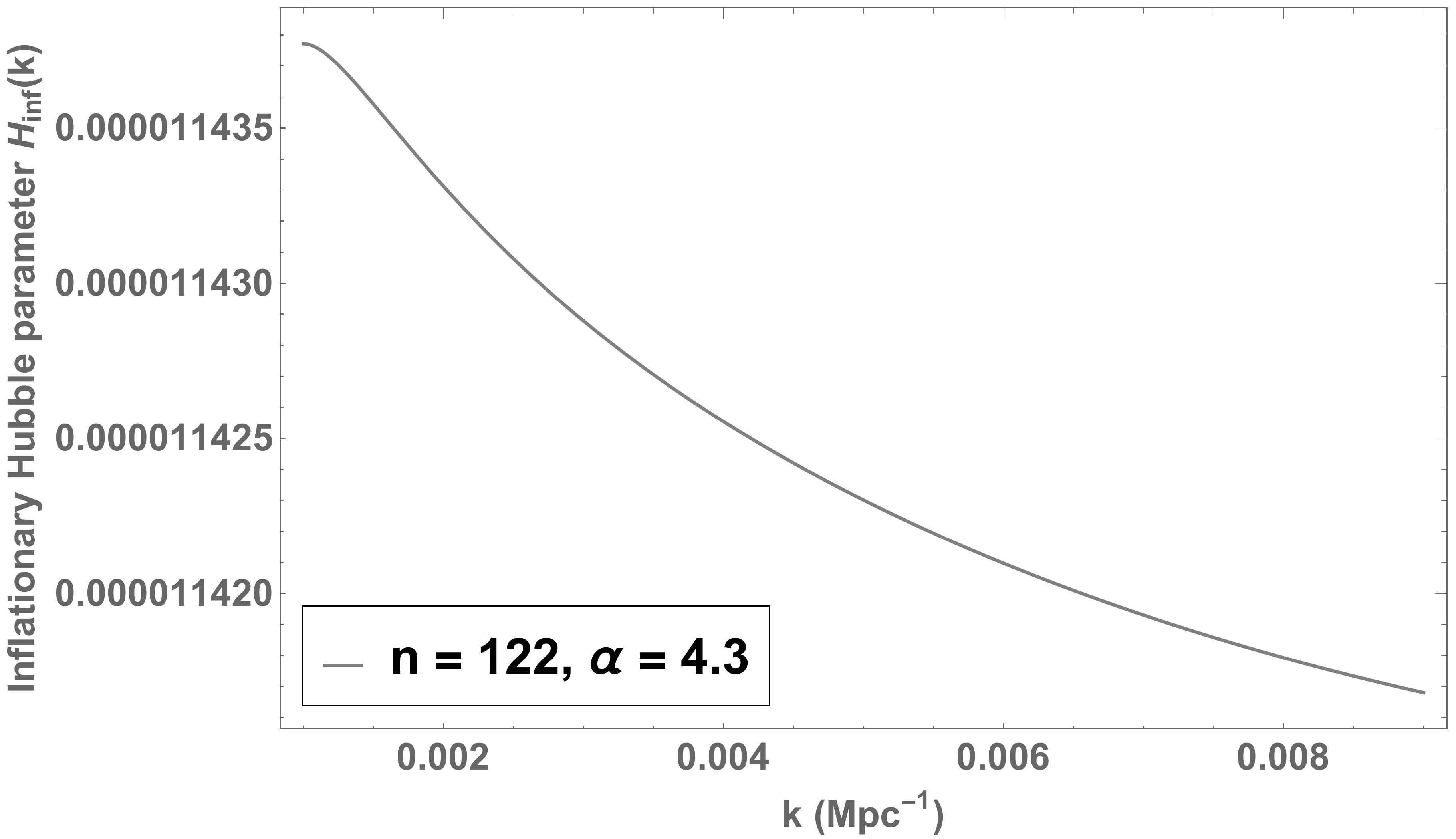}
   \subcaption{}
    \label{fig:InfHubbleParameter_4}
\end{subfigure}
\caption{Inflationary Hubble parameters for four values of $\alpha$ for a given value of $n$. The values of $H_{\mathrm{inf}}(k)$ increase on increase in $\alpha$ for a particular value of $k$.}
\label{fig:InfHubbleParameter}
\end{figure}

 Finally, in figure \ref{fig:InfHubbleParameter} we show the inflationary Hubble parameter $H(k)$ according to (\ref{eq:INFHubbleMode}). This parameter is proportional to $\sqrt{V^{(0)}(k)}$ according to (\ref{eq:DefHubbleINF}) and (\ref{eq:modeINFpot}) which we can verify by its order $10^{-5} - 10^{-6}$ shown in figure \ref{fig:InfHubbleParameter} and that of $V^{(0)}(k)\sim 10^{-12} - 10^{-10}$ in figure \ref{fig:unperturbedPOT}. Our obtained values $H_{\mathrm{inf}}(k)=1.79\times 10^{-6} - 1.14\times 10^{-5}$ or $4.3\times 10^{12} - 2.77\times 10^{13}$ GeV lie within the given range by Planck-2018: $H_{\mathrm{inf}}(k_{*})<2.5\times 10^{-5}$ or $H_{\mathrm{inf}}(k_{*})<6.07\times 10^{13}$ GeV.
\subsection{Comparison with \texorpdfstring{$\alpha$}{a}-attractors}
\label{subsec:Alphacomp}
Now we compare our results for the inflationary regime of quintessential $\alpha$-attractor shown in (\ref{eq:Eq7}) with the ordinary $\alpha$-attractors without quintessence. We choose the regular forms of $\alpha$-attractor $E$ and $T$ models
\begin{equation}
    V_E(\psi) = V_0\left(1-e^{-\sqrt{\frac{2}{3\alpha}}\psi}\right)^{2n},
\end{equation}
\begin{equation}
    V_T(\psi)=V_0\left(\tanh{\frac{\psi}{\sqrt{6\alpha}}}\right)^{2n},
\end{equation}
where the normalizing factor $V_0$ is found to satisfy the COBE/Planck condition following the equation
\begin{equation}
    V_0 = 9\pi^2 (2\alpha A_s(k_{*}))\left(N_{*}+\frac{\sqrt{3\alpha}}{2}\right)^{-2}\exp\left[\frac{3\alpha}{2}\left(N_{*}+\frac{\sqrt{3\alpha}}{2}\right)^{-1}\right]
\end{equation} by the same process described in section \ref{subsec:POWERspecSUBSEC}. Our mode equations (\ref{eq:modeKeqfinal1})-(\ref{eq:modeKeqfinal3}) are again solved for these models with the considerations described in the first part of section \ref{sec: result} with respect to the same values of $n$ and $\alpha$ given in table \ref{tab:Table1}. The relevant expressions of the cosmological parameters of $E$ and $T$ models can be found in our earlier study \cite{Sarkar:2021ird}. Only it is worthwhile to state that all the models satisfy the well known attractor equations (\ref{eq:AttractorEQ1}) - (\ref{eq:AttractorEQ3}) for large number of e-folds.\par
\begin{figure}[H]
\begin{subfigure}{0.52\linewidth}
  \centering
   \includegraphics[width=70mm,height=65mm]{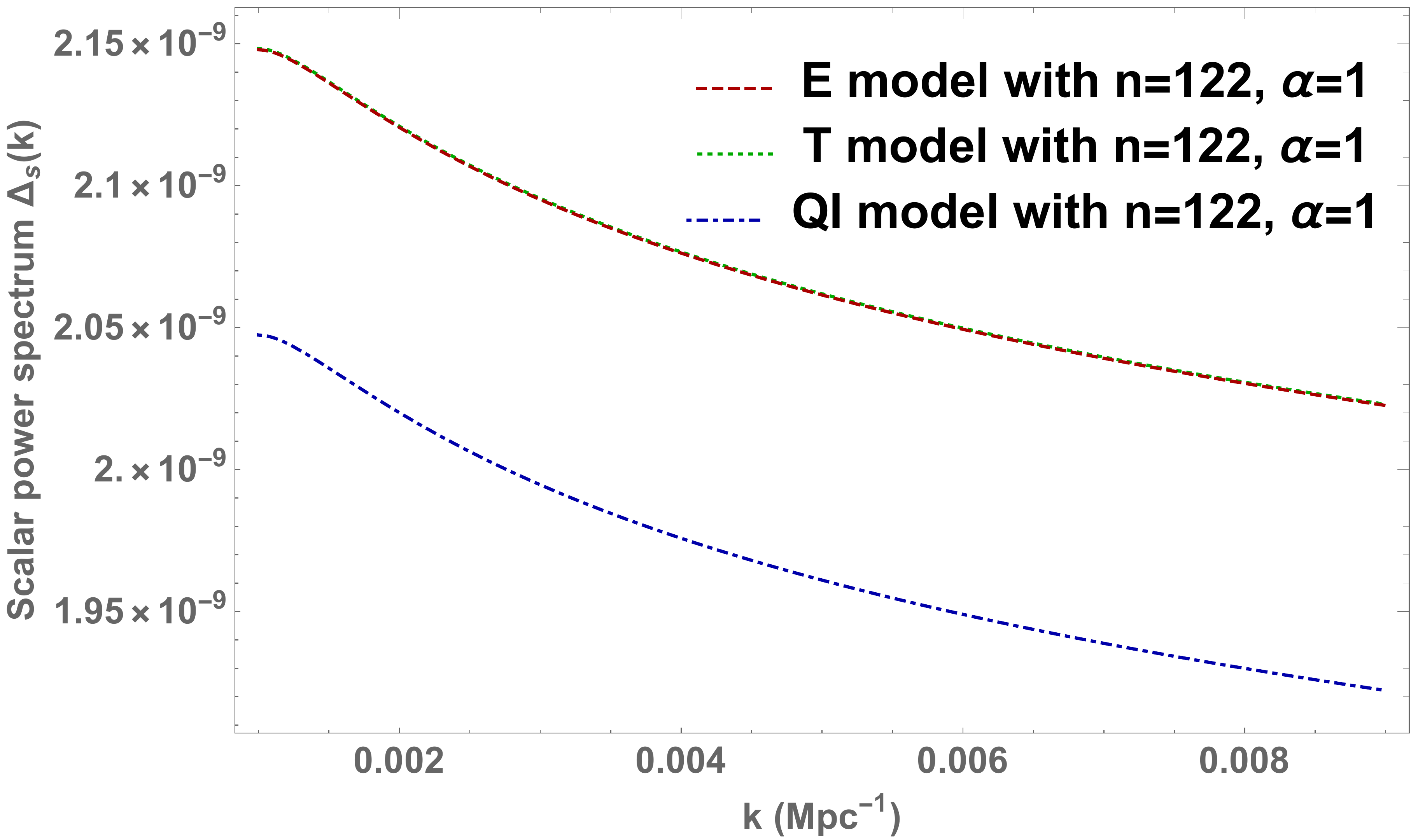} 
   \subcaption{}
   \label{fig:Comp1}
\end{subfigure}%
\begin{subfigure}{0.52\linewidth}
  \centering
   \includegraphics[width=70mm,height=65mm]{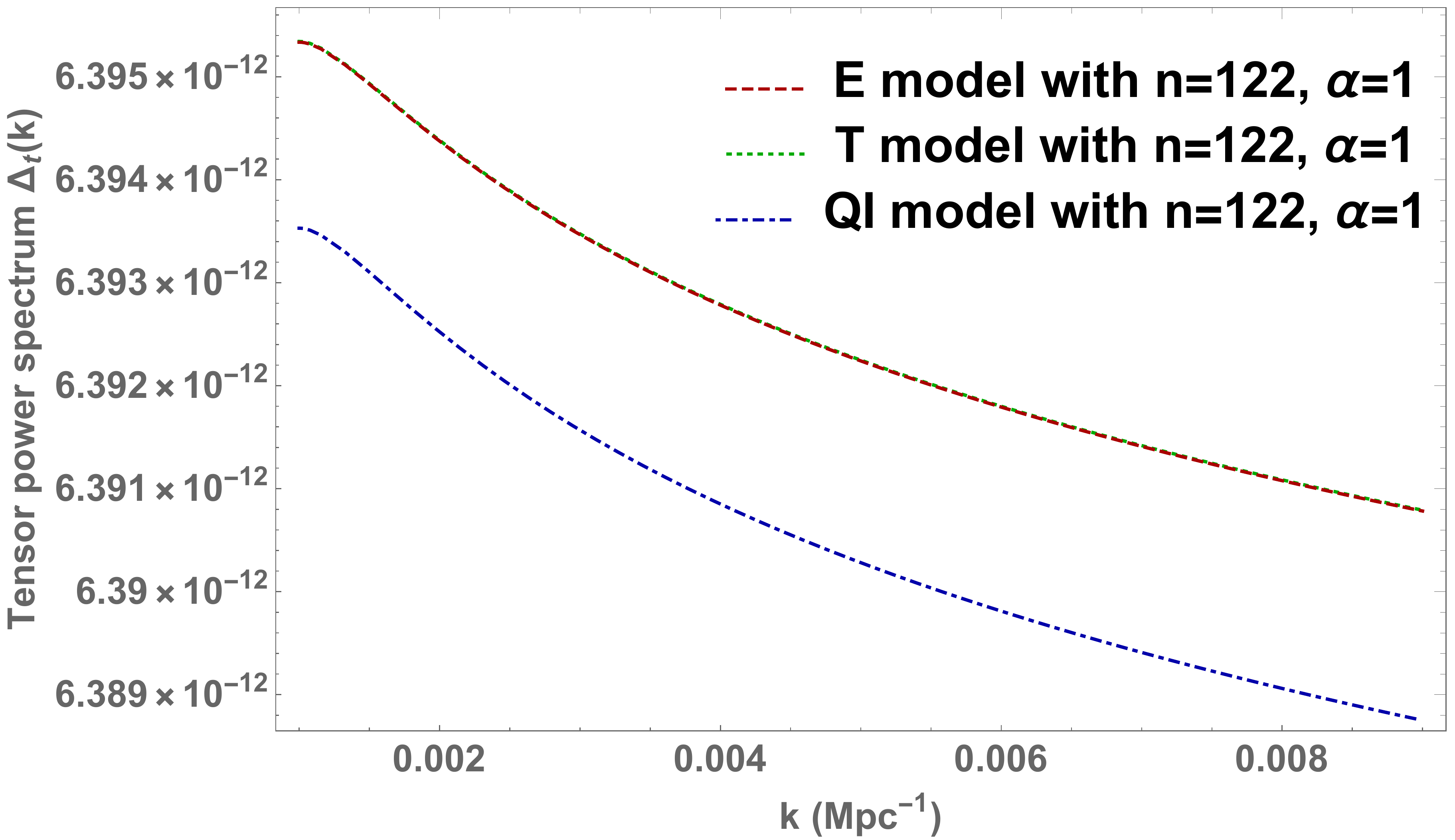}
   \subcaption{}
   \label{fig:Comp2}
\end{subfigure}%
\vspace{0.1\linewidth}
\begin{subfigure}{0.52\linewidth}
  \centering
   \includegraphics[width=70mm,height=65mm]{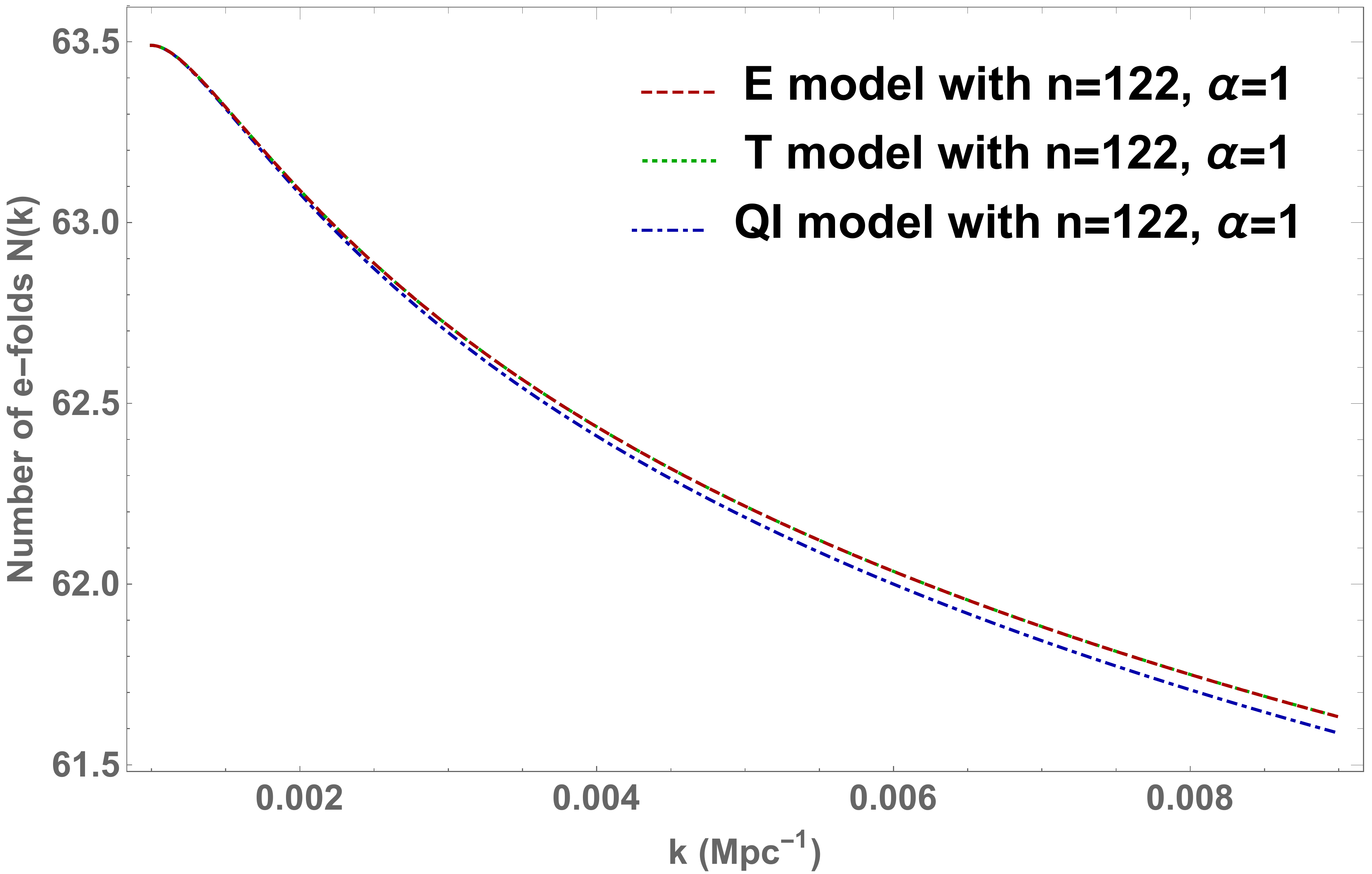}
   \subcaption{}
    \label{fig:Comp6}
\end{subfigure}%
\begin{subfigure}{0.52\linewidth}
  \centering
   \includegraphics[width=70mm,height=65mm]{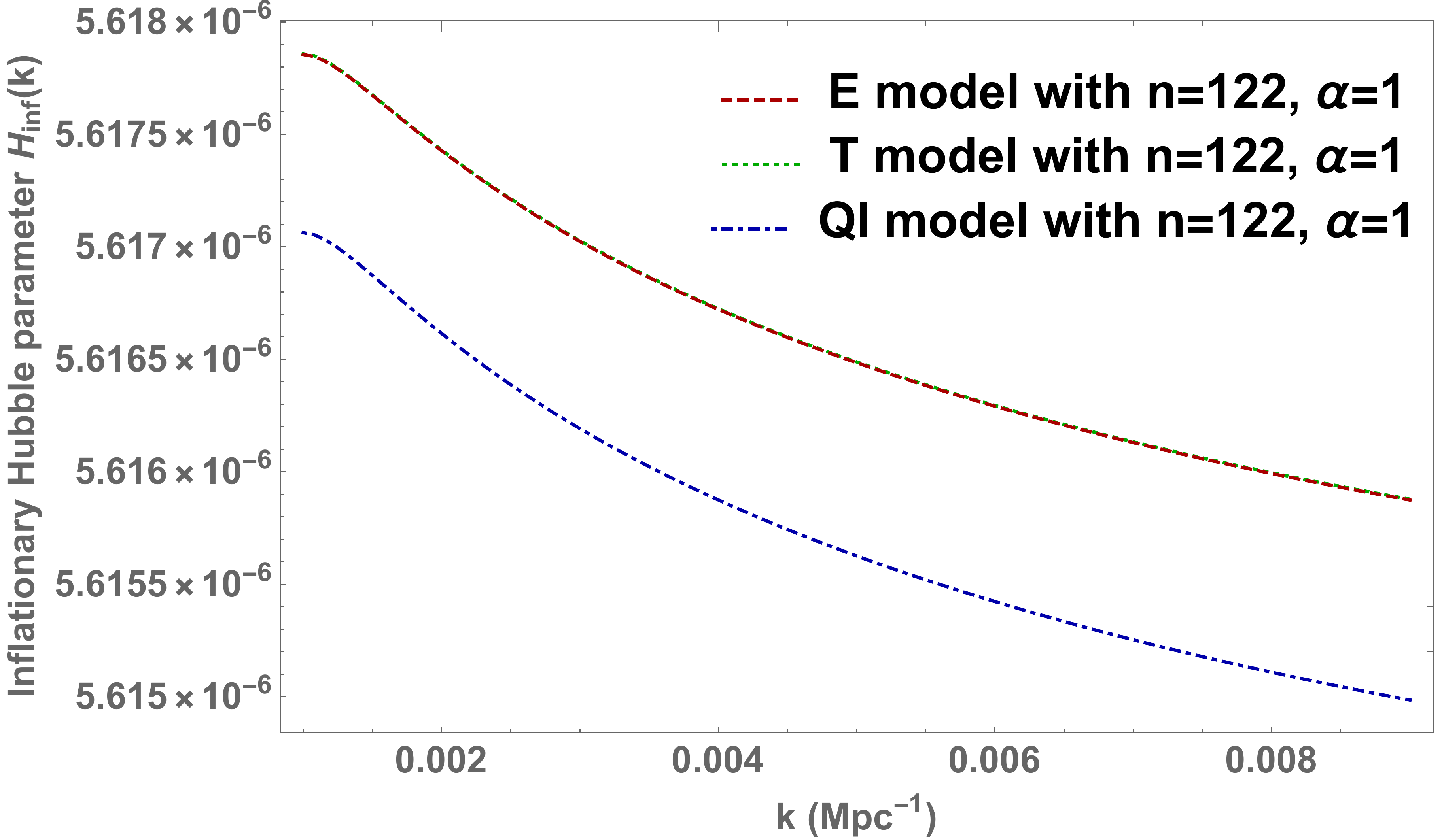}
   \subcaption{}
    \label{fig:Comp7}
\end{subfigure}
\caption{Comparisons of power spectra, number of e-folds and the inflationary Hubble parameters of quintessential $\alpha$-attractor (QI model) with that of ordinary $\alpha$-attractor $E$ and $T$ models for $n=122$ and $\alpha=1$. The results for QI model are smaller than that of $E$ and $T$ models. But the values coincide (or the differences are too small to be detected in the scale shown here) for the two $\alpha$-attractors, which is quite expected. However the $N(k)$ values are very close for all the models described here.}
\label{fig:CompAlphaData1}
\end{figure}

\begin{figure}[H]
\begin{subfigure}{0.52\linewidth}
  \centering
   \includegraphics[width=70mm,height=65mm]{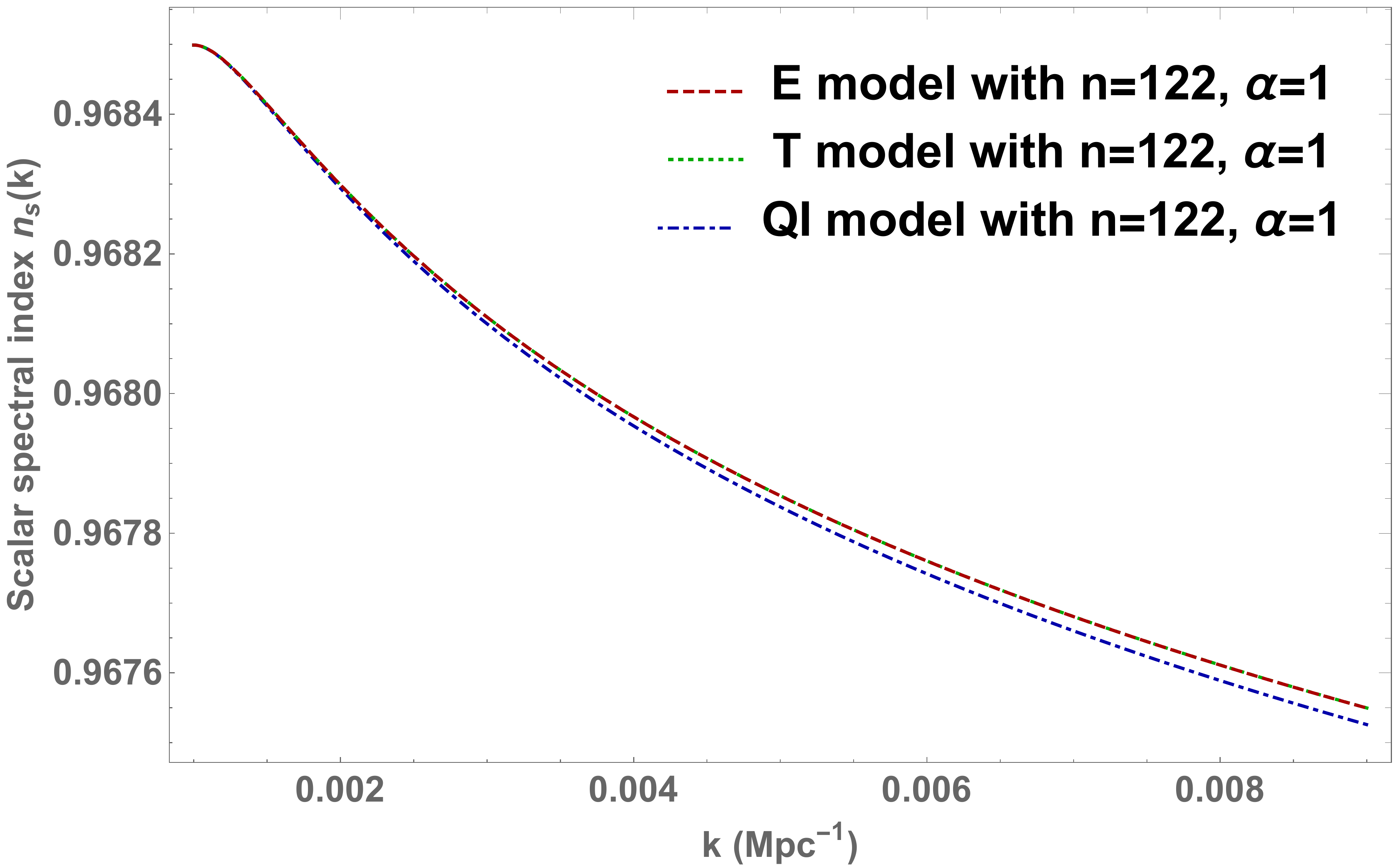} 
   \subcaption{}
   \label{fig:Comp3}
\end{subfigure}%
\begin{subfigure}{0.52\linewidth}
  \centering
   \includegraphics[width=70mm,height=65mm]{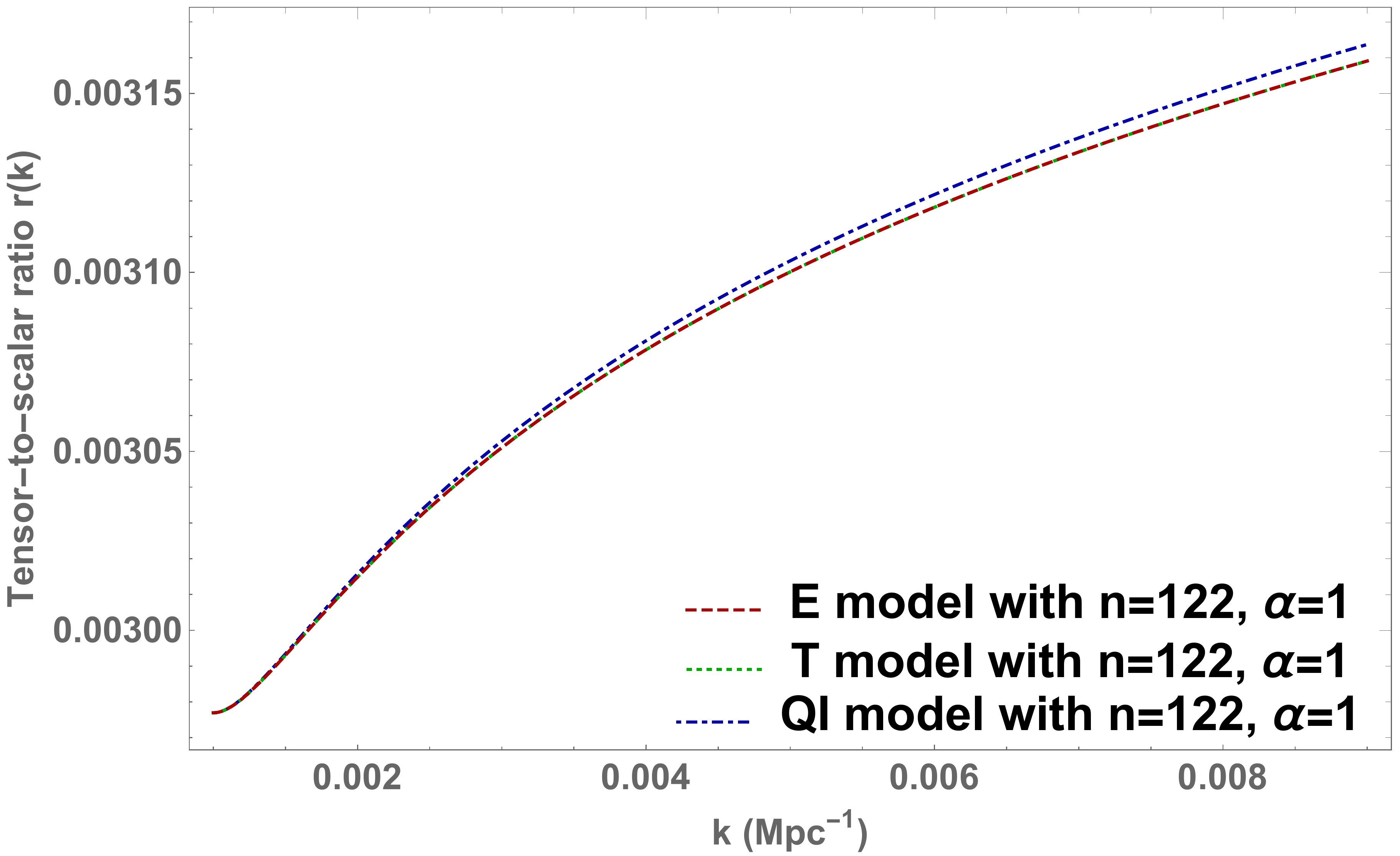}
   \subcaption{}
   \label{fig:Comp4}
\end{subfigure}%
\vspace{0.1\linewidth}
\begin{subfigure}{0.52\linewidth}
  \centering
   \includegraphics[width=70mm,height=65mm]{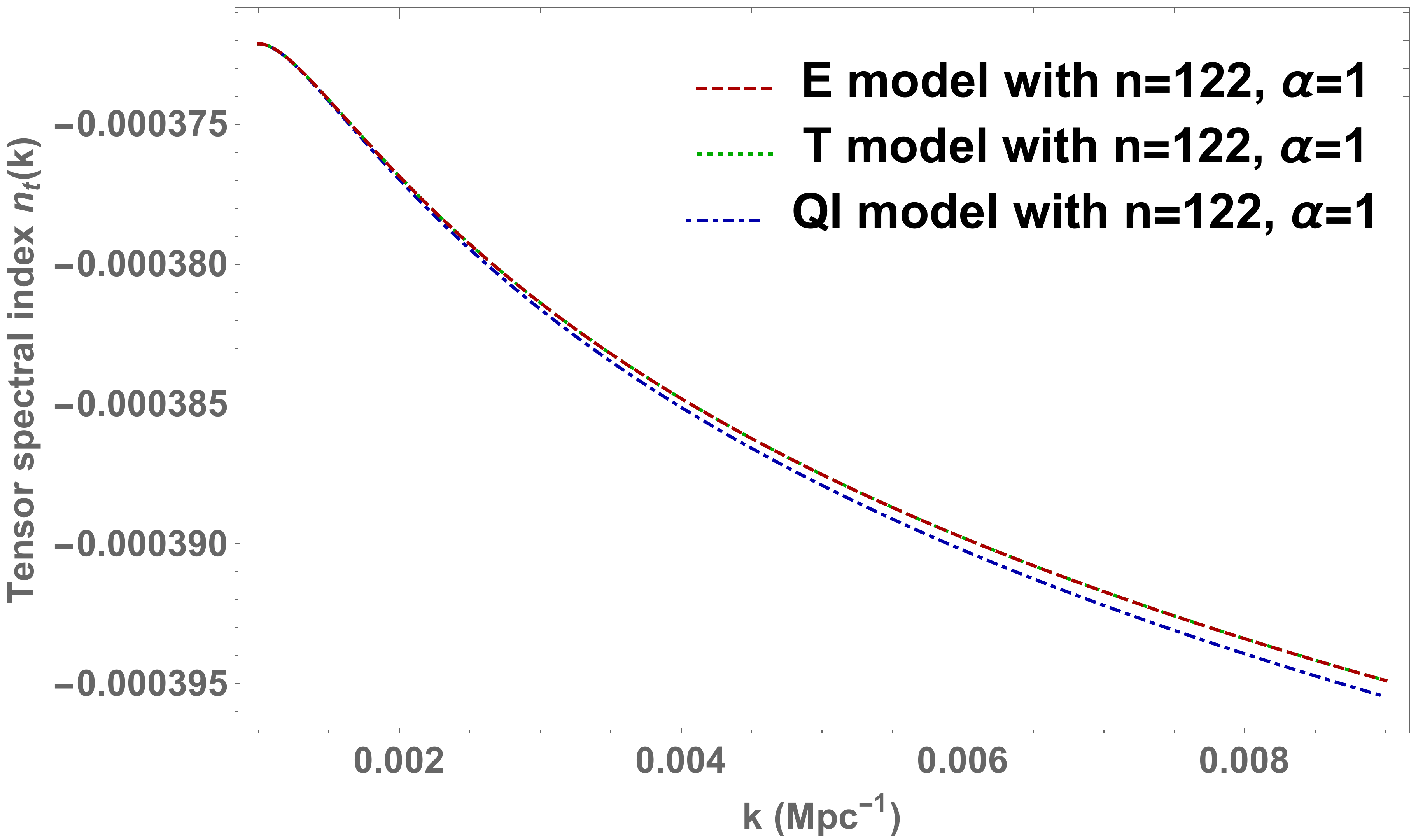}
   \subcaption{}
    \label{fig:Comp5}
\end{subfigure}%
\begin{subfigure}{0.52\linewidth}
  \centering
   \includegraphics[width=70mm,height=65mm]{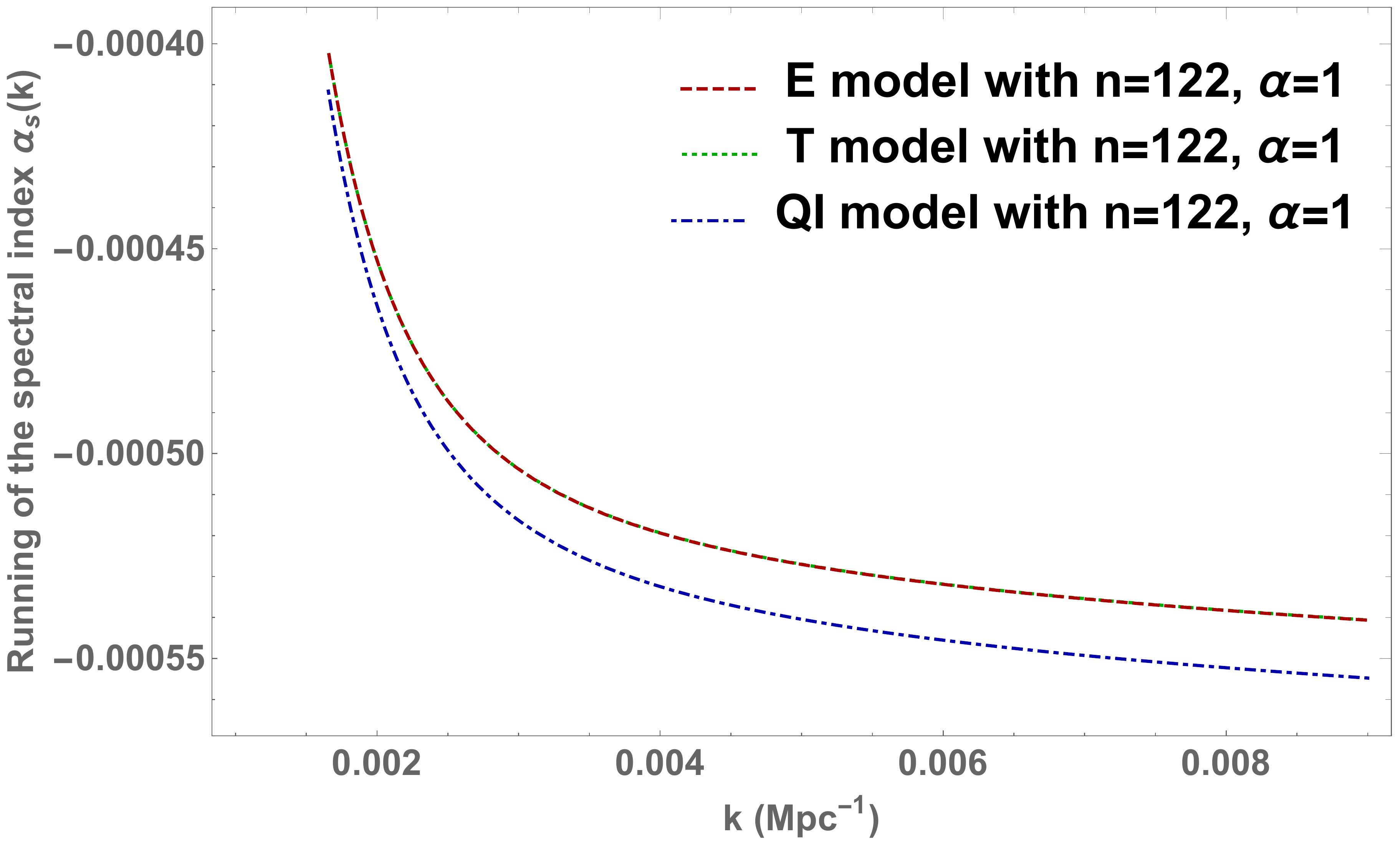}
   \subcaption{}
    \label{fig:Comp8}
\end{subfigure}
\caption{Comparisons of spectral indices, tensor to scalar ratios and the running of the spectral index of quintessential $\alpha$-attractor (QI model) with that of ordinary $\alpha$-attractor $E$ and $T$ models for $n=122$ and $\alpha=1$. The results for QI model are smaller than that of $E$ and $T$ models except for the tensor to scalar ratios, which is obvious. Here also, the values coincide (or the differences are too small to be detected in the scale shown here) for the two $\alpha$-attractors, which is quite expected. However the values of spectral indices and tensor to scalar ratios are very close for all the models described here.}
\label{fig:CompAlphaData2}
\end{figure}

 In figure \ref{fig:CompAlphaData1} we plot the power spectra ($\Delta_{s,t}(k)$), number of e-folds ($N(k)$), inflationary Hubble parameter ($H_{\mathrm{inf}}(k)$) and in figure \ref{fig:CompAlphaData2} we plot spectral indices ($n_{s,t}(k)$), tensor to scalar ratio ($r(k)$), running of spectral index ($\alpha_s(k)$) for $n=122$ and $\alpha=1$ within $k=0.001-0.009$. The results for the $E$ and $T$ models are superposed because of their nearly identical predictions or the differences are too small to be detected in the scale used for plotting. But on comparing with QI model\footnote{by QI model we mean here the inflationary model with quintessential $\alpha$-attractor potential of Eq. (\ref{eq:Eq7}).} we obtain slightly different results especially for $\Delta_{s,t}(k)$ (see figures (\ref{fig:Comp1}) and (\ref{fig:Comp2})) and $H_{\mathrm{inf}}(k)$ (see figure
(\ref{fig:Comp7})). The values of $\Delta_{s,t}(k)$ for $E$ and $T$ models are higher than that of QI model signifying the fact that the two-point correlations of scalar and tensor perturbations are decreased when one incorporate quintessence in the model. Thus quintessence diminishes the value of Hubble parameter as well as the scale of the potential in $k$ space, as shown in figure (\ref{fig:Comp7}). So far as $N(k)$ in figure (\ref{fig:Comp6}) is concerned, its variations are almost identical for all the models. Slight difference is observed at $k\gtrapprox 0.008$ that is in the higher mode regions. That is why the values of $n_{s,t}(k)$ in figures (\ref{fig:Comp3}), (\ref{fig:Comp5}) and $r(k)$ in figure (\ref{fig:Comp4}) behave identically for all the models. This can also be understood by the fact that the rate of mode variations of $\Delta_{s,t}(k)$ for $E/T$ and QI models are nearly equal (almost equal gap is maintained between the graphs). Overall, it can be said that the orders of the cosmological parameters remain same for $E/T$ and QI models, only their magnitudes vary over the specified $k$ range for the given values of $\alpha$ and $n$.
\subsection{Comparison with PLANCK-2018 bounds}
\label{subsec:PLACKcomp}
\begin{figure}[H]
\begin{subfigure}{0.5\textwidth}
  \centering
   \includegraphics[width=70mm,height=65mm]{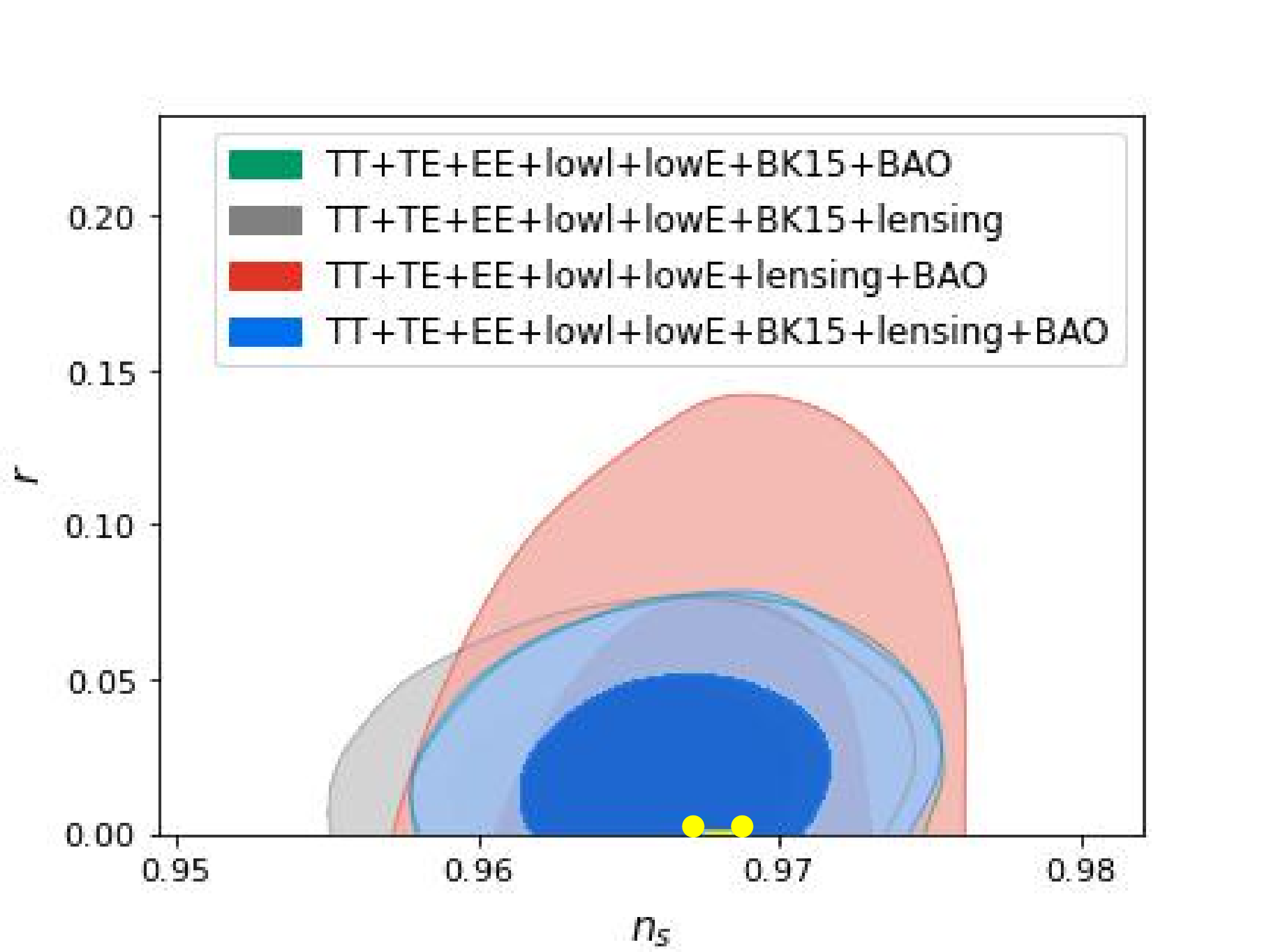} 
   \subcaption{$n=122$ and $\alpha=\frac{1}{10}$}
   \label{fig:PL_1}
\end{subfigure}%
\begin{subfigure}{0.5\textwidth}
  \centering
   \includegraphics[width=70mm,height=65mm]{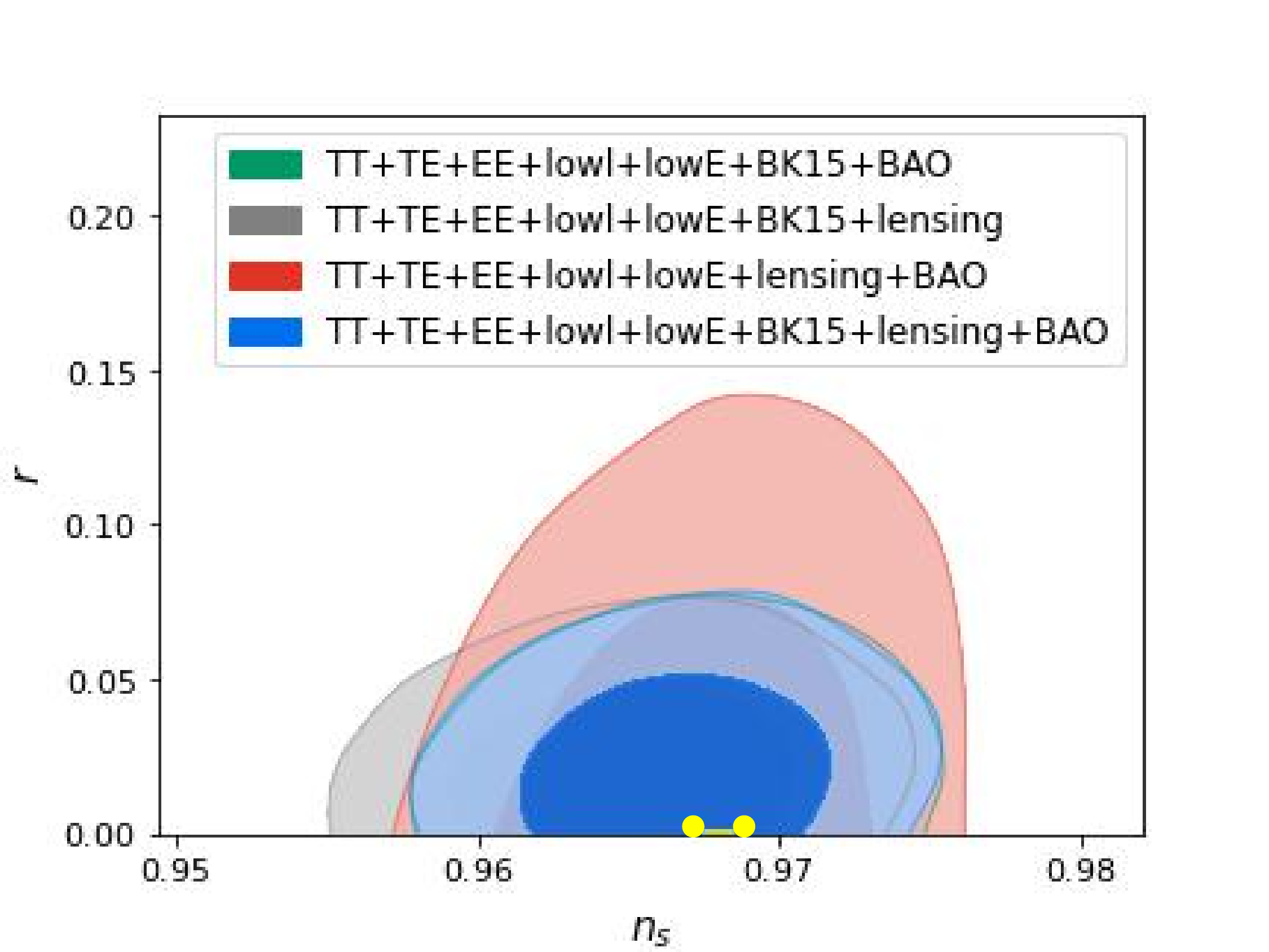}
   \subcaption{$n=122$ and $\alpha=\frac{1}{6}$}
   \label{fig:PL_2}
\end{subfigure}%

\begin{subfigure}{0.5\textwidth}
  \centering
   \includegraphics[width=70mm,height=65mm]{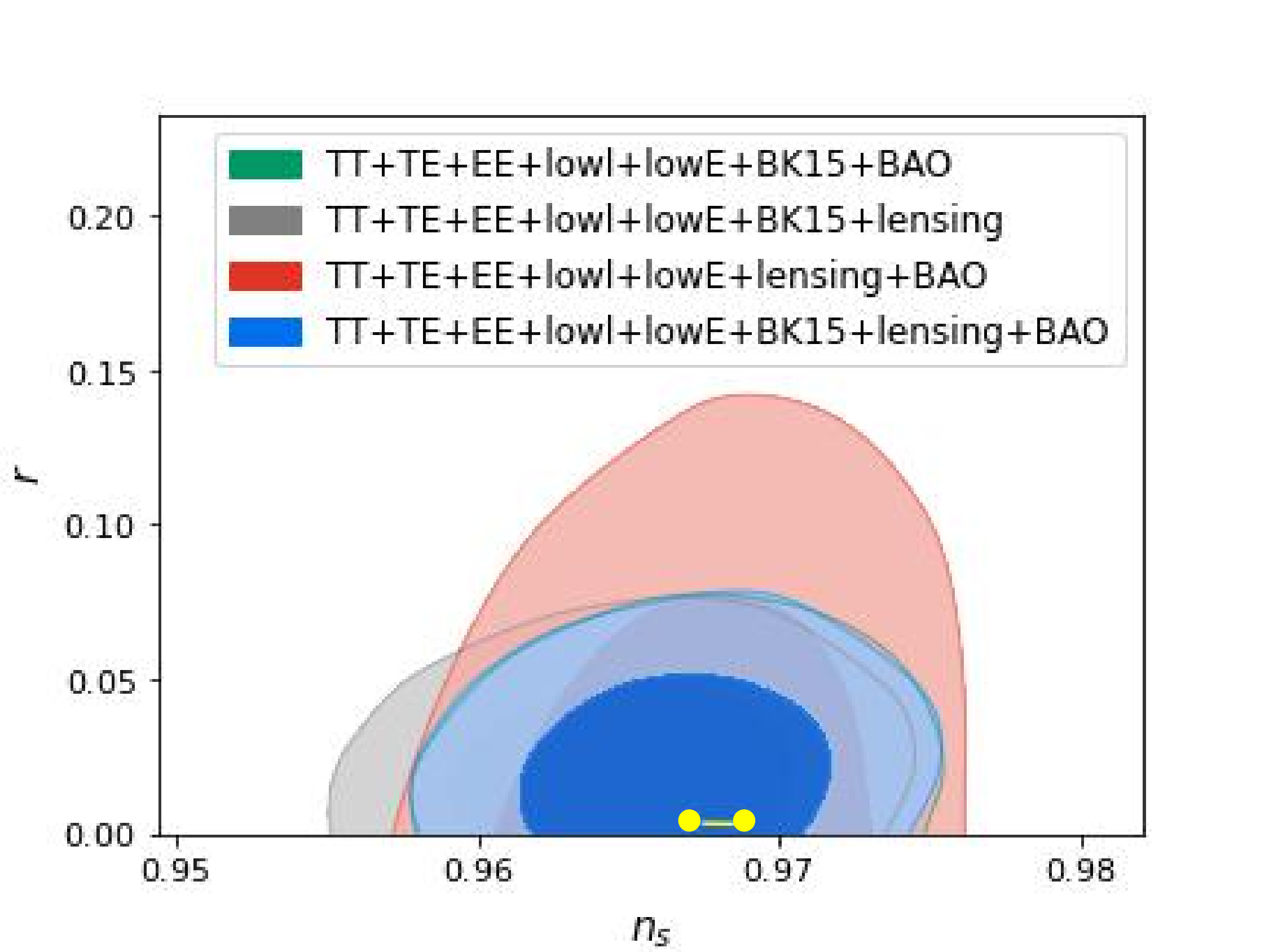}
   \subcaption{$n=122$ and $\alpha=1$}
    \label{fig:PL_3}
\end{subfigure}%
\begin{subfigure}{0.5\textwidth}
  \centering
   \includegraphics[width=70mm,height=65mm]{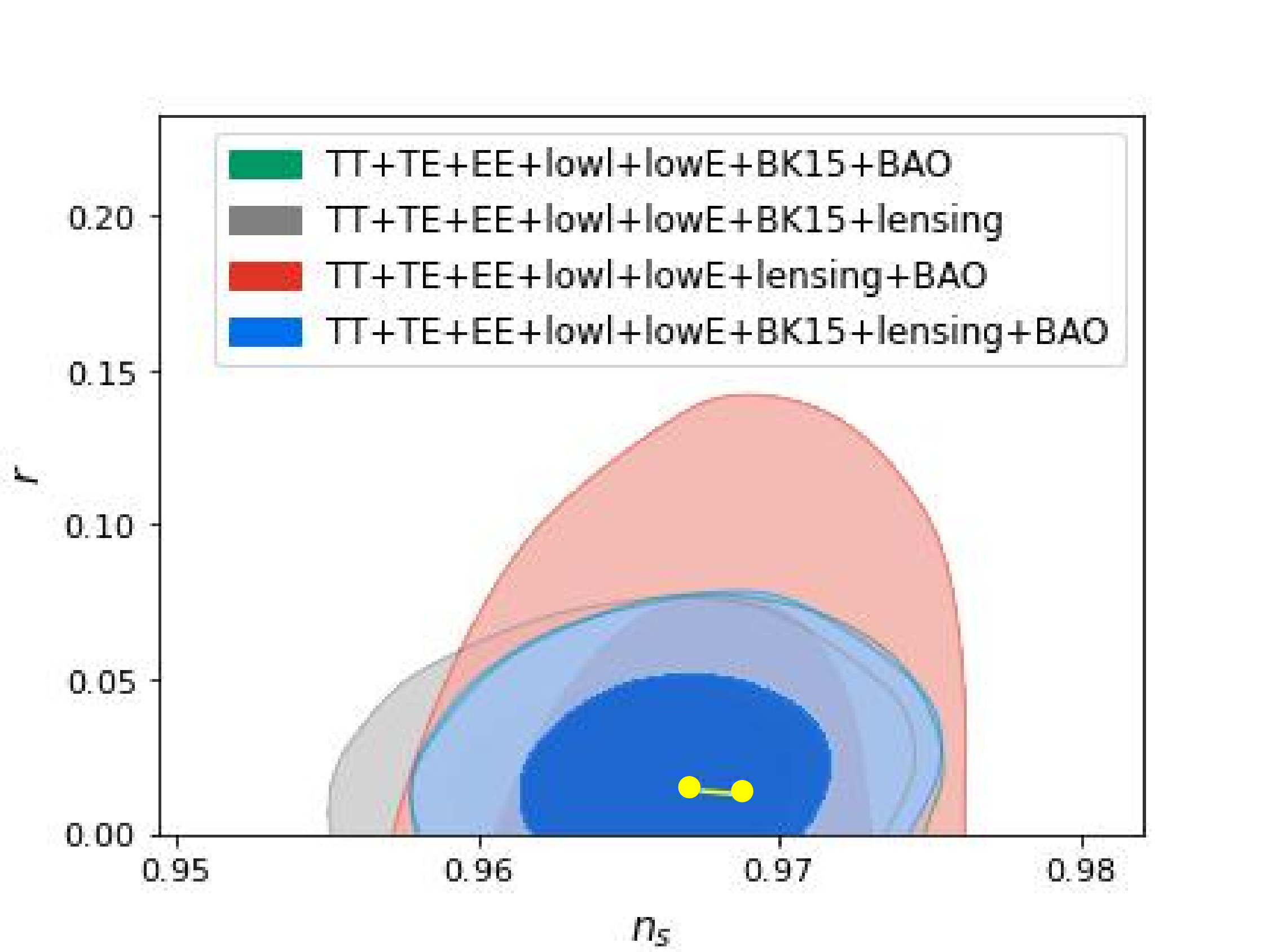}
   \subcaption{$n=122$ and $\alpha=4.3$}
    \label{fig:PL_4}
\end{subfigure}
\caption{Comparison of our calculations for $n_s$ and $r$ with that of Planck-2018 data for increasing values of $\alpha$ for a given value of $n$. All our results lie within the $68\%$ CL. The yellow dots of equal size at the ends represent the data are obtained for a fixed initial condition corresponding to the number of e-folds $N=63.49$.}
\label{fig:PlanckComparison_1}
\end{figure}

\begin{figure}[H]
	\centering
	\includegraphics[width=0.8\linewidth]{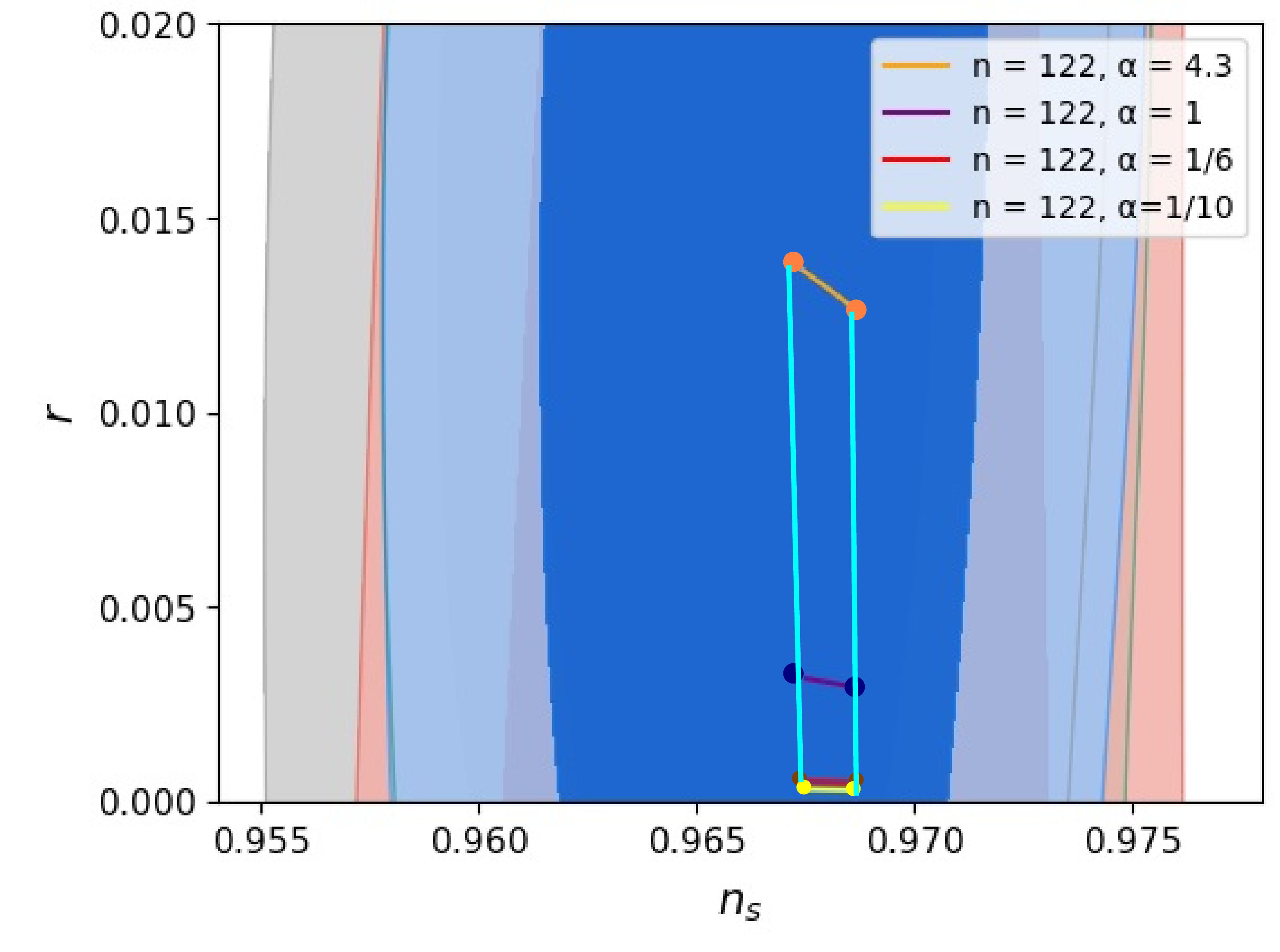}
	\caption{Combined results of figure \ref{fig:PlanckComparison_1} show an elevation of the $n_s-r$ parameteric line with increased values of $\alpha$ from $\frac{1}{10}$ through $4.3$, which is a signature of the potential distortion explained in the context of figures \ref{fig:Fig2b} and \ref{fig:Fig3} in section \ref{sec:our model}. The white line represents that all values of $\alpha$ are permitted between $\frac{1}{10}$ and $4.3$, which is a key difference from the ordinary $\alpha$-attractors explained in \cite{Sarkar:2021ird}. Another important observation is that the slopes of the $n_s-r$ lines get changed when we go from low to high values of $\alpha$. The tilts of the lines ensure that the potential in our framework is concave which is an essential requirement of Planck data for successful model building.}
	\label{fig:PlanckComparison_2}
\end{figure}

At last, in figures \ref{fig:PlanckComparison_1} and \ref{fig:PlanckComparison_2} we compare the calculated cosmological parameters discussed earlier for $\alpha = 1/10, 1/6, 1, 4.3$ and $n=122$ corresponding to $k=0.001 - 0.009$ Mpc$^{-1}$ with Planck constraints \cite{Planck:2018vyg,Planck:2018jri}. The yellow line signifies the values of $n_s$ and $r$ for the specified $k$ range from right hand side to left hand side and the dots of equal sizes at the ends indicate that the results are obtained with the initial conditions consistent with $N=63.49$. We have generated these joint marginalised contours from the data available in Planck Legacy Archive (\url{https://pla.esac.esa.int/}) as public chains by running the GetDist (\url{https://getdist.readthedocs.io/en/latest/}) plotting utility and Python Jupyter notebook (\url{https://jupyter.org/}) environment. These Planck data are based on CMB $E$ mode polarization, temperature anisotropy, Bicep/KECK-15 data \cite{BICEP2:2015nss}, CMB lensing and Baryon Acoustic Oscillation (BAO). Our calculated parameters in sub-Planckian $k$-limit match the Planck data with $68\%$ CL described by the blue zone in the figures for all values of $\alpha$.\par In the QI model, we find an interesting observation in comparison to our earlier study \cite{Sarkar:2021ird} with ordinary $\alpha$-attractors. Our previous study reveals that the parameter $\alpha$ can pick up the values from $\alpha=1$ up to $\alpha=15$. The $\alpha\rightarrow 1$ and $\alpha\rightarrow 15$ limits correspond to the Starobinsky-like plateau type model and power law type simple chaotic model respectively. For this reason the results with $\alpha\leq 10$ lie within $68\%$ CL and $\alpha=15$ lie within $95\%$ CL. But in the present case with quintessence \textit{i.e.} if the scalar field has to drive both early and late time expansions, then the scenario is strikingly different. Below $\alpha=\frac{1}{10}$ the cosmological parameters are insensitive to model parameters, whereas beyond $\alpha=4.3$ the system of mode equations do not converge to a consistent solution. Thus the parameter $\alpha$ is restricted in the limit $\frac{1}{10}\leq\alpha\leq 4.3$ continuously. This has a profound and deep implication in the cosmological perspective. Figure \ref{fig:PlanckComparison_1} shows that all the $n_s - r$ parametric lines lie within the $68\%$ CL zone and no results are found beyond that. That is, our mode analyses discover the fact that upgrading the potential with quintessence restricts the value of $\alpha$ to prevent the potential from going to the simple power-law chaotic regime, which is ruled out by Planck. Thus, a quintessential version of $\alpha$-attractor is just appropriate for the model parameters ($n$ and $\alpha$) in the range stated above. There should be no option but to obtain a potential which is Planck supported. By attaching quintessence the system selects the values of $\alpha$ in such a way that the initial conditions for both inflation and quintessence $viz.,$ the dark energy are automatically satisfied and the coincidence problem is overcome. However, as the cosmological parameters have no direct dependence on the exponent $n$, our $k$ space analysis can not constrain it directly; the only thing we can do is to take a specific value for it (as for example $n=122$ here) constrained by preheating and reheating. These post-inflationary phenomena fix the upper cut-off of $\alpha$ up to $\alpha=4.2$ which is very close to our upper bound $\alpha=4.3$. \par Lastly, we explain an important aspect of the present study in figure \ref{fig:PlanckComparison_2}. This figure shows that the tilts of the $n_s - r$ parametric lines, as we increase the value of $\alpha$ remain such that the results lie always in the regime ($68\%$ CL) where the potential is concave type. The slopes of the lines should be such that $n_s(k)$ will decrease (see figure \ref{fig:scalarSpectralIndex}) and $r(k)$ (see figure \ref{fig:tensorToScalarRatio}) will increase with $k$. Only then the cumulative effect will satisfy the tilt given by Planck data for a single field concave potential. That is exactly what we observe in our case shown in figure \ref{fig:PlanckComparison_2} \footnote{In our earlier study \cite{Sarkar:2021ird} we showed explicitly that $68\%$ CL zone is the regime of concave potentials.}.
\subsection{Calculation of \texorpdfstring{$V_{\Lambda}$}{V}}
\label{subsec:cosmological_constant}
Now, in order to complete the assessment of the efficacy of explaining inflation and quintessence/DE of the model concerned here, we need to compute the scale $V_{\Lambda}$ of the quintessential runaway potential of Eq. (\ref{eq:Eq8}) by the model parameters $n$, $\alpha$ and $M$, constrained in the sub-sections \ref{subsec:ModeBehaviours}-\ref{subsec:PLACKcomp}. This $V_{\Lambda}$ should be comparable with the value given by Planck-2018 \cite{Planck:2018jri,Planck:2018vyg}, $V_{\Lambda}^{\mathrm{Planck}}\approx 10^{-120} M_P^4$, which can explain the current DE observations, albeit being a very small quantity in the Planck scale.\par Let us recall Eq. (\ref{eq:limit2}) and write
\begin{equation}
    V_{\Lambda}=e^{-2n}M^4=10^{-x}\quad (\mathrm{say})
\end{equation} which gives
\begin{equation}
    x=\frac{2n-4\ln{M}}{\ln{10}},
    \label{eq:cosmoExpo}
\end{equation}
where $n$ and $M$ are determined by preheating and reheating and the COBE/Planck normalization (see Eq. (\ref{eq:COBEnormalization})) of the inflationary scalar perturbation as described earlier. The results of Eq. (\ref{eq:cosmoExpo}) are enlisted in the table below.
 \begin{table}[H]
    \captionsetup{justification=centering,width=1.0\textwidth}
    \caption{Constraints for $V_{\Lambda}$ for a given values of $n$ and $M$.}
    \begin{center}
        \begin{adjustbox}{width=0.6\textwidth}
        \begin{tabular}{|c|c|c|c|c|c|}
    \hline
    $n$ & $\alpha$ & $M (M_P)$ & $x$ & $V_{\Lambda} (M_P^4)$ & $V_{\Lambda}^{\mathrm{Planck}} (M_P^4)$\\
    \hline\hline
     & $\frac{1}{10}$ & $1.76\times 10^{-3}$ & $116.98$ & $10^{-116.98}$ & \\
    $122$ & $\frac{1}{6}$ & $2.00\times 10^{-3}$ & $116.76$ & $10^{-116.76}$ & $10^{-120}$\\
     & $1$ & $3.14\times 10^{-3}$ & $115.98$ & $10^{-115.98}$ & \\
     & $4.3$ & $4.57\times 10^{-3}$ & $115.33$ & $10^{-115.33}$ & \\
    \hline
    \end{tabular}
    \end{adjustbox}
    \end{center}
         \label{tab:Table2}
    \end{table}
All values of $V_{\Lambda}$ are close to Planck-bound which clearly indicates that the model of quintessential $\alpha$-attractor described in this paper can explain very well both the early and the late-time expansions of the universe.    
\section{Conclusions}
\label{sec:conclusion}
In conclusion, we have performed a sub-Planckian quantum mode analysis of linear cosmological perturbation in the quintessential $\alpha$-attractor inflaton field over a classical quasi de-Siter metric background by our novel dynamical horizon exit (DHE) method to analyse the cosmological parameters in $k$ space. Our results are found to obey Planck-2018 bounds quite satisfactorily. The cumulative effects of $n_s$ and $r$ shown by comparing with the Planck-2018 contour plots, reveal that the parametric lines lie within the $68\%$ CL zone for all values of model parameters. In this way, we constrain $\alpha$ to be confined between $\frac{1}{10}$ and $4.3$, in continuous way. The upper bound of $\alpha$ is very close to the value $\alpha=4.2$, constrained by preheating and reheating \cite{Dimopoulos:2017tud}. The newly obtained $\alpha$-values also provide extremely tiny vacuum density $V_{\Lambda}\sim 10^{-117}-10^{-115} M_P^4$ (see table \ref{tab:Table2}) which is in good agreement with Planck-value $V_{\Lambda}^{\mathrm{Planck}}\sim 10^{-120} M_P^4$ required for explaining the present accelerating expansion of the universe.\par Slight differences in power spectra and Hubble parameter have been observed in comparison with ordinary $\alpha$ attractor $E$ and $T$ models while other parameters are found to be almost concurrent.\par In view of the above statements, we conclude that the model of quintessential $\alpha$-attractor is more efficacious than ordinary $\alpha$-attractors and it is an appropriate model to accommodate early and late time expansions of our universe in the light of Planck-2018 data.\par In our QI model, we have not incorporated DE element explicitly during inflation. We have only assumed that the inflaton field resurrects as quintessence (which becomes clear from the values of $V_{\Lambda}$ in table \ref{tab:Table2}) after the completion of inflation to take the charge of DE observations of the universe. At present, attempts are being made to furnish the inflaton field by direct inclusion of the early dark energy (EDE) \cite{Brissenden:2023yko,Cruz:2023cxy,CarrilloGonzalez:2023lma,Reboucas:2023rjm,Poulin:2023lkg,Goldstein:2023gnw} in order to solve the `\textit{Hubble tension}' (HT) \cite{DiValentino:2021izs,Ben-Dayan:2023rgt}: the present value Hubble expansion rate $H_0$ is bifurcated into $H_0 = 67.44\pm 0.58$ Kms$^{-1}$Mpc$^{-1}$ from the CMB observations at redshift $z\simeq 1100$ \cite{Planck:2018vyg} and $H_0 = 73.04\pm 1.04$ Kms$^{-1}$Mpc$^{-1}$ from SH0ES observations \cite{Riess:2021jrx} using Cepheid-SN1a data. Measuring $H_0$ \textit{vis-\`{a}-vis} the dark energy more precisely is the target of many ongoing and upcoming CMB $B$ mode based and stage-III/IV LSS surveys like ACT-Pol \cite{ACTPol:2014pbf}, SPT-Pol \cite{SPT:2015htm}, AdvACT \cite{Henderson:2015nzj}, SPT-3G \cite{SPT-3G:2014dbx}, LiteBIRD \cite{LiteBIRD:2020khw}, Canada-France Hawaii Telescope Lensing Survey (CFHTLens) \cite{Heymans:2012gg}, Kilo Degree Survey (KiDS) \cite{Hildebrandt:2016iqg,Kohlinger:2017sxk}, Extended Baryon Oscillation Spectroscopic Survey (eBOSS) \cite{Dawson:2015wdb}, Dark Energy Survey (DES) \cite{DES:2015gax}, Dark Energy Spectroscopic Instrument (DESI) \cite{DESI:2016fyo,DESI:2016igz}, Large Synoptic Survey Telescope (LSST) \cite{LSSTScience:2009jmu,LSST:2017ags}, Wide Field InfraRed Survey Telescope (WFIRST) \cite{Spergel:2015sza,Hounsell:2017ejq} and Euclid \cite{Amendola:2016saw}. Therefore it should be worthwhile to implement EDE in our formalism and study the effects of such procedure.  

% \begin{itemize}
%     \item Disformal transformations and how they are more generalised than conformal transformations
%     \item What work has been done on disformal transformations so far that is relevant to our work
%     \item Why is frame invariance important and why cosmological number counts should be frame-invariant (refer to Bonvin-Durrer, Yoo etc)
%     \item Reference to our previous work on conformal transformations and making the motivation clear that we intend to generalise the results of that work 
%     \item conventions and notations
%     \item some ref's : \\
%     Physical dimensions/units and universal constants: their invariance
% in special and general relativity - HEHL
% \end{itemize}

\section*{Acknowledgments}
The authors acknowledge the University Grants Commission for the CAS-II program in the Department of Physics, The University of Burdwan. AS acknowledges the Government of West Bengal for granting him the Swami Vivekananda fellowship.

\bibliographystyle{utcaps}
\bibliography{biblio}% Produces the bibliography via BibTeX.

\providecommand{\href}[2]{#2}\begingroup\raggedright\begin{thebibliography}{100}

\bibitem{SupernovaSearchTeam:1998fmf}
{\bfseries Supernova Search Team} Collaboration, A.~G. Riess {\em et~al.},
  ``{Observational evidence from supernovae for an accelerating universe and a
  cosmological constant},'' \href{http://dx.doi.org/10.1086/300499}{{\em
  Astron. J.} {\bfseries 116} (1998) 1009--1038},
  \href{http://arxiv.org/abs/astro-ph/9805201}{{\ttfamily
  arXiv:astro-ph/9805201}}.

\bibitem{SupernovaCosmologyProject:1998vns}
{\bfseries Supernova Cosmology Project} Collaboration, S.~Perlmutter {\em
  et~al.}, ``{Measurements of $\Omega$ and $\Lambda$ from 42 high redshift
  supernovae},'' \href{http://dx.doi.org/10.1086/307221}{{\em Astrophys. J.}
  {\bfseries 517} (1999) 565--586},
  \href{http://arxiv.org/abs/astro-ph/9812133}{{\ttfamily
  arXiv:astro-ph/9812133}}.

\bibitem{WMAP:2012nax}
{\bfseries WMAP} Collaboration, G.~Hinshaw {\em et~al.}, ``{Nine-Year Wilkinson
  Microwave Anisotropy Probe (WMAP) Observations: Cosmological Parameter
  Results},'' \href{http://dx.doi.org/10.1088/0067-0049/208/2/19}{{\em
  Astrophys. J. Suppl.} {\bfseries 208} (2013) 19},
  \href{http://arxiv.org/abs/1212.5226}{{\ttfamily arXiv:1212.5226
  [astro-ph.CO]}}.

\bibitem{Planck:2015fie}
{\bfseries Planck} Collaboration, P.~A.~R. Ade {\em et~al.}, ``{Planck 2015
  results. XIII. Cosmological parameters},''
  \href{http://dx.doi.org/10.1051/0004-6361/201525830}{{\em Astron. Astrophys.}
  {\bfseries 594} (2016) A13},
  \href{http://arxiv.org/abs/1502.01589}{{\ttfamily arXiv:1502.01589
  [astro-ph.CO]}}.

\bibitem{Planck:2018vyg}
{\bfseries Planck} Collaboration, N.~Aghanim {\em et~al.}, ``{Planck 2018
  results. VI. Cosmological parameters},''
  \href{http://dx.doi.org/10.1051/0004-6361/201833910}{{\em Astron. Astrophys.}
  {\bfseries 641} (2020) A6}, \href{http://arxiv.org/abs/1807.06209}{{\ttfamily
  arXiv:1807.06209 [astro-ph.CO]}}. [Erratum: Astron.Astrophys. 652, C4
  (2021)].

\bibitem{BICEP2:2015xme}
{\bfseries BICEP2, Keck Array} Collaboration, P.~A.~R. Ade {\em et~al.},
  ``{Improved Constraints on Cosmology and Foregrounds from BICEP2 and Keck
  Array Cosmic Microwave Background Data with Inclusion of 95 GHz Band},''
  \href{http://dx.doi.org/10.1103/PhysRevLett.116.031302}{{\em Phys. Rev.
  Lett.} {\bfseries 116} (2016) 031302},
  \href{http://arxiv.org/abs/1510.09217}{{\ttfamily arXiv:1510.09217
  [astro-ph.CO]}}.

\bibitem{Huterer:1998qv}
D.~Huterer and M.~S. Turner, ``{Prospects for probing the dark energy via
  supernova distance measurements},''
  \href{http://dx.doi.org/10.1103/PhysRevD.60.081301}{{\em Phys. Rev. D}
  {\bfseries 60} (1999) 081301},
  \href{http://arxiv.org/abs/astro-ph/9808133}{{\ttfamily
  arXiv:astro-ph/9808133}}.

\bibitem{Vafa:2005ui}
C.~Vafa, ``{The String landscape and the swampland},''
  \href{http://arxiv.org/abs/hep-th/0509212}{{\ttfamily arXiv:hep-th/0509212}}.

\bibitem{Obied:2018sgi}
G.~Obied, H.~Ooguri, L.~Spodyneiko, and C.~Vafa, ``{De Sitter Space and the
  Swampland},'' \href{http://arxiv.org/abs/1806.08362}{{\ttfamily
  arXiv:1806.08362 [hep-th]}}.

\bibitem{Agrawal:2018own}
P.~Agrawal, G.~Obied, P.~J. Steinhardt, and C.~Vafa, ``{On the Cosmological
  Implications of the String Swampland},''
  \href{http://dx.doi.org/10.1016/j.physletb.2018.07.040}{{\em Phys. Lett. B}
  {\bfseries 784} (2018) 271--276},
  \href{http://arxiv.org/abs/1806.09718}{{\ttfamily arXiv:1806.09718
  [hep-th]}}.

\bibitem{Roupec:2018mbn}
C.~Roupec and T.~Wrase, ``{de Sitter Extrema and the Swampland},''
  \href{http://dx.doi.org/10.1002/prop.201800082}{{\em Fortsch. Phys.}
  {\bfseries 67} no.~1-2, (2019) 1800082},
  \href{http://arxiv.org/abs/1807.09538}{{\ttfamily arXiv:1807.09538
  [hep-th]}}.

\bibitem{Akrami:2018ylq}
Y.~Akrami, R.~Kallosh, A.~Linde, and V.~Vardanyan, ``{The Landscape, the
  Swampland and the Era of Precision Cosmology},''
  \href{http://dx.doi.org/10.1002/prop.201800075}{{\em Fortsch. Phys.}
  {\bfseries 67} no.~1-2, (2019) 1800075},
  \href{http://arxiv.org/abs/1808.09440}{{\ttfamily arXiv:1808.09440
  [hep-th]}}.

\bibitem{Kehagias:2018uem}
A.~Kehagias and A.~Riotto, ``{A note on Inflation and the Swampland},''
  \href{http://dx.doi.org/10.1002/prop.201800052}{{\em Fortsch. Phys.}
  {\bfseries 66} no.~10, (2018) 1800052},
  \href{http://arxiv.org/abs/1807.05445}{{\ttfamily arXiv:1807.05445
  [hep-th]}}.

\bibitem{Dimopoulos:2022wzo}
K.~Dimopoulos, {\em {Introduction to Cosmic Inflation and Dark Energy}}.
\newblock CRC Press, 5, 2022.

\bibitem{Copeland:2006wr}
E.~J. Copeland, M.~Sami, and S.~Tsujikawa, ``{Dynamics of dark energy},''
  \href{http://dx.doi.org/10.1142/S021827180600942X}{{\em Int. J. Mod. Phys. D}
  {\bfseries 15} (2006) 1753--1936},
  \href{http://arxiv.org/abs/hep-th/0603057}{{\ttfamily arXiv:hep-th/0603057}}.

\bibitem{Clifton:2011jh}
T.~Clifton, P.~G. Ferreira, A.~Padilla, and C.~Skordis, ``{Modified Gravity and
  Cosmology},'' \href{http://dx.doi.org/10.1016/j.physrep.2012.01.001}{{\em
  Phys. Rept.} {\bfseries 513} (2012) 1--189},
  \href{http://arxiv.org/abs/1106.2476}{{\ttfamily arXiv:1106.2476
  [astro-ph.CO]}}.

\bibitem{Peebles:2002gy}
P.~J.~E. Peebles and B.~Ratra, ``{The Cosmological Constant and Dark Energy},''
  \href{http://dx.doi.org/10.1103/RevModPhys.75.559}{{\em Rev. Mod. Phys.}
  {\bfseries 75} (2003) 559--606},
  \href{http://arxiv.org/abs/astro-ph/0207347}{{\ttfamily
  arXiv:astro-ph/0207347}}.

\bibitem{Ratra:1987rm}
B.~Ratra and P.~J.~E. Peebles, ``{Cosmological Consequences of a Rolling
  Homogeneous Scalar Field},''
  \href{http://dx.doi.org/10.1103/PhysRevD.37.3406}{{\em Phys. Rev. D}
  {\bfseries 37} (1988) 3406}.

\bibitem{Caldwell:1997ii}
R.~R. Caldwell, R.~Dave, and P.~J. Steinhardt, ``{Cosmological imprint of an
  energy component with general equation of state},''
  \href{http://dx.doi.org/10.1103/PhysRevLett.80.1582}{{\em Phys. Rev. Lett.}
  {\bfseries 80} (1998) 1582--1585},
  \href{http://arxiv.org/abs/astro-ph/9708069}{{\ttfamily
  arXiv:astro-ph/9708069}}.

\bibitem{Dimopoulos:2021xld}
K.~Dimopoulos, ``{Jointly modelling Cosmic Inflation and Dark Energy},''
  \href{http://dx.doi.org/10.1088/1742-6596/2105/1/012001}{{\em J. Phys. Conf.
  Ser.} {\bfseries 2105} no.~1, (2021) 012001},
  \href{http://arxiv.org/abs/2106.14966}{{\ttfamily arXiv:2106.14966 [gr-qc]}}.

\bibitem{Peebles:1998qn}
P.~J.~E. Peebles and A.~Vilenkin, ``{Quintessential inflation},''
  \href{http://dx.doi.org/10.1103/PhysRevD.59.063505}{{\em Phys. Rev. D}
  {\bfseries 59} (1999) 063505},
  \href{http://arxiv.org/abs/astro-ph/9810509}{{\ttfamily
  arXiv:astro-ph/9810509}}.

\bibitem{Peloso:1999dm}
M.~Peloso and F.~Rosati, ``{On the construction of quintessential inflation
  models},'' \href{http://dx.doi.org/10.1088/1126-6708/1999/12/026}{{\em JHEP}
  {\bfseries 12} (1999) 026},
  \href{http://arxiv.org/abs/hep-ph/9908271}{{\ttfamily arXiv:hep-ph/9908271}}.

\bibitem{Sen:2000ym}
A.~A. Sen, I.~Chakrabarty, and T.~R. Seshadri, ``{Quintessential inflation with
  dissipative fluid},'' \href{http://dx.doi.org/10.1023/A:1015536623936}{{\em
  Gen. Rel. Grav.} {\bfseries 34} (2002) 477--490},
  \href{http://arxiv.org/abs/gr-qc/0005104}{{\ttfamily arXiv:gr-qc/0005104}}.

\bibitem{Kaganovich:2000fc}
A.~B. Kaganovich, ``{Field theory model giving rise to 'quintessential
  inflation' without the cosmological constant and other fine tuning
  problems},'' \href{http://dx.doi.org/10.1103/PhysRevD.63.025022}{{\em Phys.
  Rev. D} {\bfseries 63} (2001) 025022},
  \href{http://arxiv.org/abs/hep-th/0007144}{{\ttfamily arXiv:hep-th/0007144}}.

\bibitem{Yahiro:2001uh}
M.~Yahiro, G.~J. Mathews, K.~Ichiki, T.~Kajino, and M.~Orito, ``{Constraints on
  cosmic quintessence and quintessential inflation},''
  \href{http://dx.doi.org/10.1103/PhysRevD.65.063502}{{\em Phys. Rev. D}
  {\bfseries 65} (2002) 063502},
  \href{http://arxiv.org/abs/astro-ph/0106349}{{\ttfamily
  arXiv:astro-ph/0106349}}.

\bibitem{Martin:2004ba}
J.~Martin and M.~A. Musso, ``{Stochastic quintessence},''
  \href{http://dx.doi.org/10.1103/PhysRevD.71.063514}{{\em Phys. Rev. D}
  {\bfseries 71} (2005) 063514},
  \href{http://arxiv.org/abs/astro-ph/0410190}{{\ttfamily
  arXiv:astro-ph/0410190}}.

\bibitem{Barenboim:2005np}
G.~Barenboim and J.~D. Lykken, ``{Slinky Inflation},''
  \href{http://dx.doi.org/10.1016/j.physletb.2005.12.041}{{\em Phys. Lett. B}
  {\bfseries 633} (2006) 453--457},
  \href{http://arxiv.org/abs/astro-ph/0504090}{{\ttfamily
  arXiv:astro-ph/0504090}}.

\bibitem{Rosenfeld:2005mt}
R.~Rosenfeld and J.~A. Frieman, ``{A Simple model for quintessential
  inflation},'' \href{http://dx.doi.org/10.1088/1475-7516/2005/09/003}{{\em
  JCAP} {\bfseries 09} (2005) 003},
  \href{http://arxiv.org/abs/astro-ph/0504191}{{\ttfamily
  arXiv:astro-ph/0504191}}.

\bibitem{Cardenas:2006py}
V.~H. Cardenas, ``{Tachyonic quintessential inflation},''
  \href{http://dx.doi.org/10.1103/PhysRevD.73.103512}{{\em Phys. Rev. D}
  {\bfseries 73} (2006) 103512},
  \href{http://arxiv.org/abs/gr-qc/0603013}{{\ttfamily arXiv:gr-qc/0603013}}.

\bibitem{BuenoSanchez:2006fhh}
J.~C. Bueno~Sanchez and K.~Dimopoulos, ``{Trapped Quintessential Inflation},''
  \href{http://dx.doi.org/10.1016/j.physletb.2006.09.045}{{\em Phys. Lett. B}
  {\bfseries 642} (2006) 294--301},
  \href{http://arxiv.org/abs/hep-th/0605258}{{\ttfamily arXiv:hep-th/0605258}}.
  [Erratum: Phys.Lett.B 647, 526 (2007)].

\bibitem{Membiela:2006rj}
A.~Membiela and M.~Bellini, ``{Quintessential inflation from a variable
  cosmological constant in a 5D vacuum},''
  \href{http://dx.doi.org/10.1016/j.physletb.2006.08.043}{{\em Phys. Lett. B}
  {\bfseries 641} (2006) 125--129},
  \href{http://arxiv.org/abs/gr-qc/0606119}{{\ttfamily arXiv:gr-qc/0606119}}.

\bibitem{Rosenfeld:2006hs}
R.~Rosenfeld and J.~A. Frieman, ``{Cosmic microwave background and large-scale
  structure constraints on a simple quintessential inflation model},''
  \href{http://dx.doi.org/10.1103/PhysRevD.75.043513}{{\em Phys. Rev. D}
  {\bfseries 75} (2007) 043513},
  \href{http://arxiv.org/abs/astro-ph/0611241}{{\ttfamily
  arXiv:astro-ph/0611241}}.

\bibitem{Neupane:2007mu}
I.~P. Neupane, ``{Reconstructing a model of quintessential inflation},''
  \href{http://dx.doi.org/10.1088/0264-9381/25/12/125013}{{\em Class. Quant.
  Grav.} {\bfseries 25} (2008) 125013},
  \href{http://arxiv.org/abs/0706.2654}{{\ttfamily arXiv:0706.2654 [hep-th]}}.

\bibitem{Bastero-Gil:2009wdy}
M.~Bastero-Gil, A.~Berera, B.~M. Jackson, and A.~Taylor, ``{Hybrid
  Quintessential Inflation},''
  \href{http://dx.doi.org/10.1016/j.physletb.2009.06.025}{{\em Phys. Lett. B}
  {\bfseries 678} (2009) 157--163},
  \href{http://arxiv.org/abs/0905.2937}{{\ttfamily arXiv:0905.2937 [hep-ph]}}.

\bibitem{Piedipalumbo:2011bj}
E.~Piedipalumbo, P.~Scudellaro, G.~Esposito, and C.~Rubano, ``{On
  quintessential cosmological models and exponential potentials},''
  \href{http://dx.doi.org/10.1007/s10714-012-1421-9}{{\em Gen. Rel. Grav.}
  {\bfseries 44} (2012) 2611--2643},
  \href{http://arxiv.org/abs/1112.0502}{{\ttfamily arXiv:1112.0502
  [astro-ph.CO]}}.

\bibitem{Wetterich:2014gaa}
C.~Wetterich, ``{Inflation, quintessence, and the origin of mass},''
  \href{http://dx.doi.org/10.1016/j.nuclphysb.2015.05.019}{{\em Nucl. Phys. B}
  {\bfseries 897} (2015) 111--178},
  \href{http://arxiv.org/abs/1408.0156}{{\ttfamily arXiv:1408.0156 [hep-th]}}.

\bibitem{Hossain:2014xha}
M.~W. Hossain, R.~Myrzakulov, M.~Sami, and E.~N. Saridakis, ``{Variable
  gravity: A suitable framework for quintessential inflation},''
  \href{http://dx.doi.org/10.1103/PhysRevD.90.023512}{{\em Phys. Rev. D}
  {\bfseries 90} no.~2, (2014) 023512},
  \href{http://arxiv.org/abs/1402.6661}{{\ttfamily arXiv:1402.6661 [gr-qc]}}.

\bibitem{Hossain:2014coa}
M.~W. Hossain, R.~Myrzakulov, M.~Sami, and E.~N. Saridakis, ``{Class of
  quintessential inflation models with parameter space consistent with
  BICEP2},'' \href{http://dx.doi.org/10.1103/PhysRevD.89.123513}{{\em Phys.
  Rev. D} {\bfseries 89} no.~12, (2014) 123513},
  \href{http://arxiv.org/abs/1404.1445}{{\ttfamily arXiv:1404.1445 [gr-qc]}}.

\bibitem{Hossain:2014ova}
M.~W. Hossain, R.~Myrzakulov, M.~Sami, and E.~N. Saridakis, ``{Evading Lyth
  bound in models of quintessential inflation},''
  \href{http://dx.doi.org/10.1016/j.physletb.2014.08.051}{{\em Phys. Lett. B}
  {\bfseries 737} (2014) 191--195},
  \href{http://arxiv.org/abs/1405.7491}{{\ttfamily arXiv:1405.7491 [gr-qc]}}.

\bibitem{Geng:2015fla}
C.-Q. Geng, M.~W. Hossain, R.~Myrzakulov, M.~Sami, and E.~N. Saridakis,
  ``{Quintessential inflation with canonical and noncanonical scalar fields and
  Planck 2015 results},''
  \href{http://dx.doi.org/10.1103/PhysRevD.92.023522}{{\em Phys. Rev. D}
  {\bfseries 92} no.~2, (2015) 023522},
  \href{http://arxiv.org/abs/1502.03597}{{\ttfamily arXiv:1502.03597 [gr-qc]}}.

\bibitem{WaliHossain:2014usl}
M.~Wali~Hossain, R.~Myrzakulov, M.~Sami, and E.~N. Saridakis, ``{Unification of
  inflation and dark energy \`a la quintessential inflation},''
  \href{http://dx.doi.org/10.1142/S0218271815300141}{{\em Int. J. Mod. Phys. D}
  {\bfseries 24} no.~05, (2015) 1530014},
  \href{http://arxiv.org/abs/1410.6100}{{\ttfamily arXiv:1410.6100 [gr-qc]}}.

\bibitem{Haro:2015ljc}
J.~Haro and S.~Pan, ``{Bulk viscous quintessential inflation},''
  \href{http://dx.doi.org/10.1142/S0218271818500529}{{\em Int. J. Mod. Phys. D}
  {\bfseries 27} no.~05, (2018) 1850052},
  \href{http://arxiv.org/abs/1512.03033}{{\ttfamily arXiv:1512.03033 [gr-qc]}}.

\bibitem{deHaro:2016ftq}
J.~de~Haro, ``{On the viability of quintessential inflation models from
  observational data},''
  \href{http://dx.doi.org/10.1007/s10714-016-2173-8}{{\em Gen. Rel. Grav.}
  {\bfseries 49} no.~1, (2017) 6},
  \href{http://arxiv.org/abs/1602.07138}{{\ttfamily arXiv:1602.07138 [gr-qc]}}.

\bibitem{deHaro:2016cdm}
J.~de~Haro, J.~Amor\'os, and S.~Pan, ``{Simple inflationary quintessential
  model II: Power law potentials},''
  \href{http://dx.doi.org/10.1103/PhysRevD.94.064060}{{\em Phys. Rev. D}
  {\bfseries 94} no.~6, (2016) 064060},
  \href{http://arxiv.org/abs/1607.06726}{{\ttfamily arXiv:1607.06726 [gr-qc]}}.

\bibitem{Guendelman:2016kwj}
E.~Guendelman, E.~Nissimov, and S.~Pacheva, ``{Quintessential Inflation,
  Unified Dark Energy and Dark Matter, and Higgs Mechanism},'' {\em Bulg. J.
  Phys.} {\bfseries 44} no.~1, (2017) 015--030,
  \href{http://arxiv.org/abs/1609.06915}{{\ttfamily arXiv:1609.06915 [gr-qc]}}.

\bibitem{Rubio:2017gty}
J.~Rubio and C.~Wetterich, ``{Emergent scale symmetry: Connecting inflation and
  dark energy},'' \href{http://dx.doi.org/10.1103/PhysRevD.96.063509}{{\em
  Phys. Rev. D} {\bfseries 96} no.~6, (2017) 063509},
  \href{http://arxiv.org/abs/1705.00552}{{\ttfamily arXiv:1705.00552 [gr-qc]}}.

\bibitem{Ahmad:2017itq}
S.~Ahmad, R.~Myrzakulov, and M.~Sami, ``{Relic gravitational waves from
  Quintessential Inflation},''
  \href{http://dx.doi.org/10.1103/PhysRevD.96.063515}{{\em Phys. Rev. D}
  {\bfseries 96} no.~6, (2017) 063515},
  \href{http://arxiv.org/abs/1705.02133}{{\ttfamily arXiv:1705.02133 [gr-qc]}}.

\bibitem{Haro:2018zdb}
J.~Haro, W.~Yang, and S.~Pan, ``{Reheating in quintessential inflation via
  gravitational production of heavy massive particles: A detailed analysis},''
  \href{http://dx.doi.org/10.1088/1475-7516/2019/01/023}{{\em JCAP} {\bfseries
  01} (2019) 023}, \href{http://arxiv.org/abs/1811.07371}{{\ttfamily
  arXiv:1811.07371 [gr-qc]}}.

\bibitem{Bettoni:2018pbl}
D.~Bettoni, G.~Dom\`enech, and J.~Rubio, ``{Gravitational waves from global
  cosmic strings in quintessential inflation},''
  \href{http://dx.doi.org/10.1088/1475-7516/2019/02/034}{{\em JCAP} {\bfseries
  02} (2019) 034}, \href{http://arxiv.org/abs/1810.11117}{{\ttfamily
  arXiv:1810.11117 [astro-ph.CO]}}.

\bibitem{Selvaganapathy:2019bpm}
J.~Selvaganapathy, ``{Pure natural quintessential inflation and dark energy},''
  \href{http://dx.doi.org/10.1142/S0217751X20500979}{{\em Int. J. Mod. Phys. A}
  {\bfseries 35} no.~19, (2020) 2050097},
  \href{http://arxiv.org/abs/1911.10466}{{\ttfamily arXiv:1911.10466
  [hep-ph]}}.

\bibitem{Lima:2019yyv}
G.~B.~F. Lima and R.~O. Ramos, ``{Unified early and late Universe cosmology
  through dissipative effects in steep quintessential inflation potential
  models},'' \href{http://dx.doi.org/10.1103/PhysRevD.100.123529}{{\em Phys.
  Rev. D} {\bfseries 100} no.~12, (2019) 123529},
  \href{http://arxiv.org/abs/1910.05185}{{\ttfamily arXiv:1910.05185
  [astro-ph.CO]}}.

\bibitem{Kleidis:2019ywv}
K.~Kleidis and V.~K. Oikonomou, ``{A Study of an Einstein Gauss-Bonnet
  Quintessential Inflationary Model},''
  \href{http://dx.doi.org/10.1016/j.nuclphysb.2019.114765}{{\em Nucl. Phys. B}
  {\bfseries 948} (2019) 114765},
  \href{http://arxiv.org/abs/1909.05318}{{\ttfamily arXiv:1909.05318 [gr-qc]}}.

\bibitem{Haro:2019peq}
J.~Haro, J.~Amor\'os, and S.~Pan, ``{Scaling solutions in quintessential
  inflation},'' \href{http://dx.doi.org/10.1140/epjc/s10052-020-7950-6}{{\em
  Eur. Phys. J. C} {\bfseries 80} no.~5, (2020) 404},
  \href{http://arxiv.org/abs/1908.01516}{{\ttfamily arXiv:1908.01516 [gr-qc]}}.

\bibitem{Benisty:2020xqm}
D.~Benisty and E.~I. Guendelman, ``{Lorentzian Quintessential Inflation},''
  \href{http://dx.doi.org/10.1142/S021827182042002X}{{\em Int. J. Mod. Phys. D}
  {\bfseries 29} no.~14, (2020) 2042002},
  \href{http://arxiv.org/abs/2004.00339}{{\ttfamily arXiv:2004.00339
  [astro-ph.CO]}}.

\bibitem{Benisty:2020vvm}
D.~Benisty, E.~I. Guendelman, E.~Nissimov, and S.~Pacheva, ``{Quintessential
  Inflation with Dynamical Higgs Generation as an Affine Gravity},''
  \href{http://dx.doi.org/10.3390/sym12050734}{{\em Symmetry} {\bfseries 12}
  (2020) 734}, \href{http://arxiv.org/abs/2003.04723}{{\ttfamily
  arXiv:2003.04723 [gr-qc]}}.

\bibitem{deHaro:2021swo}
J.~de~Haro and L.~A. Sal\'o, ``{A Review of Quintessential Inflation},''
  \href{http://dx.doi.org/10.3390/galaxies9040073}{{\em Galaxies} {\bfseries 9}
  no.~4, (2021) 73}, \href{http://arxiv.org/abs/2108.11144}{{\ttfamily
  arXiv:2108.11144 [gr-qc]}}.

\bibitem{Tian:2021cqq}
S.~X. Tian and Z.-H. Zhu, ``{Cosmological consequences of a scalar field with
  oscillating equation of state. III. Unifying inflation with dark energy and
  small tensor-to-scalar ratio},''
  \href{http://dx.doi.org/10.1103/PhysRevD.103.123545}{{\em Phys. Rev. D}
  {\bfseries 103} no.~12, (2021) 123545},
  \href{http://arxiv.org/abs/2106.14002}{{\ttfamily arXiv:2106.14002
  [astro-ph.CO]}}.

\bibitem{AresteSalo:2021wgb}
L.~Arest\'e~Sal\'o, D.~Benisty, E.~I. Guendelman, and J.~de~Haro,
  ``{$\alpha$-attractors in quintessential inflation motivated by
  supergravity},'' \href{http://dx.doi.org/10.1103/PhysRevD.103.123535}{{\em
  Phys. Rev. D} {\bfseries 103} no.~12, (2021) 123535},
  \href{http://arxiv.org/abs/2103.07892}{{\ttfamily arXiv:2103.07892
  [astro-ph.CO]}}.

\bibitem{Akrami:2020zxw}
Y.~Akrami, S.~Casas, S.~Deng, and V.~Vardanyan, ``{Quintessential
  $\alpha$-attractor inflation: forecasts for Stage IV galaxy surveys},''
  \href{http://dx.doi.org/10.1088/1475-7516/2021/04/006}{{\em JCAP} {\bfseries
  04} (2021) 006}, \href{http://arxiv.org/abs/2010.15822}{{\ttfamily
  arXiv:2010.15822 [astro-ph.CO]}}.

\bibitem{Garcia-Garcia:2019ees}
C.~Garc\'\i{}a-Garc\'\i{}a, P.~Ru\'\i{}z-Lapuente, D.~Alonso, and
  M.~Zumalac\'arregui, ``{$\alpha$-attractor dark energy in view of
  next-generation cosmological surveys},''
  \href{http://dx.doi.org/10.1088/1475-7516/2019/07/025}{{\em JCAP} {\bfseries
  07} (2019) 025}, \href{http://arxiv.org/abs/1905.03753}{{\ttfamily
  arXiv:1905.03753 [astro-ph.CO]}}.

\bibitem{Garcia-Garcia:2018hlc}
C.~Garc\'\i{}a-Garc\'\i{}a, E.~V. Linder, P.~Ru\'\i{}z-Lapuente, and
  M.~Zumalac\'arregui, ``{Dark energy from $\alpha$-attractors: phenomenology
  and observational constraints},''
  \href{http://dx.doi.org/10.1088/1475-7516/2018/08/022}{{\em JCAP} {\bfseries
  08} (2018) 022}, \href{http://arxiv.org/abs/1803.00661}{{\ttfamily
  arXiv:1803.00661 [astro-ph.CO]}}.

\bibitem{Akrami:2017cir}
Y.~Akrami, R.~Kallosh, A.~Linde, and V.~Vardanyan, ``{Dark energy,
  $\alpha$-attractors, and large-scale structure surveys},''
  \href{http://dx.doi.org/10.1088/1475-7516/2018/06/041}{{\em JCAP} {\bfseries
  06} (2018) 041}, \href{http://arxiv.org/abs/1712.09693}{{\ttfamily
  arXiv:1712.09693 [hep-th]}}.

\bibitem{Kepuladze:2021tsb}
Z.~Kepuladze and M.~Maziashvili, ``{New take on the inflationary
  quintessence},'' \href{http://dx.doi.org/10.1103/PhysRevD.103.063540}{{\em
  Phys. Rev. D} {\bfseries 103} no.~6, (2021) 063540},
  \href{http://arxiv.org/abs/2102.09203}{{\ttfamily arXiv:2102.09203
  [astro-ph.CO]}}.

\bibitem{Dimopoulos:2017zvq}
K.~Dimopoulos and C.~Owen, ``{Quintessential Inflation with
  $\alpha$-attractors},''
  \href{http://dx.doi.org/10.1088/1475-7516/2017/06/027}{{\em JCAP} {\bfseries
  06} (2017) 027}, \href{http://arxiv.org/abs/1703.00305}{{\ttfamily
  arXiv:1703.00305 [gr-qc]}}.

\bibitem{Geng:2017mic}
C.-Q. Geng, C.-C. Lee, M.~Sami, E.~N. Saridakis, and A.~A. Starobinsky,
  ``{Observational constraints on successful model of quintessential
  Inflation},'' \href{http://dx.doi.org/10.1088/1475-7516/2017/06/011}{{\em
  JCAP} {\bfseries 06} (2017) 011},
  \href{http://arxiv.org/abs/1705.01329}{{\ttfamily arXiv:1705.01329 [gr-qc]}}.

\bibitem{Agarwal:2017wxo}
A.~Agarwal, R.~Myrzakulov, M.~Sami, and N.~K. Singh, ``{Quintessential
  inflation in a thawing realization},''
  \href{http://dx.doi.org/10.1016/j.physletb.2017.04.066}{{\em Phys. Lett. B}
  {\bfseries 770} (2017) 200--208},
  \href{http://arxiv.org/abs/1708.00156}{{\ttfamily arXiv:1708.00156 [gr-qc]}}.

\bibitem{AresteSalo:2017lkv}
L.~Arest\'e~Sal\'o and J.~de~Haro, ``{Quintessential inflation at low reheating
  temperatures},'' \href{http://dx.doi.org/10.1140/epjc/s10052-017-5337-0}{{\em
  Eur. Phys. J. C} {\bfseries 77} no.~11, (2017) 798},
  \href{http://arxiv.org/abs/1707.02810}{{\ttfamily arXiv:1707.02810 [gr-qc]}}.

\bibitem{DeHaro:2017abf}
J.~De~Haro and L.~Arest\'e~Sal\'o, ``{Reheating constraints in quintessential
  inflation},'' \href{http://dx.doi.org/10.1103/PhysRevD.95.123501}{{\em Phys.
  Rev. D} {\bfseries 95} no.~12, (2017) 123501},
  \href{http://arxiv.org/abs/1702.04212}{{\ttfamily arXiv:1702.04212 [gr-qc]}}.

\bibitem{Haro:2019gsv}
J.~Haro, J.~Amor\'os, and S.~Pan, ``{The Peebles - Vilenkin quintessential
  inflation model revisited},''
  \href{http://dx.doi.org/10.1140/epjc/s10052-019-7012-0}{{\em Eur. Phys. J. C}
  {\bfseries 79} no.~6, (2019) 505},
  \href{http://arxiv.org/abs/1901.00167}{{\ttfamily arXiv:1901.00167 [gr-qc]}}.

\bibitem{deHaro:2019oki}
J.~de~Haro, S.~Pan, and L.~Arest\'e~Sal\'o, ``{Understanding gravitational
  particle production in quintessential inflation},''
  \href{http://dx.doi.org/10.1088/1475-7516/2019/06/056}{{\em JCAP} {\bfseries
  06} (2019) 056}, \href{http://arxiv.org/abs/1903.01181}{{\ttfamily
  arXiv:1903.01181 [gr-qc]}}.

\bibitem{Benisty:2020qta}
D.~Benisty and E.~I. Guendelman, ``{Quintessential Inflation from Lorentzian
  Slow Roll},'' \href{http://dx.doi.org/10.1140/epjc/s10052-020-8147-8}{{\em
  Eur. Phys. J. C} {\bfseries 80} no.~6, (2020) 577},
  \href{http://arxiv.org/abs/2006.04129}{{\ttfamily arXiv:2006.04129
  [astro-ph.CO]}}.

\bibitem{Shokri:2021zqw}
M.~Shokri, J.~Sadeghi, and S.~N. Gashti, ``{Quintessential constant-roll
  inflation},'' \href{http://dx.doi.org/10.1016/j.dark.2021.100923}{{\em Phys.
  Dark Univ.} {\bfseries 35} (2022) 100923},
  \href{http://arxiv.org/abs/2107.04756}{{\ttfamily arXiv:2107.04756
  [astro-ph.CO]}}.

\bibitem{Bettoni:2021qfs}
D.~Bettoni and J.~Rubio, ``{Quintessential Inflation: A Tale of Emergent and
  Broken Symmetries},'' \href{http://dx.doi.org/10.3390/galaxies10010022}{{\em
  Galaxies} {\bfseries 10} no.~1, (2022) 22},
  \href{http://arxiv.org/abs/2112.11948}{{\ttfamily arXiv:2112.11948
  [astro-ph.CO]}}.

\bibitem{Jaman:2022bho}
N.~Jaman and M.~Sami, ``{What Is Needed of a Scalar Field If It Is to Unify
  Inflation and Late Time Acceleration?},''
  \href{http://dx.doi.org/10.3390/galaxies10020051}{{\em Galaxies} {\bfseries
  10} no.~2, (2022) 51}, \href{http://arxiv.org/abs/2202.06194}{{\ttfamily
  arXiv:2202.06194 [gr-qc]}}.

\bibitem{Jesus:2021bxq}
J.~F. Jesus, R.~Valentim, A.~A. Escobal, S.~H. Pereira, and D.~Benndorf,
  ``{Gaussian processes reconstruction of the dark energy potential},''
  \href{http://dx.doi.org/10.1088/1475-7516/2022/11/037}{{\em JCAP} {\bfseries
  11} (2022) 037}, \href{http://arxiv.org/abs/2112.09722}{{\ttfamily
  arXiv:2112.09722 [astro-ph.CO]}}.

\bibitem{Fujikura:2022udt}
K.~Fujikura, S.~Hashiba, and J.~Yokoyama, ``{Generation of neutrino dark
  matter, baryon asymmetry, and radiation after quintessential inflation},''
  \href{http://arxiv.org/abs/2210.05214}{{\ttfamily arXiv:2210.05214
  [hep-ph]}}.

\bibitem{Karciauskas:2021fdu}
M.~Kar\v{c}iauskas, S.~Rusak, and A.~Saez, ``{Quintessential inflation and
  nonlinear effects of the tachyonic trap mechanism},''
  \href{http://dx.doi.org/10.1103/PhysRevD.105.043535}{{\em Phys. Rev. D}
  {\bfseries 105} no.~4, (2022) 043535},
  \href{http://arxiv.org/abs/2112.11536}{{\ttfamily arXiv:2112.11536
  [astro-ph.CO]}}.

\bibitem{Basak:2021cgk}
S.~Basak, S.~Bhattacharya, M.~R. Gangopadhyay, N.~Jaman, R.~Rangarajan, and
  M.~Sami, ``{The paradigm of warm quintessential inflation and spontaneous
  baryogenesis},'' \href{http://dx.doi.org/10.1088/1475-7516/2022/03/063}{{\em
  JCAP} {\bfseries 03} no.~03, (2022) 063},
  \href{http://arxiv.org/abs/2110.00607}{{\ttfamily arXiv:2110.00607
  [astro-ph.CO]}}.

\bibitem{AresteSalo:2020yxl}
L.~Areste~Salo and J.~Haro, ``{Quintessential Inflation for Exponential Type
  Potentials: Scaling and Tracker Behavior},''
  \href{http://dx.doi.org/10.1140/epjc/s10052-021-08906-2}{{\em Eur. Phys. J.
  C} {\bfseries 81} no.~2, (2021) 105},
  \href{http://arxiv.org/abs/2009.12912}{{\ttfamily arXiv:2009.12912 [gr-qc]}}.

\bibitem{Spokoiny:1993kt}
B.~Spokoiny, ``{Deflationary universe scenario},''
  \href{http://dx.doi.org/10.1016/0370-2693(93)90155-B}{{\em Phys. Lett. B}
  {\bfseries 315} (1993) 40--45},
  \href{http://arxiv.org/abs/gr-qc/9306008}{{\ttfamily arXiv:gr-qc/9306008}}.

\bibitem{Pallis:2005hm}
C.~Pallis, ``{Quintessential kination and cold dark matter abundance},''
  \href{http://dx.doi.org/10.1088/1475-7516/2005/10/015}{{\em JCAP} {\bfseries
  10} (2005) 015}, \href{http://arxiv.org/abs/hep-ph/0503080}{{\ttfamily
  arXiv:hep-ph/0503080}}.

\bibitem{Pallis:2005bb}
C.~Pallis, ``{Kination-dominated reheating and cold dark matter abundance},''
  \href{http://dx.doi.org/10.1016/j.nuclphysb.2006.06.003}{{\em Nucl. Phys. B}
  {\bfseries 751} (2006) 129--159},
  \href{http://arxiv.org/abs/hep-ph/0510234}{{\ttfamily arXiv:hep-ph/0510234}}.

\bibitem{Gomez:2008js}
M.~E. Gomez, S.~Lola, C.~Pallis, and J.~Rodriguez-Quintero, ``{Quintessential
  Kination and Thermal Production of Gravitinos and Axinos},''
  \href{http://dx.doi.org/10.1088/1475-7516/2009/01/027}{{\em JCAP} {\bfseries
  01} (2009) 027}, \href{http://arxiv.org/abs/0809.1859}{{\ttfamily
  arXiv:0809.1859 [hep-ph]}}.

\bibitem{Campos:2002yk}
A.~H. Campos, H.~C. Reis, and R.~Rosenfeld, ``{Preheating in quintessential
  inflation},'' \href{http://dx.doi.org/10.1016/j.physletb.2003.09.064}{{\em
  Phys. Lett. B} {\bfseries 575} (2003) 151--156},
  \href{http://arxiv.org/abs/hep-ph/0210152}{{\ttfamily arXiv:hep-ph/0210152}}.

\bibitem{Dimopoulos:2017tud}
K.~Dimopoulos, L.~Donaldson~Wood, and C.~Owen, ``{Instant preheating in
  quintessential inflation with $\alpha$-attractors},''
  \href{http://dx.doi.org/10.1103/PhysRevD.97.063525}{{\em Phys. Rev. D}
  {\bfseries 97} no.~6, (2018) 063525},
  \href{http://arxiv.org/abs/1712.01760}{{\ttfamily arXiv:1712.01760
  [astro-ph.CO]}}.

\bibitem{Feng:2002nb}
B.~Feng and M.-z. Li, ``{Curvaton reheating in nonoscillatory inflationary
  models},'' \href{http://dx.doi.org/10.1016/S0370-2693(03)00589-6}{{\em Phys.
  Lett. B} {\bfseries 564} (2003) 169--174},
  \href{http://arxiv.org/abs/hep-ph/0212213}{{\ttfamily arXiv:hep-ph/0212213}}.

\bibitem{BuenoSanchez:2007jxm}
J.~C. Bueno~Sanchez and K.~Dimopoulos, ``{Curvaton reheating allows TeV Hubble
  scale in NO inflation},''
  \href{http://dx.doi.org/10.1088/1475-7516/2007/11/007}{{\em JCAP} {\bfseries
  11} (2007) 007}, \href{http://arxiv.org/abs/0707.3967}{{\ttfamily
  arXiv:0707.3967 [hep-ph]}}.

\bibitem{Matsuda:2007ax}
T.~Matsuda, ``{NO Curvatons or Hybrid Quintessential Inflation},''
  \href{http://dx.doi.org/10.1088/1475-7516/2007/08/003}{{\em JCAP} {\bfseries
  08} (2007) 003}, \href{http://arxiv.org/abs/0707.1948}{{\ttfamily
  arXiv:0707.1948 [hep-ph]}}.

\bibitem{Chun:2009yu}
E.~J. Chun, S.~Scopel, and I.~Zaballa, ``{Gravitational reheating in
  quintessential inflation},''
  \href{http://dx.doi.org/10.1088/1475-7516/2009/07/022}{{\em JCAP} {\bfseries
  07} (2009) 022}, \href{http://arxiv.org/abs/0904.0675}{{\ttfamily
  arXiv:0904.0675 [hep-ph]}}.

\bibitem{Dimopoulos:2018wfg}
K.~Dimopoulos and T.~Markkanen, ``{Non-minimal gravitational reheating during
  kination},'' \href{http://dx.doi.org/10.1088/1475-7516/2018/06/021}{{\em
  JCAP} {\bfseries 06} (2018) 021},
  \href{http://arxiv.org/abs/1803.07399}{{\ttfamily arXiv:1803.07399 [gr-qc]}}.

\bibitem{Opferkuch:2019zbd}
T.~Opferkuch, P.~Schwaller, and B.~A. Stefanek, ``{Ricci Reheating},''
  \href{http://dx.doi.org/10.1088/1475-7516/2019/07/016}{{\em JCAP} {\bfseries
  07} (2019) 016}, \href{http://arxiv.org/abs/1905.06823}{{\ttfamily
  arXiv:1905.06823 [gr-qc]}}.

\bibitem{Dimopoulos:2019gpz}
K.~Dimopoulos and L.~Donaldson-Wood, ``{Warm quintessential inflation},''
  \href{http://dx.doi.org/10.1016/j.physletb.2019.07.017}{{\em Phys. Lett. B}
  {\bfseries 796} (2019) 26--31},
  \href{http://arxiv.org/abs/1906.09648}{{\ttfamily arXiv:1906.09648 [gr-qc]}}.

\bibitem{Rosa:2019jci}
J.~a.~G. Rosa and L.~B. Ventura, ``{Warm Little Inflaton becomes Dark
  Energy},'' \href{http://dx.doi.org/10.1016/j.physletb.2019.134984}{{\em Phys.
  Lett. B} {\bfseries 798} (2019) 134984},
  \href{http://arxiv.org/abs/1906.11835}{{\ttfamily arXiv:1906.11835
  [hep-ph]}}.

\bibitem{Gangopadhyay:2020bxn}
M.~R. Gangopadhyay, S.~Myrzakul, M.~Sami, and M.~K. Sharma, ``{Paradigm of warm
  quintessential inflation and production of relic gravity waves},''
  \href{http://dx.doi.org/10.1103/PhysRevD.103.043505}{{\em Phys. Rev. D}
  {\bfseries 103} no.~4, (2021) 043505},
  \href{http://arxiv.org/abs/2011.09155}{{\ttfamily arXiv:2011.09155
  [astro-ph.CO]}}.

\bibitem{AresteSalo:2021lmp}
L.~Arest\'e~Sal\'o, D.~Benisty, E.~I. Guendelman, and J.~d. Haro,
  ``{Quintessential inflation and cosmological seesaw mechanism: reheating and
  observational constraints},''
  \href{http://dx.doi.org/10.1088/1475-7516/2021/07/007}{{\em JCAP} {\bfseries
  07} (2021) 007}, \href{http://arxiv.org/abs/2102.09514}{{\ttfamily
  arXiv:2102.09514 [astro-ph.CO]}}.

\bibitem{Salo:2021vdv}
L.~A. Sal\'o and J.~de~Haro, ``{Gravitational particle production of superheavy
  massive particles in quintessential inflation: A numerical analysis},''
  \href{http://dx.doi.org/10.1103/PhysRevD.104.083544}{{\em Phys. Rev. D}
  {\bfseries 104} no.~8, (2021) 083544},
  \href{http://arxiv.org/abs/2108.10795}{{\ttfamily arXiv:2108.10795 [gr-qc]}}.

\bibitem{Salo:2021piz}
L.~A. Sal\'o and J.~de~Haro, ``{Gravitational particle production of superheavy
  massive particles in Quintessential Inflation II: $\alpha$-attractors},''
  \href{http://arxiv.org/abs/2112.12992}{{\ttfamily arXiv:2112.12992 [gr-qc]}}.

\bibitem{deHaro:2022vxc}
J.~de~Haro and L.~Arest\'e~Sal\'o, ``{The problem of calculating the
  $\beta$-Bogoliubov coefficient in Non-Oscillating models},''
  \href{http://arxiv.org/abs/2206.13854}{{\ttfamily arXiv:2206.13854 [gr-qc]}}.

\bibitem{deHaro:2022ukj}
J.~de~Haro and L.~Arest\'e~Sal\'o, ``{An analytic formula to calculate the
  reheating temperature via gravitational particle production},''
  \href{http://arxiv.org/abs/2212.01276}{{\ttfamily arXiv:2212.01276 [gr-qc]}}.

\bibitem{Sarkar:2021ird}
A.~Sarkar, C.~Sarkar, and B.~Ghosh, ``{A novel way of constraining the
  \ensuremath{\alpha}-attractor chaotic inflation through Planck data},''
  \href{http://dx.doi.org/10.1088/1475-7516/2021/11/029}{{\em JCAP} {\bfseries
  11} no.~11, (2021) 029}, \href{http://arxiv.org/abs/2106.02920}{{\ttfamily
  arXiv:2106.02920 [gr-qc]}}.

\bibitem{Brissenden:2023yko}
L.~Brissenden, K.~Dimopoulos, and S.~S\'anchez~L\'opez, ``{Non-oscillating
  Early Dark Energy and Quintessence from Alpha-Attractors},''
  \href{http://arxiv.org/abs/2301.03572}{{\ttfamily arXiv:2301.03572
  [astro-ph.CO]}}.

\bibitem{Guendelman:2002js}
E.~I. Guendelman and O.~Katz, ``{Inflation and transition to a slowly
  accelerating phase from SSB of scale invariance},''
  \href{http://dx.doi.org/10.1088/0264-9381/20/9/309}{{\em Class. Quant. Grav.}
  {\bfseries 20} (2003) 1715--1728},
  \href{http://arxiv.org/abs/gr-qc/0211095}{{\ttfamily arXiv:gr-qc/0211095}}.

\bibitem{Guendelman:2014bva}
E.~Guendelman, R.~Herrera, P.~Labrana, E.~Nissimov, and S.~Pacheva, ``{Emergent
  Cosmology, Inflation and Dark Energy},''
  \href{http://dx.doi.org/10.1007/s10714-015-1852-1}{{\em Gen. Rel. Grav.}
  {\bfseries 47} no.~2, (2015) 10},
  \href{http://arxiv.org/abs/1408.5344}{{\ttfamily arXiv:1408.5344 [gr-qc]}}.

\bibitem{Kallosh:2013yoa}
R.~Kallosh, A.~Linde, and D.~Roest, ``{Superconformal Inflationary
  $\alpha$-Attractors},'' \href{http://dx.doi.org/10.1007/JHEP11(2013)198}{{\em
  JHEP} {\bfseries 11} (2013) 198},
  \href{http://arxiv.org/abs/1311.0472}{{\ttfamily arXiv:1311.0472 [hep-th]}}.

\bibitem{Ferrara:2013rsa}
S.~Ferrara, R.~Kallosh, A.~Linde, and M.~Porrati, ``{Minimal Supergravity
  Models of Inflation},''
  \href{http://dx.doi.org/10.1103/PhysRevD.88.085038}{{\em Phys. Rev. D}
  {\bfseries 88} no.~8, (2013) 085038},
  \href{http://arxiv.org/abs/1307.7696}{{\ttfamily arXiv:1307.7696 [hep-th]}}.

\bibitem{Kallosh:2013pby}
R.~Kallosh and A.~Linde, ``{Superconformal generalization of the chaotic
  inflation model $\frac{\lambda}{4} \phi^{4} - \frac{\xi}{2} \phi^{2}R$},''
  \href{http://dx.doi.org/10.1088/1475-7516/2013/06/027}{{\em JCAP} {\bfseries
  06} (2013) 027}, \href{http://arxiv.org/abs/1306.3211}{{\ttfamily
  arXiv:1306.3211 [hep-th]}}.

\bibitem{Kallosh:2013maa}
R.~Kallosh and A.~Linde, ``{Non-minimal Inflationary Attractors},''
  \href{http://dx.doi.org/10.1088/1475-7516/2013/10/033}{{\em JCAP} {\bfseries
  10} (2013) 033}, \href{http://arxiv.org/abs/1307.7938}{{\ttfamily
  arXiv:1307.7938 [hep-th]}}.

\bibitem{Kallosh:2015lwa}
R.~Kallosh and A.~Linde, ``{Planck, LHC, and $\alpha$-attractors},''
  \href{http://dx.doi.org/10.1103/PhysRevD.91.083528}{{\em Phys. Rev. D}
  {\bfseries 91} (2015) 083528},
  \href{http://arxiv.org/abs/1502.07733}{{\ttfamily arXiv:1502.07733
  [astro-ph.CO]}}.

\bibitem{Carrasco:2015pla}
J.~J.~M. Carrasco, R.~Kallosh, and A.~Linde, ``{$\alpha $-Attractors: Planck,
  LHC and Dark Energy},'' \href{http://dx.doi.org/10.1007/JHEP10(2015)147}{{\em
  JHEP} {\bfseries 10} (2015) 147},
  \href{http://arxiv.org/abs/1506.01708}{{\ttfamily arXiv:1506.01708
  [hep-th]}}.

\bibitem{Kallosh:2013hoa}
R.~Kallosh and A.~Linde, ``{Universality Class in Conformal Inflation},''
  \href{http://dx.doi.org/10.1088/1475-7516/2013/07/002}{{\em JCAP} {\bfseries
  07} (2013) 002}, \href{http://arxiv.org/abs/1306.5220}{{\ttfamily
  arXiv:1306.5220 [hep-th]}}.

\bibitem{Maeda:2018sje}
K.-I. Maeda, S.~Mizuno, and R.~Tozuka, ``{$\alpha$-attractor-type double
  inflation},'' \href{http://dx.doi.org/10.1103/PhysRevD.98.123530}{{\em Phys.
  Rev. D} {\bfseries 98} no.~12, (2018) 123530},
  \href{http://arxiv.org/abs/1810.06914}{{\ttfamily arXiv:1810.06914
  [hep-th]}}.

\bibitem{Kallosh:2014rga}
R.~Kallosh, A.~Linde, and D.~Roest, ``{Large field inflation and double
  $\alpha$-attractors},'' \href{http://dx.doi.org/10.1007/JHEP08(2014)052}{{\em
  JHEP} {\bfseries 08} (2014) 052},
  \href{http://arxiv.org/abs/1405.3646}{{\ttfamily arXiv:1405.3646 [hep-th]}}.

\bibitem{Galante:2014ifa}
M.~Galante, R.~Kallosh, A.~Linde, and D.~Roest, ``{Unity of Cosmological
  Inflation Attractors},''
  \href{http://dx.doi.org/10.1103/PhysRevLett.114.141302}{{\em Phys. Rev.
  Lett.} {\bfseries 114} no.~14, (2015) 141302},
  \href{http://arxiv.org/abs/1412.3797}{{\ttfamily arXiv:1412.3797 [hep-th]}}.

\bibitem{Kallosh:2013tua}
R.~Kallosh, A.~Linde, and D.~Roest, ``{Universal Attractor for Inflation at
  Strong Coupling},''
  \href{http://dx.doi.org/10.1103/PhysRevLett.112.011303}{{\em Phys. Rev.
  Lett.} {\bfseries 112} no.~1, (2014) 011303},
  \href{http://arxiv.org/abs/1310.3950}{{\ttfamily arXiv:1310.3950 [hep-th]}}.

\bibitem{Ferrara:2013kca}
S.~Ferrara, R.~Kallosh, A.~Linde, and M.~Porrati, ``{Higher Order Corrections
  in Minimal Supergravity Models of Inflation},''
  \href{http://dx.doi.org/10.1088/1475-7516/2013/11/046}{{\em JCAP} {\bfseries
  11} (2013) 046}, \href{http://arxiv.org/abs/1309.1085}{{\ttfamily
  arXiv:1309.1085 [hep-th]}}.

\bibitem{Carrasco:2015rva}
J.~J.~M. Carrasco, R.~Kallosh, and A.~Linde, ``{Cosmological Attractors and
  Initial Conditions for Inflation},''
  \href{http://dx.doi.org/10.1103/PhysRevD.92.063519}{{\em Phys. Rev. D}
  {\bfseries 92} no.~6, (2015) 063519},
  \href{http://arxiv.org/abs/1506.00936}{{\ttfamily arXiv:1506.00936
  [hep-th]}}.

\bibitem{Kallosh:2015zsa}
R.~Kallosh and A.~Linde, ``{Escher in the Sky},''
  \href{http://dx.doi.org/10.1016/j.crhy.2015.07.004}{{\em Comptes Rendus
  Physique} {\bfseries 16} (2015) 914--927},
  \href{http://arxiv.org/abs/1503.06785}{{\ttfamily arXiv:1503.06785
  [hep-th]}}.

\bibitem{Carrasco:2015uma}
J.~J.~M. Carrasco, R.~Kallosh, A.~Linde, and D.~Roest, ``{Hyperbolic geometry
  of cosmological attractors},''
  \href{http://dx.doi.org/10.1103/PhysRevD.92.041301}{{\em Phys. Rev. D}
  {\bfseries 92} no.~4, (2015) 041301},
  \href{http://arxiv.org/abs/1504.05557}{{\ttfamily arXiv:1504.05557
  [hep-th]}}.

\bibitem{Kallosh:2017ced}
R.~Kallosh, A.~Linde, T.~Wrase, and Y.~Yamada, ``{Maximal Supersymmetry and
  B-Mode Targets},'' \href{http://dx.doi.org/10.1007/JHEP04(2017)144}{{\em
  JHEP} {\bfseries 04} (2017) 144},
  \href{http://arxiv.org/abs/1704.04829}{{\ttfamily arXiv:1704.04829
  [hep-th]}}.

\bibitem{Odintsov:2016vzz}
S.~D. Odintsov and V.~K. Oikonomou, ``{Inflationary $\alpha$-attractors from
  $F(R)$ gravity},'' \href{http://dx.doi.org/10.1103/PhysRevD.94.124026}{{\em
  Phys. Rev. D} {\bfseries 94} no.~12, (2016) 124026},
  \href{http://arxiv.org/abs/1612.01126}{{\ttfamily arXiv:1612.01126 [gr-qc]}}.

\bibitem{Scalisi:2018eaz}
M.~Scalisi and I.~Valenzuela, ``{Swampland distance conjecture, inflation and
  $\alpha$-attractors},'' \href{http://dx.doi.org/10.1007/JHEP08(2019)160}{{\em
  JHEP} {\bfseries 08} (2019) 160},
  \href{http://arxiv.org/abs/1812.07558}{{\ttfamily arXiv:1812.07558
  [hep-th]}}.

\bibitem{Kallosh:2017wku}
R.~Kallosh, A.~Linde, D.~Roest, A.~Westphal, and Y.~Yamada, ``{Fibre Inflation
  and $\alpha$-attractors},''
  \href{http://dx.doi.org/10.1007/JHEP02(2018)117}{{\em JHEP} {\bfseries 02}
  (2018) 117}, \href{http://arxiv.org/abs/1707.05830}{{\ttfamily
  arXiv:1707.05830 [hep-th]}}.

\bibitem{Kallosh:2021vcf}
R.~Kallosh, A.~Linde, T.~Wrase, and Y.~Yamada, ``{IIB String Theory and
  Sequestered Inflation},''
  \href{http://dx.doi.org/10.1002/prop.202100127}{{\em Fortsch. Phys.}
  {\bfseries 69} no.~11-12, (2021) 2100127},
  \href{http://arxiv.org/abs/2108.08492}{{\ttfamily arXiv:2108.08492
  [hep-th]}}.

\bibitem{Kallosh:2019hzo}
R.~Kallosh and A.~Linde, ``{CMB targets after the latest Planck data
  release},'' \href{http://dx.doi.org/10.1103/PhysRevD.100.123523}{{\em Phys.
  Rev. D} {\bfseries 100} no.~12, (2019) 123523},
  \href{http://arxiv.org/abs/1909.04687}{{\ttfamily arXiv:1909.04687
  [hep-th]}}.

\bibitem{Kallosh:2019eeu}
R.~Kallosh and A.~Linde, ``{B-mode Targets},''
  \href{http://dx.doi.org/10.1016/j.physletb.2019.134970}{{\em Phys. Lett. B}
  {\bfseries 798} (2019) 134970},
  \href{http://arxiv.org/abs/1906.04729}{{\ttfamily arXiv:1906.04729
  [astro-ph.CO]}}.

\bibitem{Ferrara:2016fwe}
S.~Ferrara and R.~Kallosh, ``{Seven-disk manifold, $\alpha$-attractors, and $B$
  modes},'' \href{http://dx.doi.org/10.1103/PhysRevD.94.126015}{{\em Phys. Rev.
  D} {\bfseries 94} no.~12, (2016) 126015},
  \href{http://arxiv.org/abs/1610.04163}{{\ttfamily arXiv:1610.04163
  [hep-th]}}.

\bibitem{Planck:2018jri}
{\bfseries Planck} Collaboration, Y.~Akrami {\em et~al.}, ``{Planck 2018
  results. X. Constraints on inflation},''
  \href{http://dx.doi.org/10.1051/0004-6361/201833887}{{\em Astron. Astrophys.}
  {\bfseries 641} (2020) A10},
  \href{http://arxiv.org/abs/1807.06211}{{\ttfamily arXiv:1807.06211
  [astro-ph.CO]}}.

\bibitem{Wetterich:2004ff}
C.~Wetterich, ``{Quintessenz \textendash{} die f\"unfte Kraft},'' {\em Physik
  J.} {\bfseries 3N12} (2004) 43--48.

\bibitem{Linde:2017pwt}
A.~Linde, ``{On the problem of initial conditions for inflation},''
  \href{http://dx.doi.org/10.1007/s10701-018-0177-9}{{\em Found. Phys.}
  {\bfseries 48} no.~10, (2018) 1246--1260},
  \href{http://arxiv.org/abs/1710.04278}{{\ttfamily arXiv:1710.04278
  [hep-th]}}.

\bibitem{Kallosh:2016gqp}
R.~Kallosh and A.~Linde, ``{Cosmological Attractors and Asymptotic Freedom of
  the Inflaton Field},''
  \href{http://dx.doi.org/10.1088/1475-7516/2016/06/047}{{\em JCAP} {\bfseries
  06} (2016) 047}, \href{http://arxiv.org/abs/1604.00444}{{\ttfamily
  arXiv:1604.00444 [hep-th]}}.

\bibitem{Chojnacki:2021fag}
J.~Chojnacki, J.~Krajecka, J.~H. Kwapisz, O.~Slowik, and A.~Strag, ``{Is
  asymptotically safe inflation eternal?},''
  \href{http://dx.doi.org/10.1088/1475-7516/2021/04/076}{{\em JCAP} {\bfseries
  04} (2021) 076}, \href{http://arxiv.org/abs/2101.00866}{{\ttfamily
  arXiv:2101.00866 [gr-qc]}}.

\bibitem{Pallis:2022cnm}
C.~Pallis, ``{Pole Inflation in Supergravity},''
  \href{http://dx.doi.org/10.22323/1.406.0078}{{\em PoS} {\bfseries CORFU2021}
  (2022) 078}, \href{http://arxiv.org/abs/2208.11757}{{\ttfamily
  arXiv:2208.11757 [hep-ph]}}.

\bibitem{Karamitsos:2019vor}
S.~Karamitsos, ``{Beyond the Poles in Attractor Models of Inflation},''
  \href{http://dx.doi.org/10.1088/1475-7516/2019/09/022}{{\em JCAP} {\bfseries
  09} (2019) 022}, \href{http://arxiv.org/abs/1903.03707}{{\ttfamily
  arXiv:1903.03707 [hep-th]}}.

\bibitem{Dias:2018pgj}
M.~Dias, J.~Frazer, A.~Retolaza, M.~Scalisi, and A.~Westphal, ``{Pole
  N-flation},'' \href{http://dx.doi.org/10.1007/JHEP02(2019)120}{{\em JHEP}
  {\bfseries 02} (2019) 120}, \href{http://arxiv.org/abs/1805.02659}{{\ttfamily
  arXiv:1805.02659 [hep-th]}}.

\bibitem{Saikawa:2017wkg}
K.~Saikawa, M.~Yamaguchi, Y.~Yamashita, and D.~Yoshida, ``{Pole inflation in
  Jordan frame supergravity},''
  \href{http://dx.doi.org/10.1088/1475-7516/2018/01/031}{{\em JCAP} {\bfseries
  01} (2018) 031}, \href{http://arxiv.org/abs/1709.03440}{{\ttfamily
  arXiv:1709.03440 [hep-th]}}.

\bibitem{Let:2022fmu}
A.~Let, A.~Sarkar, C.~Sarkar, and B.~Ghosh, ``{Single-field slow-roll effective
  potential from K\"ahler moduli stabilizations in type-IIB/F-theory},''
  \href{http://dx.doi.org/10.1209/0295-5075/ac8952}{{\em EPL} {\bfseries 139}
  no.~5, (2022) 59002}, \href{http://arxiv.org/abs/2208.06606}{{\ttfamily
  arXiv:2208.06606 [hep-th]}}.

\bibitem{Bedroya:2019tba}
A.~Bedroya, R.~Brandenberger, M.~Loverde, and C.~Vafa, ``{Trans-Planckian
  Censorship and Inflationary Cosmology},''
  \href{http://dx.doi.org/10.1103/PhysRevD.101.103502}{{\em Phys. Rev. D}
  {\bfseries 101} no.~10, (2020) 103502},
  \href{http://arxiv.org/abs/1909.11106}{{\ttfamily arXiv:1909.11106
  [hep-th]}}.

\bibitem{Brandenberger:2021pzy}
R.~Brandenberger, ``{Trans-Planckian Censorship Conjecture and Early Universe
  Cosmology},'' \href{http://dx.doi.org/10.31526/lhep.2021.198}{{\em LHEP}
  {\bfseries 2021} (2021) 198},
  \href{http://arxiv.org/abs/2102.09641}{{\ttfamily arXiv:2102.09641
  [hep-th]}}.

\bibitem{Cai:2019dzj}
R.-G. Cai and S.-J. Wang, ``{A refined trans-Planckian censorship
  conjecture},'' \href{http://dx.doi.org/10.1007/s11433-020-1623-9}{{\em Sci.
  China Phys. Mech. Astron.} {\bfseries 64} no.~1, (2021) 210011},
  \href{http://arxiv.org/abs/1912.00607}{{\ttfamily arXiv:1912.00607
  [hep-th]}}.

\bibitem{Brahma:2019vpl}
S.~Brahma, ``{Trans-Planckian censorship conjecture from the swampland distance
  conjecture},'' \href{http://dx.doi.org/10.1103/PhysRevD.101.046013}{{\em
  Phys. Rev. D} {\bfseries 101} no.~4, (2020) 046013},
  \href{http://arxiv.org/abs/1910.12352}{{\ttfamily arXiv:1910.12352
  [hep-th]}}.

\bibitem{Bedroya:2019snp}
A.~Bedroya and C.~Vafa, ``{Trans-Planckian Censorship and the Swampland},''
  \href{http://dx.doi.org/10.1007/JHEP09(2020)123}{{\em JHEP} {\bfseries 09}
  (2020) 123}, \href{http://arxiv.org/abs/1909.11063}{{\ttfamily
  arXiv:1909.11063 [hep-th]}}.

\bibitem{Andriot:2020lea}
D.~Andriot, N.~Cribiori, and D.~Erkinger, ``{The web of swampland conjectures
  and the TCC bound},'' \href{http://dx.doi.org/10.1007/JHEP07(2020)162}{{\em
  JHEP} {\bfseries 07} (2020) 162},
  \href{http://arxiv.org/abs/2004.00030}{{\ttfamily arXiv:2004.00030
  [hep-th]}}.

\bibitem{Saito:2019tkc}
R.~Saito, S.~Shirai, and M.~Yamazaki, ``{Is the trans-Planckian censorship a
  swampland conjecture?},''
  \href{http://dx.doi.org/10.1103/PhysRevD.101.046022}{{\em Phys. Rev. D}
  {\bfseries 101} no.~4, (2020) 046022},
  \href{http://arxiv.org/abs/1911.10445}{{\ttfamily arXiv:1911.10445
  [hep-th]}}.

\bibitem{Draper:2019utz}
P.~Draper and S.~Farkas, ``{Transplanckian Censorship and the Local Swampland
  Distance Conjecture},'' \href{http://dx.doi.org/10.1007/JHEP01(2020)133}{{\em
  JHEP} {\bfseries 01} (2020) 133},
  \href{http://arxiv.org/abs/1910.04804}{{\ttfamily arXiv:1910.04804
  [hep-th]}}.

\bibitem{Dasgupta:2019gcd}
K.~Dasgupta, M.~Emelin, M.~M. Faruk, and R.~Tatar, ``{de Sitter vacua in the
  string landscape},''
  \href{http://dx.doi.org/10.1016/j.nuclphysb.2021.115463}{{\em Nucl. Phys. B}
  {\bfseries 969} (2021) 115463},
  \href{http://arxiv.org/abs/1908.05288}{{\ttfamily arXiv:1908.05288
  [hep-th]}}.

\bibitem{Garg:2018reu}
S.~K. Garg and C.~Krishnan, ``{Bounds on Slow Roll and the de Sitter
  Swampland},'' \href{http://dx.doi.org/10.1007/JHEP11(2019)075}{{\em JHEP}
  {\bfseries 11} (2019) 075}, \href{http://arxiv.org/abs/1807.05193}{{\ttfamily
  arXiv:1807.05193 [hep-th]}}.

\bibitem{Dasgupta:2018rtp}
K.~Dasgupta, M.~Emelin, E.~McDonough, and R.~Tatar, ``{Quantum Corrections and
  the de Sitter Swampland Conjecture},''
  \href{http://dx.doi.org/10.1007/JHEP01(2019)145}{{\em JHEP} {\bfseries 01}
  (2019) 145}, \href{http://arxiv.org/abs/1808.07498}{{\ttfamily
  arXiv:1808.07498 [hep-th]}}.

\bibitem{Baumann:2009ds}
D.~Baumann,
  \href{http://dx.doi.org/10.1142/9789814327183_0010}{``{Inflation},''} in {\em
  {Theoretical Advanced Study Institute in Elementary Particle Physics}:
  {Physics of the Large and the Small}}, pp.~523--686.
\newblock 2011.
\newblock \href{http://arxiv.org/abs/0907.5424}{{\ttfamily arXiv:0907.5424
  [hep-th]}}.

\bibitem{Bunch:1978yq}
T.~S. Bunch and P.~C.~W. Davies, ``{Quantum Field Theory in de Sitter Space:
  Renormalization by Point Splitting},''
  \href{http://dx.doi.org/10.1098/rspa.1978.0060}{{\em Proc. Roy. Soc. Lond. A}
  {\bfseries 360} (1978) 117--134}.

\bibitem{Birrell:1982ix}
N.~D. Birrell and P.~C.~W. Davies,
  \href{http://dx.doi.org/10.1017/CBO9780511622632}{{\em {Quantum Fields in
  Curved Space}}}.
\newblock Cambridge Monographs on Mathematical Physics. Cambridge Univ. Press,
  Cambridge, UK, 2, 1984.

\bibitem{Kundu:2011sg}
S.~Kundu, ``{Inflation with General Initial Conditions for Scalar
  Perturbations},'' \href{http://dx.doi.org/10.1088/1475-7516/2012/02/005}{{\em
  JCAP} {\bfseries 02} (2012) 005},
  \href{http://arxiv.org/abs/1110.4688}{{\ttfamily arXiv:1110.4688
  [astro-ph.CO]}}.

\bibitem{Jiang:2016nok}
H.~Jiang, Y.~Wang, and S.~Zhou, ``{On the initial condition of inflationary
  fluctuations},'' \href{http://dx.doi.org/10.1088/1475-7516/2016/04/041}{{\em
  JCAP} {\bfseries 04} (2016) 041},
  \href{http://arxiv.org/abs/1601.01179}{{\ttfamily arXiv:1601.01179
  [hep-th]}}.

\bibitem{BICEP2:2015nss}
{\bfseries BICEP2, Planck} Collaboration, P.~A.~R. Ade {\em et~al.}, ``{Joint
  Analysis of BICEP2/$Keck Array$ and $Planck$ Data},''
  \href{http://dx.doi.org/10.1103/PhysRevLett.114.101301}{{\em Phys. Rev.
  Lett.} {\bfseries 114} (2015) 101301},
  \href{http://arxiv.org/abs/1502.00612}{{\ttfamily arXiv:1502.00612
  [astro-ph.CO]}}.

\bibitem{Cruz:2023cxy}
J.~S. Cruz, S.~Hannestad, E.~B. Holm, F.~Niedermann, M.~S. Sloth, and T.~Tram,
  ``{Profiling Cold New Early Dark Energy},''
  \href{http://arxiv.org/abs/2302.07934}{{\ttfamily arXiv:2302.07934
  [astro-ph.CO]}}.

\bibitem{CarrilloGonzalez:2023lma}
M.~Carrillo~Gonz\'alez, Q.~Liang, J.~Sakstein, and M.~Trodden,
  ``{Neutrino-Assisted Early Dark Energy is a Natural Resolution of the Hubble
  Tension},'' \href{http://arxiv.org/abs/2302.09091}{{\ttfamily
  arXiv:2302.09091 [astro-ph.CO]}}.

\bibitem{Reboucas:2023rjm}
J.~a. Rebou\c{c}as, J.~Gordon, D.~H.~F. de~Souza, K.~Zhong, V.~Miranda,
  R.~Rosenfeld, T.~Eifler, and E.~Krause, ``{Early dark energy constraints with
  late-time expansion marginalization},''
  \href{http://arxiv.org/abs/2302.07333}{{\ttfamily arXiv:2302.07333
  [astro-ph.CO]}}.

\bibitem{Poulin:2023lkg}
V.~Poulin, T.~L. Smith, and T.~Karwal, ``{The Ups and Downs of Early Dark
  Energy solutions to the Hubble tension: a review of models, hints and
  constraints circa 2023},'' \href{http://arxiv.org/abs/2302.09032}{{\ttfamily
  arXiv:2302.09032 [astro-ph.CO]}}.

\bibitem{Goldstein:2023gnw}
S.~Goldstein, J.~C. Hill, V.~Ir\v{s}i\v{c}, and B.~D. Sherwin, ``{Canonical
  Hubble-Tension-Resolving Early Dark Energy Cosmologies are Inconsistent with
  the Lyman-$\alpha$ Forest},''
  \href{http://arxiv.org/abs/2303.00746}{{\ttfamily arXiv:2303.00746
  [astro-ph.CO]}}.

\bibitem{DiValentino:2021izs}
E.~Di~Valentino, O.~Mena, S.~Pan, L.~Visinelli, W.~Yang, A.~Melchiorri, D.~F.
  Mota, A.~G. Riess, and J.~Silk, ``{In the realm of the Hubble
  tension\textemdash{}a review of solutions},''
  \href{http://dx.doi.org/10.1088/1361-6382/ac086d}{{\em Class. Quant. Grav.}
  {\bfseries 38} no.~15, (2021) 153001},
  \href{http://arxiv.org/abs/2103.01183}{{\ttfamily arXiv:2103.01183
  [astro-ph.CO]}}.

\bibitem{Ben-Dayan:2023rgt}
I.~Ben-Dayan and U.~Kumar, ``{Emergent Unparticles Dark Energy can restore
  cosmological concordance},''
  \href{http://arxiv.org/abs/2302.00067}{{\ttfamily arXiv:2302.00067
  [astro-ph.CO]}}.

\bibitem{Riess:2021jrx}
A.~G. Riess {\em et~al.}, ``{A Comprehensive Measurement of the Local Value of
  the Hubble Constant with 1 km s$^{−1}$ Mpc$^{−1}$ Uncertainty from the
  Hubble Space Telescope and the SH0ES Team},''
  \href{http://dx.doi.org/10.3847/2041-8213/ac5c5b}{{\em Astrophys. J. Lett.}
  {\bfseries 934} no.~1, (2022) L7},
  \href{http://arxiv.org/abs/2112.04510}{{\ttfamily arXiv:2112.04510
  [astro-ph.CO]}}.

\bibitem{ACTPol:2014pbf}
{\bfseries ACTPol} Collaboration, S.~Naess {\em et~al.}, ``{The Atacama
  Cosmology Telescope: CMB Polarization at $200<\ell<9000$},''
  \href{http://dx.doi.org/10.1088/1475-7516/2014/10/007}{{\em JCAP} {\bfseries
  10} (2014) 007}, \href{http://arxiv.org/abs/1405.5524}{{\ttfamily
  arXiv:1405.5524 [astro-ph.CO]}}.

\bibitem{SPT:2015htm}
{\bfseries SPT} Collaboration, R.~Keisler {\em et~al.}, ``{Measurements of
  Sub-degree B-mode Polarization in the Cosmic Microwave Background from 100
  Square Degrees of SPTpol Data},''
  \href{http://dx.doi.org/10.1088/0004-637X/807/2/151}{{\em Astrophys. J.}
  {\bfseries 807} no.~2, (2015) 151},
  \href{http://arxiv.org/abs/1503.02315}{{\ttfamily arXiv:1503.02315
  [astro-ph.CO]}}.

\bibitem{Henderson:2015nzj}
S.~W. Henderson {\em et~al.}, ``{Advanced ACTPol Cryogenic Detector Arrays and
  Readout},'' \href{http://dx.doi.org/10.1007/s10909-016-1575-z}{{\em J. Low
  Temp. Phys.} {\bfseries 184} no.~3-4, (2016) 772--779},
  \href{http://arxiv.org/abs/1510.02809}{{\ttfamily arXiv:1510.02809
  [astro-ph.IM]}}.

\bibitem{SPT-3G:2014dbx}
{\bfseries SPT-3G} Collaboration, B.~A. Benson {\em et~al.}, ``{SPT-3G: A
  Next-Generation Cosmic Microwave Background Polarization Experiment on the
  South Pole Telescope},'' \href{http://dx.doi.org/10.1117/12.2057305}{{\em
  Proc. SPIE Int. Soc. Opt. Eng.} {\bfseries 9153} (2014) 91531P},
  \href{http://arxiv.org/abs/1407.2973}{{\ttfamily arXiv:1407.2973
  [astro-ph.IM]}}.

\bibitem{LiteBIRD:2020khw}
{\bfseries LiteBIRD} Collaboration, M.~Hazumi {\em et~al.}, ``{LiteBIRD: JAXA's
  new strategic L-class mission for all-sky surveys of cosmic microwave
  background polarization},'' \href{http://dx.doi.org/10.1117/12.2563050}{{\em
  Proc. SPIE Int. Soc. Opt. Eng.} {\bfseries 11443} (2020) 114432F},
  \href{http://arxiv.org/abs/2101.12449}{{\ttfamily arXiv:2101.12449
  [astro-ph.IM]}}.

\bibitem{Heymans:2012gg}
C.~Heymans {\em et~al.}, ``{CFHTLenS: The Canada-France-Hawaii Telescope
  Lensing Survey},''
  \href{http://dx.doi.org/10.1111/j.1365-2966.2012.21952.x}{{\em Mon. Not. Roy.
  Astron. Soc.} {\bfseries 427} (2012) 146},
  \href{http://arxiv.org/abs/1210.0032}{{\ttfamily arXiv:1210.0032
  [astro-ph.CO]}}.

\bibitem{Hildebrandt:2016iqg}
H.~Hildebrandt {\em et~al.}, ``{KiDS-450: Cosmological parameter constraints
  from tomographic weak gravitational lensing},''
  \href{http://dx.doi.org/10.1093/mnras/stw2805}{{\em Mon. Not. Roy. Astron.
  Soc.} {\bfseries 465} (2017) 1454},
  \href{http://arxiv.org/abs/1606.05338}{{\ttfamily arXiv:1606.05338
  [astro-ph.CO]}}.

\bibitem{Kohlinger:2017sxk}
F.~K\"ohlinger {\em et~al.}, ``{KiDS-450: The tomographic weak lensing power
  spectrum and constraints on cosmological parameters},''
  \href{http://dx.doi.org/10.1093/mnras/stx1820}{{\em Mon. Not. Roy. Astron.
  Soc.} {\bfseries 471} no.~4, (2017) 4412--4435},
  \href{http://arxiv.org/abs/1706.02892}{{\ttfamily arXiv:1706.02892
  [astro-ph.CO]}}.

\bibitem{Dawson:2015wdb}
K.~S. Dawson {\em et~al.}, ``{The SDSS-IV extended Baryon Oscillation
  Spectroscopic Survey: Overview and Early Data},''
  \href{http://dx.doi.org/10.3847/0004-6256/151/2/44}{{\em Astron. J.}
  {\bfseries 151} (2016) 44}, \href{http://arxiv.org/abs/1508.04473}{{\ttfamily
  arXiv:1508.04473 [astro-ph.CO]}}.

\bibitem{DES:2015gax}
{\bfseries DES} Collaboration, T.~Abbott {\em et~al.}, ``{Cosmology from cosmic
  shear with Dark Energy Survey Science Verification data},''
  \href{http://dx.doi.org/10.1103/PhysRevD.94.022001}{{\em Phys. Rev. D}
  {\bfseries 94} no.~2, (2016) 022001},
  \href{http://arxiv.org/abs/1507.05552}{{\ttfamily arXiv:1507.05552
  [astro-ph.CO]}}.

\bibitem{DESI:2016fyo}
{\bfseries DESI} Collaboration, A.~Aghamousa {\em et~al.}, ``{The DESI
  Experiment Part I: Science,Targeting, and Survey Design},''
  \href{http://arxiv.org/abs/1611.00036}{{\ttfamily arXiv:1611.00036
  [astro-ph.IM]}}.

\bibitem{DESI:2016igz}
{\bfseries DESI} Collaboration, A.~Aghamousa {\em et~al.}, ``{The DESI
  Experiment Part II: Instrument Design},''
  \href{http://arxiv.org/abs/1611.00037}{{\ttfamily arXiv:1611.00037
  [astro-ph.IM]}}.

\bibitem{LSSTScience:2009jmu}
{\bfseries LSST Science, LSST Project} Collaboration, P.~A. Abell {\em et~al.},
  ``{LSST Science Book, Version 2.0},''
  \href{http://arxiv.org/abs/0912.0201}{{\ttfamily arXiv:0912.0201
  [astro-ph.IM]}}.

\bibitem{LSST:2017ags}
{\bfseries LSST} Collaboration, P.~Marshall {\em et~al.}, ``{Science-Driven
  Optimization of the LSST Observing Strategy},''
  \href{http://arxiv.org/abs/1708.04058}{{\ttfamily arXiv:1708.04058
  [astro-ph.IM]}}.

\bibitem{Spergel:2015sza}
D.~Spergel {\em et~al.}, ``{Wide-Field InfrarRed Survey Telescope-Astrophysics
  Focused Telescope Assets WFIRST-AFTA 2015 Report},''
  \href{http://arxiv.org/abs/1503.03757}{{\ttfamily arXiv:1503.03757
  [astro-ph.IM]}}.

\bibitem{Hounsell:2017ejq}
R.~Hounsell {\em et~al.}, ``{Simulations of the WFIRST Supernova Survey and
  Forecasts of Cosmological Constraints},''
  \href{http://dx.doi.org/10.3847/1538-4357/aac08b}{{\em Astrophys. J.}
  {\bfseries 867} no.~1, (2018) 23},
  \href{http://arxiv.org/abs/1702.01747}{{\ttfamily arXiv:1702.01747
  [astro-ph.IM]}}.

\bibitem{Amendola:2016saw}
L.~Amendola {\em et~al.}, ``{Cosmology and fundamental physics with the Euclid
  satellite},'' \href{http://dx.doi.org/10.1007/s41114-017-0010-3}{{\em Living
  Rev. Rel.} {\bfseries 21} no.~1, (2018) 2},
  \href{http://arxiv.org/abs/1606.00180}{{\ttfamily arXiv:1606.00180
  [astro-ph.CO]}}.

\end{thebibliography}\endgroup

\end{document}